\documentstyle[12pt]{article}
\begin{document}

\begin{flushright}
OUTP-96-10P \\
CERN-TH/96-162\\
hep-lat/9606021\\
\end{flushright}
\vskip 1em
\begin{center}
{\large \bf Domain walls and perturbation theory in high temperature 
gauge theory:  SU(2) in 2+1 dimensions\\}
\vspace{3ex}
\vspace{3ex}
{\large  C. Korthals Altes$^1$, A. Michels$^2$, M. Stephanov$^3$, 
M. Teper$^2$ \\}
\vspace{3ex}
{\it $^1$ Centre Physique Theorique au CNRS, Luminy,
                B.P.~907, 13288\\  Marseille, France\\}
\vspace{3ex}
{\it $^2$ Theoretical Physics, University of Oxford,
	 1 Keble Road,\\ Oxford OX1 3NP, UK\\}
\vspace{3ex}
{\it $^3$ Department of Physics, University of Illinois
                at Urbana--Champaign,\\
	 1110 West Green Street, Urbana 61801, USA\\}

\vspace{3ex}

\vspace{3ex}
{\large \bf Abstract \\}
\end{center}
\vspace{3ex}
{ We study the detailed properties of $Z_2$ domain walls in the 
deconfined high temperature phase of the $d=2+1$ $SU(2)$ gauge theory.
These walls are studied
both by computer simulations of the lattice theory
and by one-loop perturbative calculations. The latter are carried
out both in the continuum and on the lattice.
We find that leading order perturbation theory reproduces
the detailed properties of these domain walls
remarkably accurately even at temperatures where the effective
dimensionless expansion parameter, $g^2/T$, is close to unity.
The quantities studied include the surface tension, the action
density profiles, roughening and the electric screening mass. 
It is only for the last quantity that we find an
exception to the precocious success of perturbation theory.
All this shows that, despite the presence of infrared divergences
at higher orders, high-$T$ perturbation theory can be an
accurate calculational tool.} 

\newpage

\section{Introduction}

Non-Abelian gauge theories possess many surprising aspects. An
example is the linear confining potential present 
in both the three and four dimensional $SU(N)$ theories at low
temperature, $T$. At some finite $T$ there is a phase transition
and confinement is then lost\cite{Su79,Po78}; but this is not
unexpected because simple energy versus entropy arguments 
tell us that a confining `flux tube'
will condense into the vacuum at some finite value of the temperature.
These phenomena are non-perturbative and have so far defied
analytic, as opposed to numerical, approaches. However,
at sufficiently high $T$, there would appear to be an important 
theoretical simplification. All these theories are asymptotically free  
so that the effective interaction on the relevant energy scale, $T$, 
should become small at high $T$ and
the physics of the gluon plasma should become accurately 
calculable in perturbation theory. Unfortunately there are 
infrared  divergences in higher orders of perturbation theory,
which are associated with the perturbative masslessness of
the magnetic gluon. Although we expect this gluon to acquire a mass
through the non-perturbative physics of the dimensionally
reduced theory, this leaves room for uncertainty about how reliable
high-$T$ perturbation theory really is.

However even this naive picture of gauge theories at high $T$ ---
as a weakly interacting plasma of gluons --- contains surprises.
There turns out to be a symmetry associated with the centre
of the group, $Z(N)$, which is spontaneously broken at high $T$.
Separating two different $Z(N)$ vacua will be a domain wall
whose properties have been calculated in perturbation theory
for $T \to \infty$ \cite{BhGo91B}. However the reality of these domain
walls is controversial for a variety of reasons. Firstly, there
are the general doubts about high-$T$ perturbation theory that we alluded
to above. Secondly, all this is in the usual Euclidean formulation
of finite temperature field theories and there is a question of
what if anything these domain walls might correspond to in Minkowski
space-time. Finally the walls have peculiar thermodynamic properties,
which become more acute when one includes quarks into the theory 
\cite{BeKo92}.
It is important to resolve these uncertainties not only because
of the theoretical interest of these domain walls, but also
because related structures can be associated with important physical phenomena 
when one considers the Standard Model in the early universe
\cite{LeKo94}.

In this paper we address the particular problem associated with
the uncertain status of perturbation theory at high $T$. We
shall do so by calculating the properties of the domain walls
both in perturbation theory and by a fully non-perturbative
Monte Carlo computer simulation. We shall work with the $SU(2)$ 
gauge group because the problems should not be any different
for larger groups. Moreover we shall work in 2+1 dimensions
rather than in the more physical case of 3+1 dimensions. The reason
is that the computational resources needed are much less in
the former case, and only there can we perform calculations
with enough precision and control to be really useful. 
At the same time, the origin of the infrared problems is similar in
$d=2+1$ and in $d=3+1$, and their severity is, if anything, greater
in the lower dimensional case. 

We shall find in the calculations described below that 
one-loop perturbation theory does indeed work remarkably --- even 
precociously --- well for our $Z_2$ domain walls. By implication, this 
provides evidence for the general applicability of perturbation
theory at high $T$. 

At the same time we emphasise that we make no attempt,
in this paper, to address the other controversial aspects of domain 
walls and so do not attempt to settle the interesting
question of their potential role in, for example, separating bubbles
of different vacua in the early universe.

We shall now outline the contents of this paper. In Section 2 we
give a heuristic introduction to domain walls and the thermodynamics
of gauge theories at high temperatures. This will provide the
general background for the more detailed and specific calculations
of the later sections. We then turn to the perturbative calculation, 
at one-loop, of the properties of domain walls. In Section
\ref{sec:cont} we perform 
the calculation of the surface tension for the continuum theory. The
calculation is by a method that easily extends to other dimensions
and so, as well as obtaining expressions for $d=2+1$, we can
compare with previous $d=3+1$ calculations. In order to compare
with our later numerical work it is useful to have similar
results for a finite value of the lattice spacing $a$. This
calculation is carried out in Section \ref{sec:tenslat}. 
In Section \ref{sec:profs} we calculate
the action density in the domain wall, as a function of the
distance from the centre of the wall, since this is one of the
quantities we shall later calculate numerically. In 
Section \ref{sec:rough} we calculate the effects of oscillations of 
the wall 
once its length is large --- the `roughening' of the wall, which has 
been neglected in most previous studies. In Section \ref{sec:Lz} we 
show how the finite size of the volume, in the direction
orthogonal to the wall, affects the profile
and surface tension of the wall. This is important to understand
since our numerical work will necessarily be on lattices
of a finite size. We then turn, in Section 7, to a description 
of our simulations and the results we obtain thereby. We begin,
in Section 7.1, by describing how we can simulate domain
walls through the use of twisted boundary conditions.
In Section 7.2 we point out that these domain walls can be viewed as
't Hooft disorder loops that have been squashed by
the short Euclidean time direction. Section 7.3 describes
how we calculate the properties of the domain wall
and the electric screening mass. Section 7.4 lists the
large-volume raw `data' from which we will eventually
extract physical quantities. Section 7.5 describes in
detail how we control finite volume effects. This is
crucial since  the potential problems with perturbation
theory are infrared ones. Finally, in Section 7.6
we compare our numerical results with those of
perturbation theory. Section 8 contains our conclusions.

Our theoretical analysis, in Sections 3 to 6, is performed
for the general case of $SU(N)$ gauge fields in $2 < d \le 4$
dimensions. Our numerical results, on the other hand, are for 
the particular case of $SU(2)$ gauge fields in $d=2+1$. The
preliminary results of this study appeared in the Proceedings
of the 1994 Lattice Conference \cite{KoMi95}. A study of
the case of $SU(3)$ gauge fields in $d=2+1$ has recently
been reported \cite{WeWh96}. Both the $SU(2)$ and $SU(3)$
results are in agreement with our theoretical analysis.
In addition both show the same large deviation from
D'Hoker's self-consistent formula for the Debye mass.  

\section{General considerations}

In 2+1 dimensions the gauge coupling, $g^2$, is dimensionful and 
its value sets the mass scale for the theory. 
If we perform a perturbative calculation of a quantity
in which there is a dominant momentum scale, $Q$, then the effective
expansion parameter will clearly be $g^2/Q$ so that the theory 
rapidly becomes free at short distances. So at high $T$ the effective
expansion parameter, $g^2/T$, will be small and we can expect that 
we should be able to apply perturbation theory. All this is
very similar to the case in 3+1 dimensions. There the coupling
is dimensionless but this difference is only apparent: the
scale-invariance is anomalous, the coupling runs and its value
only serves to set the overall mass scale (dimensional transmutation).
The coupling becomes small at short distances, so that its value
at high temperature, $g^2(T)$, should be small enough for us to
apply perturbation theory. In other ways the two theories are
also similar: numerical simulations \cite{Tep1} show that
the $d=2+1$ theory has linear confinement, a deconfining temperature,
$T_c$, and a glueball spectrum that are similar in many ways
to that of the theory in 3+1 dimensions.

So we expect a hot gauge theory to be, to a very good approximation, 
a plasma of free gluons, with interactions given in terms of the 
small effective coupling at the ambient temperature $T$. These gluons
are screened just as are photons in a plasma of charged particles.
The Debye screening mass, $m_D$, that they acquire grows with $T$;
to lowest order ${m^2}_D$ is $O(g^2)$ so we expect, on purely 
dimensional grounds, that $m_D \sim g(T)T$ in $d=3+1$ and 
$m_D \sim g T^{1/2}$ in $d=2+1$. 

However this simple picture is not the whole story. An additional 
and important role is played by the center, $Z(N)$, of the gauge symmetry
group $SU(N)$. This sub-group has a special status because the gluons, 
which transform according to the adjoint representation, are
invariant under gauge transformations that belong to the centre. 
Since the gluons only feel $SU(N)$ gauge transformations
modulo any $Z(N)$ transformation, their symmetry group is really
$SU(N)/Z(N)$. Sources in the fundamental representation, on the
other hand, are not invariant under transformations in $Z(N)$.
Consider such a heavy source in the usual Euclidean space-time
formulation of the high-$T$ field theory where the Euclidean time
is periodic with period $1/T$. As is well known the presence of 
such a static heavy source leads to an extra factor of 
$P\equiv (1/N){\rm\ tr}{\rm\ P}\exp\{i\int A_0 dt\}$
--- a Polyakov loop --- in the partition function. In the low $T$
confining phase  $\langle P \rangle =0$ while at high $T$
$\vert \langle P \rangle \vert \simeq 1$. Since the gluons
are screened, the correlations are short range at high $T$ 
and so if, for example, $P$ is close to 1 at one point, the vacuum 
will have $P \simeq 1$ everywhere. But the $Z(N)$ symmetry tells
us that there must be $N$ such vacua, in each of which the
physics is identical, and which are differentiated by $P$ 
being close to one of the $N$ complex $N$'th roots of unity. The
picking of one of these vacua corresponds to the spontaneous 
breaking of the symmetry. That the spontaneous breaking
takes place at high $T$ rather than at low $T$ is remarkable
but not impossible; it is, for example, a commonplace in
(self-)dual theories. 

As soon as we have spontaneous symmetry breaking we have the
possibility of domain walls which will occur at the interface between
two of the vacua. (Throughout this paper we shall call these objects 
`walls', and will speak of their `surface' tension, even though
the interface is really a string when we are in 2 spatial 
dimensions. This is to avoid confusion with the confining string.)
In the case of interest to us, $SU(2)$, such
a domain wall would separate regions of our Euclidean space-time
volume that are characterised by $P \simeq +1$ on the one side
and $P \simeq -1$ on the other. One can compute the free energy 
density of the wall for small coupling and one indeed finds 
a positive excess over the free energy of the gluon plasma.

Thus we have a picture of the gluon plasma that parallels that of a 
ferro-magnetic substance below the Curie temperature: the average of
the Polyakov loop, which arises from the heavy fundamental 
source, being the order parameter. But what is the analogue
of the external field? Such a field is needed in order
to make the system choose a particular direction of magnetisation
on a macroscopic scale. Without such an analogue one cannot trigger 
a thermodynamical state where our order parameter, $P$, takes the
value $+1$  or one where it takes the value $-1$. 

As we remarked in the Introduction, the `reality' of these 
domain walls is controversial. 
Of course, the walls have been observed \cite{KaKa91}
in ($d=3+1$) Monte Carlo simulations of the theory, but this
has only been for large lattice spacings, and there have been
speculations that the walls would not survive into the
continuum limit \cite{Sm94}. In this paper
we shall address the question of whether these domain
walls do indeed exist in the continuum limit 
and, more specifically, whether high-$T$ perturbation
theory is reliable. But first, in this section, we introduce
the basic framework within which we work.

\subsection{Yang-Mills fields: the basic parameters}

We consider a pure $SU(N)$ gauge theory in $2$ or $3$ space dimensions. 
We will concentrate in this subsection on the thermodynamic 
quantities that one can define by enclosing the system in a box of 
size $L_yL_z$. (For the three-dimensional case we add an 
$x$-direction.) The Hamiltonian for this system reads:
\begin{equation}
\label{eqno2.1}
\hat H\equiv{1\over 2}\int_{\mbox{\boldmath $x$}}
\mbox{ tr} \bigg (
g^2\hat{\mbox{\boldmath $E$}}^2
+ {1\over g^2}\hat{\mbox{\boldmath $B$}}^2\bigg ),
\end{equation}
where $\hat{\mbox{\boldmath $E$}}$ is the canonical momentum for 
$\hat{\mbox{\boldmath$A$}}$ and 
$\hat B_{mn} \equiv \partial_m \hat A_n - \partial_n \hat A_m +
i[\hat A_m,\hat A_n]$.
We use the standard notation for the fields as $N\times N$ matrices 
in the Lie algebra of the defining representation of $SU(N)$.

We introduce the free energy $F$ of the system in a heat bath at 
temperature $T$ through the Gibbs trace over the physical states
of the system. By `physical' we mean that the states obey the 
Gauss constraint
\begin{equation}
\label{eqno2.2}
\left(\mbox{\boldmath $\nabla$} \hat{\mbox{\boldmath $E$}}+ 
i[\hat{\mbox{\boldmath$A$}},
\hat{\mbox{\boldmath $E$}}]\right) |\psi\rangle = 0.
\end{equation}
The free energy is defined by
\begin{equation}
\label{eqno2.3}
\exp\left({-{F\over T}}\right)={\rm Tr}_{\rm phys}\exp
\left({-{\hat H\over T}}\right).
\end{equation}
As is well known, the Gibbs trace can be related to the Feynman 
path integral by
\begin{equation}
\label{eqno2.4}
{\rm Tr}_{\rm phys}\exp\left({-{\hat H\over T}}\right)=\int {\cal D}A_0 
{\cal D}\mbox{\boldmath$A$} \exp\left\{{-{1\over {g^2}}S(A)}\right\}.
\end{equation}
Here the integration over $A_0$ implements the Gauss law, the action $S(A)$ 
is equal to 
$(1/ 4)\int d\mbox{\boldmath $x$} dt \mbox{ tr} 
F_{\mu\nu}F_{\mu\nu}$, 
and the temperature enters the formalism through the fact that
we make the potentials periodic in the Euclidean time $t$ with
period $1/T$. The relationship (\ref{eqno2.4}) is fundamental
in that it allows us to calculate the free energy using the
whole panoply of methods available for calculating path integrals;
such as perturbation theory(see Sections 3 and 4) and Monte Carlo
methods (see Section 7).  

Throughout this paper we suppose the potentials to be periodic in 
the spatial directions. However, this does not mean that any gauge 
transformation ${\Lambda}$ has to be periodic, only that it should
be periodic modulo an element, $z_k=\exp\{ik{2\pi\over N}\}$,
of the centre $Z(N)$ of $SU(N)$. That this is so is easily seen from 
the transformation properties of the potential:
\begin{equation}
\label{eqno2.5}
A_{\mu}^{{\Lambda}}={\Lambda}^{\dagger} A_{\mu} {\Lambda}-i 
{\Lambda}^{\dagger}\partial_{\mu}{\Lambda}.
\end{equation}
If say ${\Lambda}(L_z)={\Lambda}(0) z_k$, then the $z_k$ will 
commute with all the matrices in (\ref{eqno2.5}) and so will disappear 
from the right hand side, leaving the transformed potential still 
periodic \cite{tH79}. We shall denote such gauge transformations
by $\hat{\Lambda}_{\mbox{\boldmath $k$}}$. If 
$\mbox{\boldmath $k$} = (k_y,k_z)$ then this is a
gauge transformation which is periodic up to $z_{k_y}$ in the
$y$-direction and up to $z_{k_z}$ in the $z$-direction.

The interest of these extended gauge transformations lies in two facts:
\begin{itemize}
\item{they leave the Hamiltonian (\ref{eqno2.1}) invariant;}
\item{they serve to distinguish subspaces, in the space of physical states, 
which possess a given number, $e_y$ and $e_z$, of electric fluxes in the 
$y$ and $z$ directions respectively. Clearly this distinction is modulo $N$.}
\end{itemize}
The above notion of electric flux is developed in detail in \cite{tH79}. 
The following remarks represent no more than a heuristic outline. Suppose 
first that we have opposite fundamental sources at 
$\mbox{\boldmath $x$}_1$ and $\mbox{\boldmath $x$}_2$. 
To make this system gauge invariant we need to join the sources
by a finite string ${\rm\ P}\exp\{
i\int_{\mbox{\boldmath $x$}_1}^{\mbox{\boldmath $x$}_2}
\hat{\mbox{\boldmath$A$}}
d\mbox{\boldmath $x$}\}$ running between them. Such a string 
operator creates the unit fundamental (electric) flux 
that must flow between the sources.
Suppose we are now in the purely gluonic system with no sources and 
suppose we wish to add a unit of electric flux across the whole 
volume, running in the $z$-direction. From the above we expect that we 
can do so by applying to our state the periodic string operator
${\rm\ tr}{\rm\ P}\exp\{i\int_0^{L_z}dz \hat A_z\}$. 
Unlike a contractible string operator loop, which would represent some
local excitation, this operator will clearly
feel the centre element of the gauge transformation 
$\hat{\Lambda}_{\mbox{\boldmath $k$}}$ if $k_z \not= 0$. 
Indeed it is easy to see that it
will acquire a factor of $z_{k_z}$. If we create a state with
$n$ such units of electric flux in the $z$-direction, then it
will acquire a factor of $(z_{k_z})^n$. So if we label a state
with electric flux $\mbox{\boldmath $e$} 
= (e_y,e_z)$ by $\vert \mbox{\boldmath $e$} \rangle$,
it will be an eigenstate of $\hat {\Lambda}_{\mbox{\boldmath $k$}}$ 
with eigenvalue
$\exp\{i{\mbox{\boldmath $k$}}\mbox{\boldmath $\cdot$}
{\mbox{\boldmath $e$}}{2\pi\over N}\}$, (assuming trivial
transformation properties for the fluxless state $\vert{\bf 0}\rangle$).
Clearly the state with $N+1$ fluxes has the same transformation
property as that with 1 unit of flux; as one would expect
in a non-Abelian theory.  
 
Since the Hamiltonian is invariant under 
$\hat {\Lambda}_{\mbox{\boldmath $k$}}$, the
energy eigenstates can be simultaneously labeled by $\mbox{\boldmath $e$}$
and we can define a free energy $F_{\mbox{\boldmath $e$}}$ by restricting the
Gibbs trace in (\ref{eqno2.3}) to a given electric flux sector:
\begin{equation}
\label{eqno2.5a}
\exp\left( -\frac{F_{\mbox{\boldmath $e$}}}{T}\right) 
={\rm Tr}_{\mbox{\boldmath $e$}}
\exp\left({-{\hat H\over T}}\right)
\end{equation}
With this definition relation (\ref{eqno2.3}) can be rewritten
as \cite{tH79}:
\begin{equation}
\label{eqno2.6}\label{Fe-Zk}
\exp\left({-{F_{\mbox{\boldmath $e$}}\over T}}\right)={1\over
N^{d-1}}\sum_{\mbox{\boldmath $k$}}\exp\left\{-i\mbox{\boldmath
$k$}\mbox{\boldmath $\cdot$}\mbox{\boldmath$e$}{2\pi\over
N}\right\}Z_{\mbox{\boldmath $k$}},
\end{equation}
where
\begin{equation}
\label{Zk-HOmega}
Z_{\mbox{\boldmath $k$}} \equiv 
{\rm Tr}_{\rm phys}\exp\left({-{\hat H\over
T}}\hat{\Lambda}_{\mbox{\boldmath $k$}}\right).
\end{equation}
The Gibbs traces $Z_{\mbox{\boldmath $k$}}$ in (\ref{Zk-HOmega}) can
be expressed as ``twisted'' path integrals and can be computed using
Monte Carlo methods. These path integrals have the following defining
property. Suppose, for example, that ${\mbox{\boldmath $k$}} =
(0,k_z)$.  Then due to the occurrence of the transformation
$\hat{\Lambda}_{\mbox{\boldmath $k$}}$ in (\ref{Zk-HOmega}) one picks up
a factor of $\exp\{ik_z{2\pi/N}\}$ in the gauge transform relating
$A(t,y,0)$ to $A(t,y,L_z)$ and $A(0,y,z)$ to $A(L_t,y,z)$, after going
around the boundary of the box in the $z$-$t$ directions. This
multivaluedness does not affect the gluon field $A_\mu$.  In Section 7
we will perform Monte Carlo simulations of such a twisted partition
function.

\subsection{The two important parameters: string and surface tension}

Now we are ready to give a thermodynamic characterisation  of 
the various phases of the gauge theory, through the behaviour of 
the flux free energies $F_{\mbox{\boldmath $e$}}$  as the temperature 
is varied.
To obtain simple formulas we will restrict ourselves to $SU(2)$, 
but the end results will be valid for any $SU(N)$. The interesting 
quantity is the flux free energy $F_{01}$ in the elongated
direction. We want to compare it to the flux free energy $F_{00}$ 
and see how the difference behaves for low and high temperatures. 
This is straightforward using (\ref{eqno2.6}) for the flux free 
energies. But we still need some theoretical input on how the twisted 
functionals behave. What one finds for low $T$ is that
\begin{equation}
\label{eqno2.7}
1-{Z_{01}\over {Z_{00}}}=C \exp\left( -\frac{\rho(T)}{T}L_z\right),
\end{equation}
while $1-(Z_{k_y1}/{Z_{00}})$ is exponentially smaller for
any $k_y\ne 0$. On the other hand for high enough $T$ 
one finds:
\begin{equation}
\label{eqno2.8}
Z_{01}=D\exp\left(-\frac{\sigma(T)}{T}L_y\right)Z_{00}
\end{equation}
and $Z_{k_y1}$ is exponentially smaller for any 
$k_y\ne 0$. The $C$ and $D$ are some pre-exponential factors.
The evidence for (\ref{eqno2.7}) and (\ref{eqno2.8}) 
comes from Monte Carlo simulations, as well as analytic Hamiltonian analyses
of gauge Potts models~\cite{GoSh82}. So one gets for the free energy 
difference:
\begin{equation}
\label{eqno2.9}
F_{01}-F_{00} \sim \rho(T)L_z\qquad\mbox{if $T$ is small}
\end{equation}
\begin{equation}
\label{eqno2.10}
F_{01}-F_{00} \sim L_z
\exp\left\{-\frac{\sigma(T)}{T}L_y\right\} \qquad \mbox{if $T$ is large}
\end{equation}
Clearly what (\ref{eqno2.9}) is telling us is that at low temperatures
we are in a confining regime, where imposing unit electric flux
across the lattice costs us an energy that is proportional to
the length traversed by the flux; and the tension of this flux `string' is 
$\rho$. So the free energy difference becomes very large at large $L_z$, 
but is insensitive to the transverse spatial dimensions. At some critical 
temperature $T_c$ this behaviour changes into that of (\ref{eqno2.10}). 
The free energy difference now becomes exponentially small with the 
transverse size. This behaviour 
suggests that there is a wall, with an energy density independent of 
the transverse direction, and a total energy proportional to $\sigma$. 
This quantity has been computed in a semiclassical approximation 
\cite{BhGo91B,BhGo92} at very high temperatures, where perturbation 
theory should apply. The purpose of the present study is to check 
the validity of this approximation using Monte Carlo methods. 

The way the surface tension enters the free energy difference is 
through the exponential. This was first noticed by Bhattacharya et
al.  \cite{BhGo91} and was recently discussed in a $Z(2)$ gauge 
model \cite{Ki95}. It is reminiscent of an energy difference induced 
by tunneling. As we will see in the next section this is indeed a  
tunneling through a potential that arise from quantum one-loop effects.

\subsection{Effective Action and Polyakov Loops}

As we have seen above, the fundamental quantity of interest is 
the ratio of twisted path integrals $Z_{\mbox{\boldmath $k$}}$. 
These are computed 
by converting from the vector potentials, as integration
variables, to Polyakov loops:
\begin{equation}
\label{eqno2.13}\label{Omega}
{\Omega}(\mbox{\boldmath $x$})\equiv {\rm\ P}
\exp\left\{i\int_0^{1/T}dt A_0(t,\mbox{\boldmath $x$})\right\}
\end{equation}
and by integrating out the remaining variables to get an effective 
action for ${\Omega}(\mbox{\boldmath $x$})$. In a suggestive notation:
\begin{equation}
\label{eqno2.15}
\exp\{-S_{\rm eff}({\Omega})\}
=\int {\cal D}A\:\delta\left[{\Omega}-{\rm\ P}
\exp\left\{i\int_0^{1/T}dt A_0\right\}\right]\:
\exp\left\{-{1\over g^2}S(A)\right\}
\end{equation}
The reader, in looking at this equation, should keep in mind that
only the eigenvalues $\lambda_i$ of the loop ${\Omega}$ are gauge
invariant. So it is only
these that should appear in the delta function constraint and in
$S_{\rm eff}$. Note also the relation between ${\Omega}$ and $P$:
\begin{equation}
\label{P-Omega}
P = {1\over N} {\rm\ tr} {\Omega} = {1\over N} \sum_i \lambda_i.
\end{equation}
The  $S_{\rm eff}$ has been worked out \cite{BhGo91,BhGo92,Ko94} to two 
loop order in the $d=3+1$ case. In Section \ref{sec:cont} we will derive 
the one loop result for any $d$, and in particular for
the case of interest in this paper, $d=2+1$. 

It is important to note 
that the effective action does not depend on the boundary conditions
if the volume is large enough. It is also easy to see that the twist
$\mbox{\boldmath $k$}=(0,k_z)$
of the previous section corresponds to the
following boundary conditions in the effective theory for ${\Omega}(y,z)$:
\begin{equation}
{\Omega}(y,L_z) = \exp\{ik_z{2\pi\over N}\}{\Omega}(y,0).
\end{equation}
With such boundary conditions $Z_{\mbox{\boldmath $k$}}$ is given by:
\begin{equation}
\label{eqno2.14}\label{Z-Omega}
Z_{\mbox{\boldmath $k$}}
=\int_{(\mbox{\boldmath $k$})}{\cal D}{\Omega}\:
\exp\left(-S_{\rm eff}({\Omega})\right).
\end{equation}

We also note that the path integral has a formal resemblance to 
that of a spin model partition function. The effective action, when 
evaluated in perturbation theory, will start with a classical kinetic 
term. The first non-zero contribution for constant
${\Omega}(\mbox{\boldmath $x$})$
appears at one-loop. So to this order we can write:
\begin{equation}
\label{eqno2.16}\label{Seff-Omega}
S_{\rm eff}({\Omega})
= \int_{\mbox{\boldmath $x$}} {\rm tr} \left[
{T\over{2g^2}}|\mbox{\boldmath $\nabla$} {\Omega}|^2
+ V_{\rm eff}({\Omega})\right],
\end{equation}
At high $T$ the coefficient of the gradient term is large and we can
expect that the path integral will be saturated by smooth 
configurations of Polyakov loops. We shall see in the next
section that $V_{\rm eff}$ reaches its minima when all $N$ eigenvalues
of ${\Omega}$ coincide. Due to a condition $\det{\Omega}=1$ there are
$N$ such minima: ${\Omega}\in Z(N)$. The order parameter
distinguishing between these $N$ degenerate phases is the value of 
the Polyakov loop $P$ (\ref{P-Omega}). The $Z(N)$ symmetry 
in the effective theory of Polyakov loops is due to the existence of
gauge transformations which are periodic in the Euclidean $t$ direction
up to an element of $Z(N)$: ${\Lambda}(1/T)={\Lambda}(0)z_k$. 
These transformations leave $S(A)$
invariant but multiply  Polyakov loops by $z_k$.

In spin model language, we are in an ordered phase when $T$ is large. 
This is a  
phase in which the entropy of the spin system is low. However 
the spin system is, at the same time, supposed to be describing 
a very hot gauge system 
with a large entropy in terms of quantum states. It is this 
complementary feature of the spin and gauge systems, that 
has given rise to a lot of confusion in the past few years. 
For example, one can ask the following question: what is the meaning 
of a localised surface in the spin model in terms of
the gauge model? In this paper we will not go into this question, 
but take the pragmatic point of view that we just want to calculate
the exponent in the decay law (\ref{eqno2.10}) for the hot
fluxes. Nonetheless, one should bear in mind, that if one does 
assume the existence of a well localised interface, this leads
to unusual thermodynamic properties; as we shall see below.

\subsection{Some Thermodynamic Properties of the Wall in Gauge Theory}

The study of surface effects requires a careful specification of boundary
conditions. Only when that is done can we separate the free energy of the
system into well defined bulk and surface free energies. 
Let us suppose we have in our two dimensional `box' a domain wall, 
separating two domains. Then the free energy $F$ will be
for large $L_y$ and $L_z$:
\begin{equation}
\label{eqno2.17}
F=f L_y L_z+\sigma L_y
\end{equation}
Here we assume that the size of the box is much larger than any of 
the microscopic quantities in the system.

What this equation describes is simply a wall with a constant free 
energy density in the $y$- direction, and a non-trivial free energy
profile in the $z$-direction. When integrated over $z$, this profile 
gives the interface tension $\sigma$. This profile is like a soliton.

In our three dimensional $SU(2)$ gauge theory we have a dimensionful 
coupling constant $g$, with dimension $\sqrt{\rm{mass}}$. The only 
other dimensionful quantity in our problem is the temperature $T$. 
We will typically work in the regime where ${g^2/ T}$ is a small number.
So on the basis of dimensions alone the interface tension has a simple form:
\begin{equation}
\label{eqno2.18}
\sigma=a({{g^2}/T}) T^2
\end{equation}
with function $a$ positive and dimensionless. We can apply semi-classical 
methods to calculate $a$. These methods typically give, when 
applied to solitons such as monopoles and sphalerons,
\begin{equation}
\label{eqno2.19}
E_{\rm sol}\sim{{\rm scale}\over {\rm coupling}}
\end{equation}
for the energy $E_{\rm sol}$ in terms of the scale 
(Higgs expectation value).
So it is not surprising that we obtain, as described in detail in 
the next section, a similar result for our profile: 
\begin{equation}
\label{eqno2.20}\label{sigma-alpha}
\sigma=\alpha_0{T^2\over {{g\over {\sqrt T}}}}
=\alpha_0{T^{5/2}\over g}.
\end{equation}
where $\alpha$ is a numerical factor of geometrical origin.
The expression (\ref{sigma-alpha}) can be easily understood
if one realizes that the domain wall has a width $w$ of the order
of the screening length $w\sim 1/(gT^{1/2})$, carries an excess
of free energy density $\Delta f\sim T^3$ and that 
$\sigma\sim w\Delta f$.

We are now ready for the following thermodynamical observations.
First consider a droplet of radius $R$ of `minus' phase in a 
sea of `plus' phase. The free energy excess due to the presence 
of this droplet equals
\begin{equation}
\Delta F=2\pi R\sigma.
\end{equation} 
Now, the probability for the appearance of such a droplet is 
$\exp\{-{\Delta F/T}\}$. From the explicit dependence of 
$\sigma$ on T we learn that this probability becomes exponentially 
small for {\it large } $T$. So we find that an ordered phase
prevails at high temperature{\footnote{A study of discrete gauge models 
reveals the same phenomenon: they can be  mapped onto a two dimensional
Ising model with a coupling $\sim \log T$ for $T$ large. That is, for $T$ 
large both models start to order.}}.

The second observation is a simple consequence of the positivity 
and the explicit temperature dependence of the interface tension. 
The interface entropy equals
\begin{equation}
\label{eqno2.21}
{\rm entropy}=-{\partial\over \partial T}\sigma=-{5\over2}
\alpha_0{{T^{3/2}}\over g}
\end{equation}
So the interface entropy is negative! As is the interface internal 
energy $\epsilon$, although the free energy given by the difference 
$-T\sigma$ is positive. Such a negative interface entropy 
is to be expected quite generally for models that order
at high temperature: the free energy starts to {\em grow} at the critical 
temperature, and consequently the entropy is negative. So the presence
of the wall diminishes the entropy of the system.

\section{Continuum calculation}
\label{sec:cont}

In this section we calculate $S_{\rm eff}$, the surface tension and the
profile of the order parameter (Polyakov loop) of the $Z(N)$ wall.
When the temperature is high the effective gauge coupling becomes
small and one can use perturbation theory to calculate the effective
action (\ref{Seff-Omega}). This has been done by several
authors \cite{We81,GrPi81} for $d=4$. Here we derive similar results
for $d=3$. Our method is simpler and is trivially generalized to
arbitrary $d$. The lattice version of these results is presented in 
Section \ref{sec:tenslat}.

We start from the partition function of the pure gauge theory at
finite temperature:
\begin{eqnarray}\label{z}
\lefteqn{
Z=\int {\cal D}A_\mu 
\exp\left\{
-{1\over 4g^2} \int_0^{1/T}dt\int_{\mbox{\boldmath $x$}}  
{\rm tr}\,F_{\mu\nu}F_{\mu\nu}
\right\}
}
\nonumber \\ &&
=
\int {\cal D} A_0 {\cal D} \mbox{\boldmath $A$}
\exp \left\{ 
-{1\over 2g^2} \int_0^{1/T}dt \int_{\mbox{\boldmath $x$}}
{\rm tr}
\left[ 
(\partial_0\mbox{\boldmath $A$} + \mbox{\boldmath $D$}A_0)^2 +
\mbox{\boldmath $B$}^2
\right]
\right\}
\end{eqnarray}
We use some obvious short hand notations. The gauge
potentials: $A_\mu \equiv i A_\mu^a T^a$ and the hermitian generators
of the $SU(N)$ group are normalized as: ${\rm tr}\,T^a T^b = \delta^{ab}$. 
The covariant derivative is:
\begin{equation}
\mbox{\boldmath $D$}A_0 \equiv \mbox{\boldmath $\nabla$}A_0
+ i [ \mbox{\boldmath $A$} , A_0 ].
\end{equation}
Bold symbols
denote vectors ($d-1$ components). The exception is {\boldmath $B$}
which is an antisymmetric tensor ($(d-1)(d-2)/2$ components):
\begin{equation}
B_{ik}=\partial_i A_k - \partial_k A_i + i [A_i , A_k]
\end{equation}

In contrast to \cite{We81,GrPi81} we do not begin by fixing the gauge.
First we rewrite the partition function in the form of an integral
over physical fluctuations at very high $T$. These are the transverse
components of the vector potential {\boldmath $A$}. To achieve this
we integrate over $A_0$ which is clearly an auxiliary field.
For every given configuration $\mbox{\boldmath $A$}(\mbox{\boldmath
$x$},t)$ we can do this easily: the integral is Gaussian. We obtain:
\begin{eqnarray}\label{z-det}
\lefteqn{
Z=\int {\cal D} \mbox{\boldmath $A$}
\det(-\mbox{\boldmath $D$}^2)^{-1/2} 
}
\nonumber \\ &&
\times
\exp \left\{
-{1\over 2g^2} \int_0^{1/T}dt \int_{\mbox{\boldmath $x$}}
{\rm tr}
\left[
(\partial_0\mbox{\boldmath $A$} -
\mbox{\boldmath $D$}\mbox{\boldmath $D$}^{-2}\mbox{\boldmath $D$}
\partial_0\mbox{\boldmath $A$})^2 +
\mbox{\boldmath $B$}^2
\right]
\right\}
\end{eqnarray}

Thus far our manipulations have been exact. Now we use the smallness
of $g$.  In the leading order in $g$ (saddle point approximation) we
need to keep only terms quadratic in {\boldmath $A$} in the exponent
and can neglect the pre-exponent. As expected the effect of $A_0$ is to
project out the longitudinal fluctuations of {\boldmath $A$} from the
kinetic term in (\ref{z-det}). The {\boldmath $B$}$^2$ term does not
contain them already. We see that the integral over {\boldmath $A$}
factorises into integrals over $d-2$ (times $N^2-1$) transversal
fluctuations and a trivial infinite factor.

Each of these integrals represents a very well known partition
function of the photon gas, the logarithm of which is:
\begin{eqnarray}\label{z-zero}
\lefteqn{
\ln Z_0=
(d-2)\ln\det(-\partial_0^2 - \mbox{\boldmath$\nabla$}^2)^{-1/2} 
} \nonumber \\ &&
=
- (d-2)\int {{\cal V} d^{d-1}\mbox{\boldmath $k$} \over (2\pi)^{d-1}} 
\ln (1-\exp\{-|\mbox{\boldmath $k$}|/T\}),
\end{eqnarray}
where ${\cal V}$ is the volume of space.  Thus we obtain the
Stefan-Boltzmann law for the free energy density of the hot gluon gas
(with a multiplicity of `photons' $N^2-1$):
\begin{equation}
{\cal F}=-(N^2-1){T \over {\cal V}}\ln Z_0=-cT^d,
\end{equation}
where $c$ is:
\begin{equation}\label{c}
c=(N^2-1)(d-2){\Gamma(d/2)\over\pi^{d/2}} \zeta(d).
\end{equation}

Next, we want to find the dependence of the free energy density on the
value of the Polyakov loop. To this end we calculate the partition
function (\ref{z}) about a constant background field $\underline
A_0$. We make a shift: $A_0=A^\prime_0 + \underline A_0$ and integrate
over $A^\prime_0$. This results in simply replacing
$\partial_0${\boldmath $A$} with
\begin{equation}\label{shift}
\partial_0\mbox{\boldmath $A$}
+ i [ \underline A_0 , \mbox{\boldmath $A$} ] 
\end{equation}
in equation (\ref{z-det}). For a given matrix $\underline A_0$ there
will be at least $N-1$ generators that commute with it (`neutral') and
at most $N(N-1)$ that do not (`charged'). The `charges' are given by:
$q_i-q_j$, where $q_i$ are the eigenvalues of the matrix $\underline
A_0/(2\pi T)$. These are related to the eigenvalues $\lambda_i$ of the
Polyakov loop ${\Omega}$ (\ref{Omega}) by: $\lambda_i=\exp\{i2\pi
q_i\}$. The contribution of `neutral' bosons is unchanged by
(\ref{shift}) and is still given by (\ref{z-zero}).  The partition
function of the `charged' gluons is given by:
\begin{eqnarray}
\lefteqn{
\ln Z_q=
(d-2)\ln\det(-(\partial_0 + i2\pi qT)^2 
- \mbox{\boldmath$\nabla$}^2)^{-1/2} 
}\nonumber \\&&
=
-(d-2)\int {{\cal V} d^{d-1}\mbox{\boldmath $k$}\over (2\pi)^{d-1}} 
\ln (1-\exp\{-|\mbox{\boldmath $k$}|/T + i2\pi q\}).
\end{eqnarray}
where $q=q_i-q_j$. The periodicity in $q$ reflects the
$Z(N)$ symmetry.

Following \cite{BhGo91} we choose $q_1=...=q_{N-1}=q/N$.
Only $N-1$ eigenvalues are independent: $\sum_i q_i=0$ and
thus $q_N=q/N-q$. As $q$ varies from 0 to 1 the Polyakov
loop:
\begin{equation}
P={1\over N} {\rm\ tr} \exp 
\left\{
{i\over T}\, \underline A_0
\right\}
\end{equation}
changes from 1 to $\exp(i2\pi/N)$. In the $SU(2)$ case $q$ parameterises
the only path between the two minima of the effective potential for
the Polyakov loop. In $SU(3)$ it parametrises the lowest action path
\cite{BhGo92}. It is possible that it does the same for $N>3$.  

With our choice of $\underline A_0$ there are $(N-1)^2$ `neutral'
gluons and $2(N-1)$ gluons with `charges' equal to $\pm q$.  The free
energy density as a function of the Polyakov loop order parameter
(parameterised by $q$) is now given by:
\begin{equation}
{\cal F}(q) = - cT^d + V(q)T^d
\end{equation}
All the dependence on $q$ comes from the `charged' gluons and is given by
the universal function:
\begin{eqnarray}                    \label{v-def}
\lefteqn{
V(q)=-2(N-1)(d-2)
\int { d^{d-1}\mbox{\boldmath $k$} \over (2\pi T)^{d-1}}
\ln (1-\exp\{-|\mbox{\boldmath $k$}|/T\}(1-\cos2\pi q))
}
\nonumber \\ &&
=2(N-1)(d-2){\Gamma(d/2)\over \pi^{d/2}} 
\sum_{\nu=1}^{\infty} {1\over \nu^d}
(1 - \cos2\pi q \nu).
\mbox{\hspace{10em}}
\end{eqnarray}
For $d=4$ this function is expressed via the Bernoulli polynomial 
$B_4(q)$. For $d=3$ only a numerical evaluation is possible.

For $q$ varying slowly on the scale of $1/T$ one can use the tree
expression for the gradient term \cite{BhGo92}. Then the free
energy density reads:
\begin{equation} \label{free-gamma}
{\cal F}(q) = { N-1 \over N }
\left(
{2\pi T \over g} \mbox{\boldmath$\nabla$} q
\right)^2 
+ (V(q) - c) T^d 
\equiv 
T^d \left[\,
{\gamma^2\over2} (\mbox{\boldmath$\nabla$} q)^2
+ V(q) - c
\,\right]
\end{equation}
where we defined a dimensionful parameter $\gamma$:
\begin{equation}
\label{gamma}
\gamma=\sqrt{N-1\over N}\sqrt{8\pi^2\over g^2T^{d-2}}.
\end{equation}
This completes the calculation of $S_{\rm eff}$ of Section 2.3:
$S_{\rm eff} = \int_{\mbox{\boldmath $x$}} {\cal F}/T$.

The profile of the wall between phases with $q(-\infty)=0$ and
$q(\infty)=1$ is given by the function $q_0(z)$ which minimizes
(\ref{free-gamma}) under corresponding boundary conditions. It
satisfies:
\begin{equation} \label{diff-eq}
\gamma{dq(z)\over dz} = \sqrt{ {2}  V(q) }.
\end{equation}
The width of the wall is controlled by $\gamma$:
$w\sim\gamma\sim 1/(gT^{d/2-1})$. The exact solution $q_0(z)$ for
$d=4$ is known \cite{BhGo92}. For $d=3$ we solve (\ref{diff-eq})
numerically. The surface tension $\sigma$ is given by the integral of
the excess of the free energy density inside the wall and is 
proportional to the action on the trajectory $q_0(z)$:
\begin{equation}   \label{sigma}
\sigma=T^d\int dz\, 2V(q_0(z)) 
= \gamma T^d \int_0^1 dq\, \sqrt{2 V(q)} =
\alpha_0 {T^{d/2+1} \over g},
\end{equation}
where we defined:
\begin{equation}
\alpha_0 \equiv 4\pi \sqrt{N-1\over N} \int_0^1 dq \sqrt{V(q)}.
\end{equation}
For $d=3$ and $N=2$ we find using (\ref{v-def}): 
$\alpha_0=5.104$ .

An interesting property of $d=3$ is that the second
derivative of $V(q)$ diverges at $q=0 \bmod 1$ (see (\ref{v-def})).
In fact, $V(q)\propto q^2 \ln (1/q)$ rather than $q^2$ for small $q$.
One can see from (\ref{diff-eq}) that this results in a Gaussian
rather than exponential fall-off in the tails of the solution. This
is related to the fact that the Debye mass in the Thomas-Fermi
approximation is divergent in $d=3$ if the charged particles are
massless: the well-known $\int d\mbox{\boldmath $k$}^2/\mbox{\boldmath
$k$}^2$. On the other hand, D'Hoker argued \cite{DH82} that {\em
all\/} infrared divergences in $QCD^T_3$ are cut off by a Debye
mass $m_D$ of order $g\sqrt{ T \ln(T/g^2) }$ (see
eq. (\ref{D'Hoker})). This means that the fast
Gaussian fall-off should saturate at the exponential when $q$ gets
small enough ($2\pi q T \ll m_D$) , i.e., sufficiently far from the
center of the wall.%
\footnote{\label{foot:gauss} This happens, however, beyond 
the applicability of our formulas for $q(z)$.} 
This behavior of the tail does indeed occur in our
Monte Carlo study (see Section 7.5).

\section{Lattice version of the interface tension calculation}
\label{sec:tenslat}

To compare numerical lattice data with theory requires 
a lattice version of the calculation of the previous section.

In this section we develop a one loop expression on a
lattice
with $L_t$ sites in the temperature direction and an infinite number
in all space directions.
Quantities computed in this section are the effective potential, the 
profile of the wall, and the surface tension. In Section
\ref{sec:profs} we calculate the expectation values
of electric and magnetic plaquettes.
In Section \ref{sec:size} the reader will find estimates of finite size 
corrections. 
 In the dimensionality we are interested in we have to go to higher
loops
to get a finite Debye mass. We follow to this end the ideas of
D'Hoker \cite{DH82}, which amount to a very simple prescription.

Let us first fix our notation.
The continuum action of Section \ref{sec:cont} becomes on the lattice
\cite{Wi74}:
\begin{equation}
\label{eqno4.1}\label{Scont-Slat}
{1\over g^2}S_{\rm cont}\to \beta S_{\rm lat}
=\beta \sum_p {\rm Re}[1-{1\over N}{\rm tr}U_p]
\end{equation}
Here every plaquette $p$ is summed over only once; in contrast to 
the continuum action where the sum over $\mu,\nu$ included both orders.

The lattice spacing $a$ is related to the temperature $T$ by
\begin{equation}
\label{eqno4.2}
a L_t={1\over T}
\end{equation}
which follows immediately from the fact that the length of the system 
in the fourth direction is $1/T$.

The lattice action should become the continuum action in the 
limit where the dimensionless quantity $aA_{\mu}$  becomes
very small:
\begin{equation}
U_P\sim \exp\{ia^2F_{\mu,\nu}\}.\nonumber
\end{equation}
The lattice coupling becomes in that limit:
\begin{equation}
\label{beta-g}
{\beta\over 2N}={a^{d-4}\over g^2} \nonumber
\end{equation}
or 
\begin{equation}
\label{eqno4.3}
{\beta\over 2N}L_t^{4-d}={T^{4-d}\over g^2}
\end{equation}

This is the relation between lattice and continuum parameters
in any number of dimensions $d$. For $d=4$ all reference to dimensionful
properties drops out.
Perturbation theory is defined by taking the dimensionless
parameter ${g^2 T^{4-d}}$ small, or $\beta$ large at fixed $L_t$. 

First we need a definition of the surface tension on the lattice.
This can be done by taking the lattice analogues of twisted and
untwisted boxes and their corresponding partition functions.
This corresponds precisely to the way we measure the
surface tension on the lattice (see section 7):
\begin{equation}
\exp\left\{-{\sigma\over T}L^{d-2}\right\}
\equiv {Z({\rm twisted})\over Z({\rm untwisted})}.
\end{equation}

Although our aim is to study a continuum theory, all our Monte Carlo
results come from a lattice with finite lattice spacing $a$. The value
of $a$ has a meaning only in comparison with physical length
parameters. For our system there are two such parameters: the inverse
temperature $1/T$ and the Debye length, or the width of the wall $w$.
The  width $w\sim 1/(gT^{d/2-1})$ is large compared to $1/T$ for
small $g$. To the order we are interested in we can neglect
corrections due to finiteness of the ratio of $w$ over $1/T$ or
$a$. We choose large $\beta$ so that the ratio $w/a$ is big:
$10-20$, and we neglect corrections due to its finiteness.%
\footnote{We study the corrections due to the finiteness of $L_z/w$ in
Section \ref{sec:Lz}.}

However, to save computer time we choose to keep the ratio of $1/T$
to $a$ relatively small. This ratio
is the number of lattice sites in the temporal
direction: $L_t$. To compare with our MC data we calculate the quantities
discussed in Section \ref{sec:cont} for {\em any\/} $L_t$ 
to the leading order in $g$. 

To achieve this we notice, that to the order we are working
in we have a theory of free gluons interacting only with the background. All
path integrals are Gaussian and factorise into integrals over
momentum modes. We need only to substitute the lattice periodic 
momenta for continuum ones:
\begin{equation}
k_i \longrightarrow {2\over a} \sin {{ak_i}\over2} \equiv \hat k_i; 
\end{equation}
\begin{equation}
k_0 + 2\pi qT \longrightarrow
{2\over a} \sin a{k_0+2\pi qT\over2} \equiv  \hat k_0(q). 
\end{equation}
The lattice momentum varies inside the Brillouin zone: $-\pi/a < k_i <
\pi/a$. The component $k_0$ is discrete and takes values:
$2\pi T n$, $n=0,1,\ldots,L_t-1$. 

The partition function for a `charged' gluon normalized by the
`neutral' one is given by:
\begin{equation}    \label{z-lat}
\ln{Z_q \over Z_0} =
-{d-2 \over 2}\, \left[\,
\ln\det(- \hat k_0(q)^2 - \mbox{$\sum_i$}\hat k_i^2) -
\ln\det(- \hat k_0^2 - \mbox{$\sum_i$}\hat k_i^2)
\,\right].
\end{equation}
We use the proper time trick to calculate (\ref{z-lat}). To simplify
formulas we work in lattice units $a=1$. The r.h.s. of (\ref{z-lat})
becomes:
\begin{eqnarray}        \label{z-prop-time}
{d-2\over 2}
{{\cal V}\over T}
\sum_{k_0} \int_{-\pi}^{\pi} \prod_i {dk_i\over2\pi}
\int_0^\infty {dt \over t}
\left[
\exp\{
-(\hat k_0(q)^2 + \mbox{$\sum_i$}\hat k_i^2)t
\}
\right.\nonumber\\ \left.
- \exp\{
-(\hat k_0^2 + \mbox{$\sum_i$}\hat k_i^2)t
\}
\right]
\end{eqnarray}
To rewrite the sum over $k_0=0,2\pi/ L_t,\ldots,2\pi(1 - 1/L_t)$
(remember that $T=1/L_t$ in our units) we use the Poisson summation
formula, which for a function $f(k_0)$ periodic with a period $2\pi$
has the form:
\begin{equation}    \label{poisson}
\sum_{k_0=0}^{2\pi(1 - 1/L_t)} f(k_0) = \sum_{\nu=-\infty}^{\infty} 
\int_{-\pi}^\pi {dk_0\over2\pi} e^{iL_t\nu k_0} f(k_0).
\end{equation}
We substitute (\ref{poisson}) into (\ref{z-prop-time}), shift the
variable $k_0$ and obtain for $\ln Z_q / Z_0$:
\begin{eqnarray}        \label{z-poisson}
{d-2\over 2}
{{\cal V}\over T}
\sum_{\nu=-\infty}^{\infty}  \int_{-\pi}^{\pi} \prod_\mu {dk_\mu\over2\pi}
\int_0^\infty {dt \over t}
\left[
\exp\{
-(\hat k_0^2 + iL_t\nu k_0 - i2\pi q\nu + \mbox{$\sum_i$}\hat k_i^2)t
\}
\right.\nonumber\\ \left.
-\exp\{-(\hat k_0^2 + iL_t\nu k_0 + \mbox{$\sum_i$}\hat k_i^2)
\} 
\right]. 
\end{eqnarray}
Integration over $k_\mu$ produces modified Bessel functions:
\begin{equation}
I_n(2t)=\int_{-\pi}^{\pi} {dk\over 2\pi} \exp\{2t\cos k + ink\},
\end{equation}
and $t$ can be rescaled after that.
Finally, we obtain the expression which replaces 
$V(q)$, eq.~(\ref{v-def}), at finite $L_t$:
\begin{equation}             \label{v-lat}
V_{\rm lat}(q) = 2(N-1)(d-2) L_t^d
\int_0^\infty {dt \over t}
\sum_{\nu=1}^{\infty} 
e^{-dt} I_{\nu L_t} (t) [ I_0(t)]^{d-1}
(1 - \cos2\pi q\nu)
\end{equation}  

It is instructive to see how the continuum limit $L_t\to\infty$ (i.e.,
$aT\to 0$) is recovered. For large $n$ and large
$t\raisebox{-.3em}{$\stackrel{>}{\sim}$}n$ the Bessel function behaves
as:
\begin{equation}
e^{-t}I_n(t) \approx {1\over \sqrt{2\pi t}} \exp\{-{n^2\over 2t}\}
\quad \mbox{and} \quad
e^{-t}I_0(t) \approx {1\over \sqrt{2\pi t}}.
\end{equation}
Only large $t$ of order $(\nu L_t)^2$ will contribute to the integral in
(\ref{v-lat}). After rescaling $t$ by $(\nu L_t)^2$ and integrating 
we get the same expression as in (\ref{v-def}).

The variation of $q$ with $z$ can be also taken into account. The
gradient term on the lattice becomes:
\begin{equation}
{ \gamma^2\over 2 } ( q(z+1) - q(z) )^2,
\end{equation}
so that the profile equation (\ref{diff-eq}) is replaced with:
\begin{equation}
q(z+1) = q(z) + \sqrt{{2\over\gamma^2}V_{\rm lat}(q)},
\end{equation}
where similarly to the continuum case 
\begin{equation}\label{gamma2}
\gamma^2=8\pi^2 {N-1\over N} {T^{2-d} \over g^2} = 
4\pi^2{N-1\over N^2} {\beta L_t^{d-2}}. 
\end{equation}
The interface tension is given by an expression similar to (\ref{sigma})
except for $\alpha_0$ being replaced by its lattice version:
\begin{equation}\label{alpha-lat}
\alpha = 4\pi \sqrt{N-1\over N} \int_0^1 dq \sqrt{V_{\rm lat}(q)}\quad.
\end{equation}

\section{Action density profiles}
\label{sec:profs}

An important quantity that we measure on the lattice is the
expectation value of the plaquette action. In the continuum limit
this corresponds to $\langle \mbox{\boldmath $E$}^2 
+ \mbox{\boldmath $B$}^2 \rangle$, where the Euclidean {\boldmath $E$}
is:
\begin{equation}
\mbox{\boldmath $E$} = - \partial_0 \mbox{\boldmath $A$}
- \mbox{\boldmath $D$} A_0.
\end{equation}
We also measure separately the expectation values of the plaquettes of
each orientation, which correspond to $\langle E_y^2 \rangle$,
$\langle E_z^2 \rangle$, $\langle B^2 \rangle$ in our $d=3$ case.

An interesting result that we find in our Monte Carlo study is that
these expectation values display nontrivial profiles, correlated with
the position of the domain wall. One expects that at high $T$ a
perturbative calculation of these quantities is possible. In this
section we perform such a calculation.%
\footnote{Such profiles were also measured in \cite{KaRu91}. Here, we
show that at high $T$ one can actually calculate them analytically.}

We start again with the partition function (\ref{z}) on a constant
background $\underline A_0$. At high $T$ the effective interaction
is weak and we apply a saddle point approximation. We write:
\begin{equation}    \label{z-eb}
Z(\underline A) = \int {\cal D} A_\mu \exp 
\left\{
-{1\over 2g^2} \int_0^{1/T}\!dt\int_{\mbox{\boldmath $x$}}
( \mbox{\boldmath $E$}^2 + \mbox{\boldmath $B$}^2 )
\right\},
\end{equation}
where we linearize $E$ and $B$:
\begin{equation}
\mbox{\boldmath $E$} = -\partial_0\mbox{\boldmath $A$} 
- i [ \underline A_0 , \mbox{\boldmath $A$}] 
- \mbox{\boldmath $\nabla$} A_0;       
\qquad
\mbox{\boldmath $B$} 
=  \mbox{\boldmath $\nabla$} \times \mbox{\boldmath $A$}.
\end{equation}

We have already calculated the integral (\ref{z-eb}) in
Section \ref{sec:cont}:
\begin{equation}        \label{z-v}
Z(\underline A) \equiv Z(q)
= \exp \left\{ 
- T^{d-1} \int_{\mbox{\boldmath $x$}} (V(q) - c) 
\right\}.
\end{equation}

Now, to find the action density profiles we want to calculate things
like $\langle \mbox{\boldmath $E$}^2 \rangle$ and $\langle
\mbox{\boldmath $B$}^2 \rangle$, where the average is understood in terms
of the probability distribution given by the integrand in
(\ref{z-eb}). These averages depend on $\underline A$, or $q$.  The
easiest way to calculate $\langle \mbox{\boldmath $E$}^2 \rangle$ is
to introduce a parameter, say $\epsilon$, in front of this term in the
exponent, do the integral and differentiate the logarithm of the
result over $\epsilon$. We shall do this shortly, but before 
that let us improve
a bit on the formula (\ref{z-v}). What we need to include is the
contribution of zero point energies of the modes of the fields
$A$. This contribution does not depend on the temperature and is not
present in (\ref{z-v}), but it gives an overwhelmingly dominant
contribution to $\langle \mbox{\boldmath $E$}^2 \rangle$ and $\langle
\mbox{\boldmath $B$}^2 \rangle$. Indeed, this contribution is of order
$(1/a)^d$, where $a$ is the lattice spacing (the UV cutoff), and is much
larger than $T^d$ - the thermal contribution. To see what we are missing
consider calculating $\langle \mbox{\boldmath $E$}^2 + \mbox{\boldmath
$B$}^2 \rangle$ from (\ref{z-v}). For that one needs only to
differentiate (the logarithm of) the right hand side over
$(1/g^2)$. But the right hand side does not depend on $g$! This means
that the quantity $\langle \mbox{\boldmath $E$}^2 + \mbox{\boldmath
$B$}^2 \rangle$ (which is obviously not zero) gives the sum of zero
point energies and has no thermal contribution at that order. To make
this fact explicit imagine rescaling $g^2$ in (\ref{z-v}) by a factor
$1/\epsilon$.  One can absorb this factor by rescaling the fields $A$
by $\sqrt{\epsilon}$. This will change the measure by a factor
$(\sqrt{\epsilon})^{(N^2-1)(d-1){\cal N}}$, where $\,{\cal
N}=T^{-1}{\cal V}/a^d$ is the total number of the lattice sites.  The
number $(N^2-1)(d-1){\cal N}$ is simply the number of non-zero modes, or
non-zero eigenvalues of the matrix of the quadratic form in the
exponent (\ref{z-eb}).  Eventually we get instead of (\ref{z-v}):
\begin{equation}    \label{z-gv}
Z(q) 
= g^{(N^2-1)(d-1){\cal N}}
\exp \left\{
 - T^{d-1} \int_{\mbox{\boldmath $x$}} (V(q) - c) 
\right\}.
\end{equation}
Using (\ref{z-gv}) we get for any $T$:
\begin{equation}           \label{e-plus-b}
{1\over 2g^2} \langle 
\mbox{\boldmath $E$}^2 + \mbox{\boldmath $B$}^2
\rangle 
= {(N^2-1)(d-1)\over 2}\,a^{-d}.
\end{equation}
It means that each space component of the vector {\boldmath$E$} or the
tensor {\boldmath$B$} has the average at $T=0$ (due to Euclidean
invariance):
\begin{equation}
{1\over 2g^2} \langle E_x^2 \rangle = {N^2-1\over d}\,a^{-d}.
\end{equation}
This is the dominant contribution and should be compared to $\beta
(1-{\rm tr}U_P/N)$ per plaquette which we measure (in our case
$N=2$, $d=3$). It is indeed equal to 1 in lattice units up to a small
correction. Part of this correction is the thermal effect.

To calculate the thermal part of the action density, multiply only the term
$\mbox{\boldmath$E$}^2$ in (\ref{z-eb}) by a factor $\epsilon$. 
Then consider the
following transformation: $t'= t/\sqrt\epsilon$, $A_0'=A_0\sqrt
\epsilon$, $T'=T\sqrt\epsilon$ and $g'^2=g^2/\sqrt\epsilon$. In terms
of new variables the integral is the same up to a Jacobian factor and thus:
\begin{eqnarray}         
\label{z-epsilon}  
Z(q) &=& \int DA_\mu \exp\left\{-{1\over 2g^2} 
\int_0^{1/T}dt\int_{\mbox{\boldmath $x$}}
(\epsilon\mbox{\boldmath$E$}^2 + \mbox{\boldmath$B$}^2 ) 
\right\} \nonumber\\
     &=& \left( g\over\epsilon^{1/d}  \right)^{(N^2-1)(d-1){\cal N}}
\exp \left\{ 
- (\sqrt\epsilon T)^{d-1} \int_{\mbox{\boldmath $x$}} (V(q) - c) 
\right\}
\end{eqnarray}

Therefore,
\begin{equation}   \label{e-av}
{1\over 2g^2} \langle \mbox{\boldmath$E$}^2 \rangle 
= (d-1){N^2-1\over d}\,a^{-d}
 - {d-1\over2} ( c - V(q) ) T^d.
\end{equation}
The $\mbox{\boldmath$E$}^2$ is a sum of $d-1$ components each
contributing equally.  Using (\ref{e-plus-b}) and (\ref{e-av}) we
get also:
\begin{equation}   \label{b-av}
{1\over 2g^2} \langle \mbox{\boldmath $B$}^2 \rangle
= {(d-1)(d-2)\over2}{N^2-1\over d}\,a^{-d}
+ {d-1\over2} ( c - V(q) ) T^d.
\end{equation}
Note that while $(1/2g^2)\langle \mbox{\boldmath$E$}^2 +
\mbox{\boldmath $B$}^2 \rangle$ is totally due to the vacuum zero
point energy, it is $(1/2g^2)\langle \mbox{\boldmath$B$}^2 -
\mbox{\boldmath $E$}^2 \rangle$ that contains the thermal energy. The
vacuum term in this quantity cancels in $d=4$, because there is the
same number of {\boldmath $E$} and {\boldmath $B$} components.

So far we have  neglected the variation of $q$ with $z$. This can be
easily corrected for. It contributes only to $\langle E_z^2 \rangle$
the amount $(\gamma^2/2) (\partial_z q)^2 T^d $, which is equal to
$V(q)T^d$ due to (\ref{diff-eq}).

Finally, we write down expressions for the plaquette action densities in
our case ($N=2$, $d=3$). We use lattice units: $a=1$, in which
$T=1/L_t$, and $\beta$ is given by (\ref{beta-g}): 
\begin{eqnarray}                      \label{act-profs} 
{1\over2g^2} \langle B^2 \rangle
&=& 1 + (c - V(q)) {1\over L_t^3} + O(1/\beta);
\nonumber\\
{1\over2g^2} \langle E_y^2 \rangle
&=& 1 - {1\over2}(c - V(q)) {1\over L_t^3} + O(1/\beta);
\nonumber\\
{1\over2g^2} \langle E_z^2 \rangle
&=& 1 - {1\over2}(c - 3V(q)) {1\over L_t^3} + O(1/\beta).
\end{eqnarray}
The corrections of order $1/\beta$ are due to non-quadratic terms and
are beyond our approximation. However, they should be the same for all
plaquettes at that order. They cancel in, e.g.: $\langle B^2 \rangle -
\langle E_y^2 \rangle $ or $\langle B^2 \rangle (q) - \langle B^2
\rangle (0)$.  The formulas (\ref{act-profs}) are in good agreement with
our MC data (see Section 7).

What do we learn from all this? One can see, for example, that the
thermal part in the fluctuations of $\langle B^2 \rangle$ becomes
negative at some $q$ near $1/2$, i.e., inside the wall. This means
that the $\langle B^2 \rangle$ becomes smaller than the contribution
of the vacuum fluctuations to that quantity!%
\footnote{In other words, $\langle B^2 \rangle - \langle E_y^2 \rangle$ 
becomes negative.} This is just another side of the old puzzle with
negative entropy and thermal energy density \cite{BeKo92,Sm94}.

\section{Finite size corrections}
\label{sec:size}

\subsection{Roughening}
\label{sec:rough}

Due to the long wavelength thermal fluctuations of its shape, an 
interface in 3 spatial dimensions oscillates from its central position
by a distance which grows as the logarithm of its area.
This roughening also occurs in 2 dimensions, where the effect is
proportional to $\sqrt L_y$. Here we calculate the effect of the
roughening on our measurements of the profile of the wall and
the interface tension and show that this effect is rather small.

The roughening is due to long wavelength fluctuations of the shape of
the interface. These fluctuations are therefore essentially classical:
$\omega_k \ll T$. We can consider our interface in 2 spatial dimensions
as a classical string of length $L_y$ with tension and mass per unit 
length equal to $\sigma$. The string is a set of free oscillators, the
normal modes. The amplitudes $f_k$ of these are the Fourier components
of $f(y)$, the shape of the string at a given instance. Each
oscillator (mode) has energy $T$ in the heat bath (equipartition). On
the other hand, the mean energy of such an oscillator is $\sigma \omega_k^2
\langle f_k^2 \rangle$.  The mean square of the fluctuation of the
string is then:
\begin{equation}
\langle  \int_0^{L_y} dy f^2(y)\rangle 
= \langle \sum_k  f_k^2 \rangle 
= \sum_k {T \over \sigma \omega_k^2}
\end{equation}
The dispersion law is: $\omega_k=k$. If we take $L_y=\infty$ and
replace the sum with an integral it will diverge linearly: the soft
modes get out of hand if $L_y$ does not cut them off. For
finite $L_y$ the values of momenta $q$ are given by the periodic b.c.:
\begin{equation}
k={2\pi\over L}n,\quad n=1,2,3,\ldots,
\end{equation}
where $k=0$ (translational mode) is removed by our procedure of
shifting the center of the interface. In principle, our string approximation
will break down at some large $k_{\rm max}$, but the value of this UV cutoff
is not essential for the long wavelength effect we are interested
in. We put $k_{\rm max}=\infty$.
Thus we have:
\begin{equation}
\langle  f^2(y)\rangle = {2\over L_y}\sum_{n=1}^{\infty} 
{T\over\sigma}{L_y^2\over(2\pi n)^2}.
\end{equation}
The factor of 2 is because for each $n$ there is a cosine and a
sine mode. The sum over $n$ can be evaluated and we get for
the mean square deviation of the wall from a straight line:
\begin{equation}
\Delta w^2 = \langle  f^2(y)\rangle = {T\over\sigma}{L_y\over12}.
\end{equation}

Let us estimate this effect in our case. Take, for example, $L_t=3$
and $\beta=75$: 
\begin{equation}
\Delta w = \left({2\over
\alpha}\sqrt{L_t^3\over\beta}{L_y\over12}\right)^{1/2}\approx
\sqrt{L_y/50},
\end{equation}
where we used the formula (\ref{sigma}) for $\sigma$
and the relation (\ref{beta-g}). For $L_y=12-60$ this varies 
from 0.5 to 1 or so, as compared to $w\sim17$. We observe a slight
variation of the width of the wall of roughly this size in our 
MC data, although it is no doubt optimistic to be applying
a string formalism in a situation where the length of the
wall is comparable to its width!

We conclude that roughening does not affect our estimates of the
profile of the wall. This effect is small, because the wall is stiff,
and $L_y$ is not too large. So what we measure numerically is 
really the intrinsic profile of the interface.

One can also calculate the correction to the interface tension from
the string-like fluctuations of the interface  using the same idea that
the string is a set of oscillators in a heat bath.  A difficulty lies
in the fact that unlike $\langle f^2 \rangle$ which is a convergent
sum of $\langle f_k^2 \rangle$, the amplitudes of the oscillators, the
sum of their free energies, $-T\ln(T/\omega_k) + {\rm const}$, is
divergent.  This divergence is ultraviolet, however, and can be
subtracted when computing the finite size dependence, similarly to the
Casimir effect.

Another, more illuminating way of deriving this correction is to
consider the ${\cal F}(q)$ in (\ref{free-gamma}) as an effective
potential energy for the long wavelength classical thermal fluctuations of
the wall. The profile $q_0(z)$ satisfying (\ref{diff-eq}) is a minimum
of ${\cal F}(q)$ for the corresponding boundary conditions. The leading
exponential behavior of the partition function is then:
\begin{equation}    \label{z-wall}
Z_{\rm wall} \sim \exp\left\{- { 1\over T } \int_{\mbox{\boldmath$x$}} 
[\, {\cal F}(q_0(\mbox{\boldmath$x$})) - {\cal F}(0) \,] \right\}
= \exp \left\{ - {\sigma L_y \over T} \right\},
\end{equation}
where we subtracted the bulk free energy.

The pre-exponential correction to (\ref{z-wall}) is due to fluctuations
of $q(y,z)$ around $q_0(z)$ and is given by the determinant:
\begin{equation}    \label{det-prime}
{\det}'(-\gamma^2 \mbox{\boldmath $\nabla$}^2  
+ V'' (q_0(z)))^{-1/2},
\end{equation}
where the prime on `det' denotes the fact that we omitted the
translational mode.  This mode is proportional to
$\partial_zq_0$. Properly normalized it produces a factor
$\sqrt{\sigma L_y/\gamma^2 T}$ and an integration over the position of
the center.  The spectrum of the operator in (\ref{det-prime}) can be
written as
\begin{equation}
\lambda = \gamma^2 k^2 + \lambda_m 
\end{equation}
where $k=2\pi n/L_y$ and $\lambda_m$ are the eigenvalues of
$-\gamma^2\partial_z^2 + V''(q_0(z))$. The $\lambda=\gamma^2 k^2$  band 
attached to the zero eigenvalue $\lambda_m=0$, corresponds to
fluctuations of the wall as a whole. The roughening effects are due to
these gapless fluctuations. The corresponding determinant is
${\det}'(-\gamma^2\partial_y^2)$. We can regularize it in the UV by 
dividing it by a similar determinant with $L_y=\infty$. On dimensional
grounds:
\begin{equation}
{{\det}'(-\gamma^2\partial_y^2)_{L_y}
\over
\det(-\gamma^2\partial_y^2)_\infty}=
{L_y^2\over\gamma^2} \cdot {\rm const},	
\end{equation}
Other eigenvalues are not related to the roughening and we neglect their
contribution in our estimate. Collecting all the factors we obtain:
\begin{equation}         \label{z-rough}
Z_{\rm wall} \approx{\rm const} \,\int dz\,
\sqrt{\sigma\over TL_y}
\exp\left\{-{\sigma L_y\over T} \right\} 
= 
{\rm const} \,
\sqrt{\sigma L_z^2\over TL_y}
\exp\left\{-{\sigma L_y\over T} \right\}.
\end{equation}
 From (\ref{z-rough}) we can read off the correction to the interface
tension%
\footnote{This correction is analogous to the Luscher's 
correction~\cite{Lu81}, but for a classical string in a thermal bath,
rather than a strip (quantum string at $T=0$). The universal
coefficient of the Coulomb correction $1/L$ to the free energy in
Luscher's case becomes in our case the coefficient of the $\ln L$
correction. It is especially obvious in the interpretation given 
by Stack and Stone \cite{StSt81}.}%
:
\begin{equation}
\Delta \sigma = {T\over 2L_y}\ln{TL_y\over \sigma L_z^2} + O(T/L_y)
\end{equation}
This means that the correction to $\alpha$, in the
way we measure it in our simulations (see Section 7), is given by:
\begin{equation}
\label{alpha-rough}
\Delta \alpha = { 4 L_t^{3/2}\over L_y } \sqrt\beta 
{\partial\over\partial\beta} \left(\Delta \sigma L_y\over T \right)
\approx -{L_t^{3/2}\over L_y\sqrt\beta}.
\end{equation}
For example, at $\beta=75$, $L_t=3$ and $L_y \sim 50$ this correction
is only about 0.2\%.

\subsection{Finite $L_z$ corrections}
\label{sec:Lz}

In the previous section we discussed the roughening effect which
introduces corrections of the type $(\ln L_y)/L_y$ to $\sigma$.
In this section we discuss another source of finite size effects:
the finiteness of $L_z$. There are two ways the finiteness of $L_z$
affects the free energy of the wall. First, there is a correction to
$V(q)$ itself because it is given by one-loop integrals which depend
on $L_z$ through the quantization of momenta running in the loop.%
\footnote{There is, of course, a similar dependence on $L_y$.} 
This correction should be of order $(T/L_z)^2$ and is relatively
small. For $L_z \raisebox{-.3em}{$\stackrel{>}{\sim}$} w$ this
correction is beyond our leading order in $g$ approximation, because
$(T/w)^2$ is of order $g^2$.

The second source of corrections is due to the finiteness of $L_z/w$.
It is obtained by calculating the action of a particle with a
`mass' $\gamma^2$ which, moving in the potential $-V(q)$, returns
to its starting value of $q$ in a period of `time' $2L_z$ 
This finite size effect we estimate here and show that, for large 
enough $L_z$, it is exponential in $d>3$, i.e., $\exp\{-L_z/w\}$,
and Gaussian in $d=3$, i.e., $\exp\{-L_z^2/w^2\}$.

So we consider a trajectory which starts at rest at $q=\varepsilon$ and
arrives at $q=1-\varepsilon$ precisely after a given `time' $L_z$ so
as to satisfy the boundary condition on the Polyakov loop.
The trajectory satisfies a Lagrange-Euler equation which can be
integrated to give (`energy' conservation):
\begin{equation}  \label{le-int}
{ \gamma^2 \over 2 } (q')^2 - V(q) = - V(\varepsilon) \equiv  E.
\end{equation}
We can use it to relate $L_z$ to $\varepsilon$:
\begin{equation}            \label{l-eps}
L_z(\varepsilon) = \int_{\varepsilon}^{1-\varepsilon} 
{ dq \over \sqrt{{2\over\gamma^2} (V(q)-V(\varepsilon))}}
\end{equation}

The action can be cast into the form:
\begin{equation}             \label{s-eps}
S(\varepsilon) = \int_0^{L_z}  dz\, [ {\gamma^2 \over 2} (q')^2 + V(q) ]
= \int_{\varepsilon}^{1-\varepsilon}  dq\,
\sqrt{2\gamma^2(V(q)-V(\varepsilon))} + V(\varepsilon)L_z(\varepsilon),
\end{equation}
which  is convenient for the numerical
evaluation of $S(\varepsilon)$. Also note, that this form is familiar in
theoretical mechanics as $dS=pdq-Edt$. The equations (\ref{l-eps}) and 
(\ref{s-eps}) give a parametric representation of $S$ as a function of
$L_z$ which we use to evaluate the correction numerically.

To find the asymptotic dependence of $S$ on $L_z$ one can use the
relation:
\begin{equation}               \label{ds-dl}
{dS \over dL_z} = V(\varepsilon),
\end{equation}
which is a consequence of $dS=pdq-Edt$, and can be also derived explicitly
from (\ref{l-eps}), (\ref{s-eps}).

Now we use the asymptotic form of $V(q)$ at small $q$.
For $d>3$ it is:
\begin{equation}        \label{v-4}
V(q)  = b q^2 + O(q^4) \quad (d>3),
\end{equation}
where $b$ is a constant which depends on $N$ and $d$ (\ref{b-4}).
Using this form we get from (\ref{l-eps}):
\begin{equation}
L_z(\varepsilon) = 2\sqrt{\gamma^2\over 2b}\ln{1\over\varepsilon} + O(1),
\quad (d>3).
\end{equation}
Integrating the equation:
\begin{equation}
{dS \over dL_z} = b\varepsilon^2 + O(\varepsilon^4) 
\approx \mbox{\ const\ } b \exp\left\{
-\sqrt{2b\over\gamma^2} L_z \right\},
\quad (d>3),
\end{equation}
we obtain the asymptotic form of the large $L_z$ correction to $S$:
\begin{equation}
\delta S \sim 
- \mbox{\ const\ }\sqrt{b\gamma^2\over 2}\exp\left\{
-\sqrt{2b\over\gamma^2}L_z \right\},
\quad (d>3).
\end{equation}

For $d=3$, however, we have:
\begin{equation}             \label{v-3}
V(q) = b q^2\ln{1\over q} + O(q^2), \quad  (d=3).
\end{equation}
where $b$ is given by (\ref{b-3}). This leads to
\begin{equation}
L_z(\varepsilon) = \sqrt{8\gamma^2\over b}\sqrt{\ln{1\over \varepsilon}}
+ O(1),    \quad  (d=3);
\end{equation}
and
\begin{equation}
{dS \over dL_z}
\approx \mbox{\ const\ } {b^2 L_z^2\over 8\gamma^2} 
\exp\{-{bL_z^2\over4\gamma^2}\},   \quad  (d=3);
\end{equation}
Integrating we obtain:
\begin{equation}
\delta S \sim - \mbox{\ const\ } 
{b L_z\over 4} \exp\{-{b\over4\gamma^2}L_z^2\},   \quad    (d=3).
\end{equation}
We see that the asymptotic form of the correction is related to the
way the tail of the wall decays: exponential in $d>3$ and Gaussian in
$d=3$.%
\footnote{See, however, footnote \ref{foot:gauss} 
on page \pageref{foot:gauss} and the related discussion.}

 From (\ref{v-def}) we find for $b$ in (\ref{v-4}):
\begin{equation}   \label{b-4}
b = (2\pi)^2 (N-1)(d-2){\Gamma(d/2)\over\pi^{d/2}}\zeta(d-2),
\quad  (d>3);
\end{equation}
while in $d=3$ the value of $b$ defined as in (\ref{v-3}) equals:
\begin{equation}     \label{b-3}
b=2\pi (N-1),   \quad    (d=3).       
\end{equation}

To get an idea of the size of this correction in our case let us
estimate it for the case of $\beta=100$, $L_t=4$, $L_z=120$. We get
$\gamma^2=\pi^2\beta L_t=400\pi^2\approx 60^2$. The exponent is
$bL_z^2/4\gamma^2\approx (L_z/50)^2\approx 5.8$ and $e^{-5.8}\approx
1/300$.  The pre-exponent $bL_z/4\approx 200$. Thus $\delta S \sim 1$.
This should be compared to:
\begin{equation}
S_0=\int_0^1 dq\, \sqrt{2\gamma^2 V(q)} \approx 50.
\end{equation}
Thus the correction, $\delta S/S_0$, is of the order of a (few) percent.
In the following section we shall evaluate these corrections,
numerically, for all $L_z$ and not just large $L_z$ as herein.

\section{Numerical Simulations of Domain Walls}

As we have seen, at high temperatures the theory appears to
have degenerate vacua which are separated by domain walls.
At asymptotic temperatures many properties of these
domain walls can be calculated in perturbation theory;
indeed the existence of these interfaces can only be
seen when one goes beyond tree level. However, as we remarked earlier,
the presence of infra-red divergences in higher-orders
has raised doubts about the applicability of perturbation
theory and, indeed, about the actual existence of the
interface. To address these doubts we have performed
accurate computer simulations of the domain walls,
and have compared what we find with the results of the perturbative
calculations. These computer simulations will be
described in this section.

If we simulate the high temperature $SU(2)$ gauge theory in a 
finite but large spatial volume, with periodic boundary conditions,
then we expect some fraction of the field configurations
to contain both $Z(2)$ phases in different portions of the torus.
Such a configuration will contain domain walls separating the
two phases and in principle one could study the domain walls
by focusing on these particular field configurations.
However the relative probability of such
configurations is very small for the temperatures of interest
and they would not be encountered in a typical Monte Carlo calculation.
So we have to use an alternative less direct method. What
we do is to impose twisted boundary conditions on our system,
so enforcing the existence of at least one domain wall. 
This will be described in Section 7.1. In Section 7.2
we show how the domain wall can be
interpreted as a 't Hooft disorder loop. We then specify
the physical quantities that we plan to calculate and
describe the methods by which we do so in Section 7.3. 
Section 7.4 summarises our Monte Carlo results.
Of course, it is crucial to demonstrate that we have
all finite-volume effects under control --- after all, it is
infrared effects that are the potential problem here --- and
this we do in Section 7.5. Finally, in Section 7.6, we
will take our raw `data' and use it to extract quantities
that are of direct physical interest in the present context
and compare them to the perturbative predictions.

\subsection{Twisted Boundary Conditions}

We work on lattices of size $L_y \times L_z \times L_t$ in
lattice units. The Euclidean time extent determines the
temperature, $aT=1/L_t$, of the field theory. The partition 
function contains the factor $\exp(-\beta S)$ where 
$\beta=4/(ag^2)$ and the lattice action is as in (\ref{Scont-Slat}):
\begin{equation}
\label{eqn(1)}\label{Slat}
S = \sum_{p} [1 - {1 \over 2} {\rm\ tr} U_p],
\end{equation}
where $U_p$ is the path ordered product of the $SU(2)$ matrices, 
$U_l$, on the links, $l$, that form the boundary of the plaquette $p$.  

The simplest and most usual way to introduce twisted
boundary conditions is as follows \cite{KaRu91}. We change the above
action to a twisted action, $S_{\rm tw}$,
by replacing  ${\rm\ tr}U_p$ with $-{\rm tr}U_p$ 
for those plaquettes
in the $zt$ plane that emanate from the sites $(y,z,t)$ 
where $z$ and $t$ are fixed to some particular values, say
$z=j$ and $t=k$, while $y$ takes all values from 1 to $L_y$
(see Fig. 1).

%
%

The system with this altered action and with periodic
boundary conditions is equivalent to the system with
the original action but with twisted boundary conditions \cite{GJA}.
This we see from the following argument. 
Firstly, let us choose labeling $z$ of the sites so that $j=L_z$,
i.e. the twist is between $z=L_z$ and $z=1$.
Secondly, to include the possibility of boundary conditions
that are not periodic it is convenient to extend our
labeling to include $z=0$, as well as $z = 1,\ldots,L_z$.
If the system is periodic then corresponding sites and links 
with $z=0$ and $z = L_z$ are identified (and similarly
for other directions). Then the system with a twisted action
can be viewed as a system with the original action but with fields
which are not periodic. To be more specific
they are periodic except that for $y=1,\ldots,L_y$ and $t=k$
the time-like link at $z=0$ is mapped into the negative of
itself at $z=L_z$. This is the lattice version
of the twisted boundary conditions described
in Section 2.1. 

One can move the $zt$ position
of the line of twisted plaquettes by flipping the sign of all 
$U_l$ which bound these plaquettes from one of the sides; but
one cannot undo the twist completely. It should be clear that
the position of the twist does not carry any physical significance
since it can be moved by such a redefinition of the variables 
$U_l$.

How does the twist lead to the presence of a domain wall?  To see this
consider the same labeling of sites as we have just used. With
free boundary conditions the system would spend most of the time in
one of the two phases where Polyakov loops are all near $+1$ or all
near $-1$. With the twisted boundary condition a Polyakov loop at
$z=0$ is mapped onto negative of itself at $z=L_z$. Therefore a
homogeneous configuration is frustrated and the Polyakov loops must
create a nontrivial profile in the $z$ direction to interpolate between
$z=0$ and $z=L_z$.

So to study the high $T$ properties of domain walls
we perform Monte Carlo calculations on lattices  with periodic
boundary conditions but with a twisted action. How
well defined is the domain wall in practice? To answer
this question we show in Fig. 2a the distribution of Polyakov
loops on a typical field configuration taken from a
$30\times 80 \times 2$ lattice at $\beta =100$. In physical
units this corresponds to a temperature $T \sim 30 T_c$
where $T_c$ is the deconfining temperature. We see that
the domain wall is very well defined, with relatively
small fluctuations around a smooth background distribution.
This is in fact the highest value of $T$ at which we work.
The lowest value is on a $12 \times 30 \times 2$ lattice 
at $\beta =7$ corresponding to $T \sim 2T_c$. There a typical
field configuration looks as in Fig. 2b. The fluctuations 
are now much larger, but the domain wall can still be 
unambiguously located. So it is clear that, for the range of $T$
we study, there is no ambiguity in identifying the domain
wall.

\subsection{Domain Walls as Disorder Loops}

Before going on to the details of the calculations, we
address the following natural question. Since the twist
is entirely symmetric in $z$ and $t$ why should the
`domain wall' separate regions in $z$ rather than regions
in $t$? 
This question can be plausibly answered, 
in a way that highlights the physics, by first
considering the twisted system at very low $T$ where 
the system is manifestly rotationally invariant. Here
introducing a twist introduces into the system a
't Hooft disorder loop \cite{tH78} which is closed
through the boundary in the
$y$-direction. This loop will presumably be a flux tube whose
width will be on the order of the characteristic length
scale of the theory, which here is $1/g^2$. Its
special property is that it if one considers the gauge 
potential on a closed path that encircles the disorder loop 
far from its centre, then the presence of this loop leads
to the potential acquiring a gauge transformation that goes
from 1 to a non-trivial element of the centre as we go
once around this closed path. So if we take a large Wilson
loop and pierce it once (or an odd number of times) by this
disorder loop, then the value of the Wilson loop is changed by a 
factor of $-1$ as compared to the value it would possess in the
absence of the disorder loop. Hence the name `disorder loop'.
It is clear that if the vacuum contained a condensate
of such loops, then these would be sufficient to ensure
that large Wilson loops varied as the exponential of their area,
so that we had linear confinement. Suppose we now increase
$T$ by reducing the extent of the system in the 
$t$-direction. Clearly at some point
the time extent will become smaller than the width of
the flux tube, the tube will become squeezed 
so that it extends right across the
time direction while still extending over a finite region
in the $z$-direction. This will occur once $T$ is
sufficiently large compared to $g^2$; presumably around
the deconfining transition. So at high temperatures our
`domain wall' is actually a squeezed disorder loop
that closes upon itself through the $y$-direction. It is
indeed symmetric in $z$ and $t$ except for the deformation
induced by the limited $t$ extent. The fact that the
Polyakov loops on either side of the wall have opposite
signs is what one might expect from such a squeezed 't Hooft
disorder loop. 

We have simplified the above argument by assuming that
the disorder loops exist as definite field fluctuations
in the low-$T$ theory. This is assuming a great deal of course.
Whether they do so exist is one of the central questions in
the still unresolved problem of colour confinement.
This makes the connection between these loops and the
domain walls at high-$T$ of added interest.

\subsection{Quantities Calculated}

We perform calculations on lattices with and without a twist.
The simplest and most interesting quantity we extract is 
the extra action, $S_w$, associated with the presence of a
domain wall. If both twisted and untwisted lattices are of the 
same size, then 
\begin{equation}
\label{eqn(2)}
S_w = \langle S_{\rm tw}\rangle - \langle S_{\rm nt}\rangle
\end{equation}
where $\langle S_{\rm tw}\rangle$, $\langle S_{\rm nt}\rangle$ 
are the average values of the twisted 
and untwisted actions, as defined in Section 7.1.
The extra action is related to the free energy of the wall, 
$F_w  \equiv F_{\rm tw} - F_{\rm nt}$, by
\begin{equation}
\label{eqn(3)}\label{Fw-Sw}
{{\partial}\over{\partial\beta}} \left({F_w}\over{T}\right) = 
 S_w
\end{equation}
where the derivative is taken at constant values of $L_y,L_z,L_t$. 
Using this relationship we shall test the one-loop prediction
for $F_w$.

It might appear that changing the action for a line of
$L_y$ parallel plaquettes could introduce some additional {\em local}
contribution to $\langle S_{\rm tw}\rangle 
- \langle S_{\rm nt}\rangle$ which is not related to the
free energy of the wall. That this is not so
one can see by considering a system with 2 parallel twists. This
system is equivalent to a system without a twist after a 
redefinition of variables $U_l$ which move the twists to a
single position where they cancel each other.

In addition to making predictions for the domain wall free energy,
perturbation theory can also be used to 
predict the detailed shape of the
domain wall as it interpolates between the two $Z(N)$ vacua.
In the Monte Carlo calculations the domain wall is free to move
and so if we are to obtain an average profile, we need to
shift our origin, in each Monte Carlo generated configuration,
to the centre of the domain wall. We also need to take into
account the presence of the twist, since the Polyakov loops
change sign as one moves through it. Our algorithm is as follows.
Consider a single Monte Carlo generated field configuration.
First we average Polyakov loops over $y$. We write this average
as $p(z)$. We now want to identify the location, $z=z_c$,
of the centre of the domain wall. This is defined operationally 
as follows. We first identify the values of $z$ where $p(z)$
changes from positive to negative values (factoring out, of course,
the trivial change at the twist itself). Clearly the number of
such changes must be odd. In practice the domain wall is
very smooth at high T --- as one can see in Fig. 2 --- and 
it is almost always the case that there is only one place 
where there is a sign change. This occurs between sites and we 
shift our origin in $z$ so that the sites where the sign changes 
are labeled by $z=0$ and $z=1$. We now ensure that $p(z=1) > 0$
by multiplying the whole profile by -1, if necessary. Our range
of $z$ is now from $-L_z/2+1$ to $+L_z/2$. Somewhere in this range
there is the twist and the value of $p(z)$ will flip sign there.
If this occurs for $z \geq 1$ then we flip the signs of $p(z)$
for values of $z$ beyond the twist; if it occurs for $z \leq 0$
then we flip the signs for $z$ before the twist. In this way we
obtain a wall profile with $p(z\geq 1) > 0$ and
$p(z \leq 0) < 0$. We can now average this profile over many 
configurations to obtain an average profile. This will
be symmetric about $z=1/2$ so we can fold the profile over
(with a sign flip) so that it is defined for $1 \leq z\leq L_z/2$ and
is positive. This is our final averaged profile. Note that in
any individual configuration the centre of the domain wall
may be closer to $z=0$ than to $z=1$. That is to say, our
profile is `smeared' over distances $\delta z \sim 1/2$.

In rare cases a given configuration contains more than one sign change
in $p(z)$ (always factoring out the trivial sign change at the twist).
The number of these sign changes is clearly odd. There are two
possibilities. One is that we have a configuration with
more than one domain wall, i.e. the one enforced by the twist plus
pairs that are genuine quantum fluctuations. In this case we
would typically expect at least one large gap in $z$ between the 
walls. The more trivial possibility is that we might be simply
seeing a large fluctuation of the values of $p(z)$ near the
centre of the wall (where the values are small on the average).
This would be characterised by very small gaps between the
locations of the sign changes. The first type of configuration, which
we should not include in our average, did not occur in any of
the calculations that we include below. (It does occur if we
approach the deconfining transition or if we make the
extent in $y$ of the lattice, and hence of the wall, sufficiently
small.) The second type of configuration we should include
and we do so by taking its centre to be located in the middle
sign change. In practice these configurations are so rare
that there is no visible change in any extracted quantities
whether we include them or not.

Having obtained a centre for the wall from the 
Polyakov loop distribution, we can also 
define an action profile for the wall, and we can clearly do this 
separately for the different $\mu\nu$ components of the action.

A quite different but equally interesting quantity is the electric
screening mass, $m_D$. This can be obtained from the lightest mass,
$m_{P}$, that couples to Polyakov loops, and hence from the tail of the
wall profile. We expect that for large enough $z$,
\begin{equation}
\label{eqn(4)} 
p(\infty) - p(z) \propto e^{-a m_{P} z}
\end{equation}
where $p(\infty)$ can be obtained either by working with very large
lattices, or by performing simulations on a lattice without a twist
and using the average value of the Polyakov loop obtained therein.
So if we define an effective mass by 
\begin{equation}
\label{eqn(5)}
a m_{\rm eff}(z) = \ln {{p(\infty) - p(z-1)}
\over {p(\infty) - p(z)}}
\end{equation}
then
\begin{equation}
\label{eqn(6)}
a m_{P} =  \lim_{z\to\infty} a m_{\rm eff}(z)
\end{equation}
In practice we would extract $m_{P}$ from $m_{\rm eff}$ once we were
at large enough $z$ that the latter had become independent of $z$.
The electric screening mass, $m_D$, should then be given by 
$m_{P}=2m_D$ (see below).

On a finite lattice the above needs to be altered because we expect 
contributions going both ways around the $z$-torus. So instead of
(\ref{eqn(4)}) we use
\begin{equation}
\label{eqn(7)}
p(\infty) - p(z) \propto e^{-a m_{P} z} + e^{-a m_{P}(L_z - z)}
\end{equation}
and alter (\ref{eqn(5)}) correspondingly.

Since we need to calculate the average action without a twist,
we can also calculate the screening mass on these 
untwisted field configurations. Here we follow 
standard techniques for such mass calculations \cite{Tep1}.
We construct $p_y=0$ sums of Polyakov loops at each value
of $z$ and then obtain the vacuum-subtracted correlation function,
$C(z_1-z_2)$, as a function of their separation, $z=z_1-z_2$. For large 
separation $z$ we have $C(z) \propto \exp(-a m_{P} z)$. We define
an effective mass $a m_{\rm eff}(z) = \ln[C(z-1)/C(z)]$, and we increase
$z$ until $m_{\rm eff}$ becomes independent of $z$. At this point
we can estimate $m_{P} = m_{\rm eff}$. This calculation has
the advantage that we can prove that 
$m_{\rm eff}(z) \geq m_{\rm eff}(\infty)$.
In practice we modify this formalism for the periodicity
in $z$ as described above. In addition
we calculate with a range of smeared Polyakov loops and
use the correlation function that minimizes $m_{\rm eff}(z=1)$; in the
spirit of a variational calculation. However, this turns out not
to be really necessary here; unlike the situation at $T=0$.
As we shall see below, this
method turns out to be much more efficient for the calculation of 
screening masses than using the tails of domain walls.

\subsection{Monte Carlo simulations}

Our Monte Carlo simulations were performed on a variety of
periodic lattices with and without a twist. We used a
standard heat-bath update algorithm mixed with 
over-relaxation steps. 

The control of finite volume effects is particularly important 
in these calculations  since it is infra-red effects 
that are usually seen as being at the root of 
any possible breakdown of perturbation theory at 
high temperatures. We have therefore performed
extensive numerical checks of finite volume effects and
these will be described in detail in Section 7.5.
In this section we shall confine ourselves to a presentation
of those results that have been obtained on lattices 
which are sufficiently large that 
any finite-volume corrections are much smaller than
our (very small) statistical errors. This will, of
course, need to be demonstrated and we shall do so later.  

Now, let us estimate how large the required volumes must be
in lattice units.
If we use a periodic $L_y \times L_z \times L_t$ lattice, this
corresponds to the fields in the spatial $L_y \times L_z$
volume being at a temperature $aT=1/L_t$. The dimensionless inverse
coupling $\beta$, as we have seen, is related to the dimensionful coupling,
$g^2$ by $\beta = 4/ag^2$. Thus in physical units the temperature is
$T/g^2 = \beta/4L_t$, and, at a fixed value of $L_t$,
$\beta \propto T$. Perturbation theory is expected to be
most reliable at very high $T$, so we want to study the theory for very 
large $\beta$. The characteristic length scale 
at high $T$ is of the order of $1/gT^{1/2}$:
the inverse of the Debye mass $1/m_D$ and the width of the wall $w$ are
of that order. Therefore the spatial sizes in units of $a$ must satisfy:
\begin{equation}
\label{LyLz}
L_y,L_z \gg {1\over ag\sqrt T} = L_t \sqrt{T\over g^2}
\end{equation}
 From (\ref{LyLz}) we see that for a fixed value of the
temperature in physical units, $T/g^2$, the required volumes will be 
smallest, in lattice units, for $L_t=2$ (since $L_t=1$ is not sensible).
Since the perturbative properties of the domain wall can be calculated
on the lattice, the minimal calculation one might perform is
to do everything at $L_t=2$. However in this case $a$ is as large
as possible in units of $1/T$ ($aT=1/L_t=1/2$)
and since there have been suggestions \cite{Sm94}
that high-T perturbation theory might break down as $a \to 0$
we choose to perform calculations
for several values of $L_t$. (This will have other advantages
that will become apparent below.) We shall cover a range
of temperatures for $L_t =2,3,4$ and we shall perform 
a calculation at one reasonably high value of $T$ for the
case $L_t =6$. In this latter case $aT=1/6$ which is surely small
enough that any breakdown of perturbation theory, as $a \to 0$, should
have become prominent.

In Table 1 we list the average values of the plaquette, 
$1-s_{\rm{nt}} \equiv {1\over 2}\langle{\rm tr}U_p\rangle$, 
for the calculations without a twist. In Table 2 we
do the same for the corresponding quantity, 
$1-s_{\rm{tw}}$, with a twist. We show 
the values of $\beta$, the lattice sizes, the number of Monte Carlo
sweeps and the average plaquette action. In the twisted case we 
perform `measurements' 
every Monte Carlo sweep; in the untwisted case every four sweeps.
The typical number of thermalisation sweeps prior to
taking any measurements is between 25000 and 50000. The errors,
given in brackets, are typically based on 40 or 50 bins. In
a few of the lower statistics cases we use as few as 25 bins.
The reader will note that at some values of the parameters
we have several different lattices. These arose during
the finite volume studies that will be described in detail
later on. The measurements that we list here are those that
do not suffer significant finite-size corrections (and 
are statistically accurate enough to be useful). 
At the different values of $L_t$ we have
chosen values of $\beta$ such that the temperatures, in units
of $g^2$, are roughly the same, although for higher $L_t$ we
are forced to cover more limited ranges of $T$. (The reader
may be puzzled that in some cases $\beta$ has not been chosen
exactly proportional to $L_t$; for the purposes of the
work in this paper no particular significance should be 
attached to these choices.) If we now multiply
$s_{\rm tw}-s_{\rm nt}$ by the number of plaquettes in the twisted
lattice, which is the one containing the domain wall, we obtain a
value for $S_w$ and hence, from (\ref{Fw-Sw}), information on $F_w$.
We shall see later on what this comparison tells us about
the accuracy of perturbation theory.

As described in the previous section, the lightest mass
that couples to Polyakov loops, is of particular interest
because it is related to the Debye screening mass. It can be
calculated either from correlations of Polyakov loops in 
the system without a twist, or from the way the tail of the 
domain wall merges into  the vacuum once we are far enough away 
from the centre of the wall. In Fig. 3 we show the effective masses
as obtained by the two methods. In Fig. 3a we have chosen
our highest value of $T$ for $L_t=3$ while in Fig. 3b we
show what one obtains for a medium value of $T$ with
$L_t=2$. We see that in both cases the values of $am_{\rm eff}(z)$
as obtained from Polyakov loop correlations do become
independent of $z$ at larger $z$, and that these `plateaux'
occur early enough for the errors to be very small. Since
in this case $m_{\rm eff}(z)$ is always an upper bound on $m_{P}$,
we can extract an accurate estimate of $m_{P}$ using the
first value of the effective mass that is, within errors,
on the plateau. The effective masses obtained from the
domain walls are clearly consistent with being asymptotic 
to these mass values. However it is equally clear
that they would give us much less accurate estimates of $m_{P}$. 
(We would need to do fits with at least two masses, since there
are no convenient plateaux, and so the errors on $m_{P}$ would be 
perhaps an order of magnitude greater. Moreover the assumption that
the effective masses asymptote from below, while reasonable,
introduces a difficult to quantify extra systematic error.)
So from now on we shall only use the values of $m_{P}$ as extracted 
from Polyakov loop correlations. These are listed in Table 3,
for those lattice volumes which do not suffer significant
finite-volume corrections.

To obtain the extra action of the domain wall, $S_w$, at a particular
value of $\beta$, we take the difference $s_{\rm tw}-s_{\rm nt}$ at that
$\beta$ and multiply by the number of plaquettes on the twisted lattice,
which contains the domain wall. We expect that this extra action
will be proportional to the length of the domain wall, i.e. to 
$L_y$, as long as $L_y$ is not very small. (This and the
related question of roughening will be addressed when we discuss
finite volume corrections.) So we form the quantity $S_w/L_y$,
which is the extra action of the wall per unit length (in units of
the lattice spacing). If we have values of this quantity for
several values of $L_y$ at a given value of $\beta$ and $L_t$, we
can average them to obtain our best overall estimate. In Table 4
we present our final averages for this quantity and for the
mass $am_{P}$, as obtained by averaging the values given in
Tables 1-3. These will form the basic raw material for our
later comparisons with perturbation theory.

\subsection{Finite Volume Corrections}

We shall be using our values of $S_w$ to test perturbation
theory. The details of the $T$ and $L_t$ dependence will be important 
in this comparison. Since the size of the domain wall varies with $T$, 
it is important that we control any finite volume corrections
at all values of our parameters. Otherwise part of the $T$
dependence we observe might be due to such corrections.
In this section we describe in detail how we control finite size
effects. We begin with effects of finite $L_z$ and then consider
finite $L_y$.

To establish how the finite periodicity in the $z$-direction 
affects the action of the domain wall, we  perform 
numerical calculations for a large range of values of $L_z$.
Since these effects may well vary with the lattice spacing,
i.e. with $aT=1/L_t$, we perform such calculations for two
different values of $L_t$, but at the same value of
physical temperature $T/g^2$. 
Since the finite-size corrections may differ for the 
contributions that are leading and non-leading in $g^2/T$, we 
also perform the calculations for two different values
of $T/g^2$ at the same value of $L_t$. The parameter
values and the corresponding values of $S_w / L_y$
are displayed in Table 5. As we discussed previously, see
(\ref{LyLz}),
the natural scale for the domain wall should be of the order of
$1/agT^{1/2} = \sqrt{\beta L_t}/2$. This is the scale 
that appears in perturbation theory (\ref{gamma},\ref{gamma2}):
$\gamma=\pi\sqrt{\beta L_t}$.
We therefore plot in Fig. 4 the values shown in Table 5
against the scaled lattice length $L_z/\gamma$. We see
that to a good approximation the finite size effects 
are indeed just functions of this scaled length. We
also see that the finite size effects vary from being
very large to being very small over a narrow range of
values of $L_z/\gamma$. Indeed, the domain wall 
effectively disappears for $L_z/\gamma \le 0.8$, and while the 
corrections are still large for $L_z/\gamma \sim 0.9$ to 1.1, they 
have become invisible, within our statistical errors
by the time $L_z/\gamma \sim 1.35$. We therefore see that
the values in Tables 2-4, which all correspond to lattices
satisfying $L_z/\gamma \ge 1.65$, are effectively for
$L_z = \infty$. As a final precaution against the unexpected,
we show in Table 6 some further calculations obtained
for a wide range of values of $\beta$ and $T$. Taking
these together with the values in Tables 1-3 confirms that
the pattern of finite size effects we see in Fig. 4 is
indeed characteristic of  the range of $T$ and $a$
covered by the calculations in this paper.
 
Since the finite size effects appear to be insensitive to the value of
$T$, it is interesting to ask whether they can be reproduced in
leading order perturbation theory, which, after all, is supposedly
exact in the $T = \infty$ limit. The appropriate formalism is that of
the `ball rolling in the inverse potential' as described in Section
6. To that order the finite size effects are functions of the scaled length
$L_z/\gamma$, as can be seen from eqs. (\ref{l-eps},\ref{s-eps}).
In Section \ref{sec:Lz} we solved the equations analytically in
the limit of large $L_z/\gamma$.  In this section we solve them
numerically for all $L_z/\gamma$, and different $L_t$, using $V_{\rm
lat}(q)$. We note that solution does not
exist if the `time' $2L_z$ after which the particle has to come back
oscillating around $q=1/2$ is smaller than the period of small
harmonic oscillations around this point. 
This gives for the minimal value of $L_z$:
$L_z/\gamma=\sqrt{\pi/4\ln2}=1.06$ (for $L_t=\infty$).
This fits
in well with what we observe in Fig. 4. For large
$L_z$ the correction is exponential in ${L_z}^2$ in this order.
This would be difficult to see given our finite
statistical errors. Moreover we know that in the
full theory the correction must ultimately be
exponential in $L_z$.
 From Fig. 4 we see that leading
order perturbation theory describes the observed
finite size effects reasonably well. We also see
that our criterion, $L_z/\gamma \geq 1.65$, should
be a safe one to use for all values of $L_t$.  

We now turn to the finite-size effects associated with the
transverse length of the domain wall, $L_y$. If $L_y$ is small
enough then extra domain walls will be produced as
quantum fluctuations since the main factor in the suppression
of domain wall excitation is $\sim \exp(-\beta S_w)$
and $S_w \propto L_y$. When $L_y$ becomes 
sufficiently large we expect the leading correction to be that due
to roughening, as discussed in Section \ref{sec:rough}. 
As we see in (\ref{alpha-rough})
these corrections should be very small for the parameters 
we use. Of course at asymptotically large $L_y$ the profile of 
the domain wall, defined by
averaging the Polyakov loops over $y$, will broaden as
$\sqrt L_y$. For our values of $L_y$ the broadening of the profile
is small (see Section \ref{sec:rough}). 

To find out what are the corrections at finite $L_y$, we have
performed calculations for a range of values of $L_y$. As
in our study of the $L_z$ dependence we do so
for two values of $L_t$ at the same value of $T/g^2$
and for two values of $T$ at the same value of $L_t$. These
are presented in Table 7. We see that the surface tension
shows no variation with $L_y$ at, say, the $2\sigma$ level
except for $L_y =4$ at $\beta=25$. Here there appears to be
a $\sim 4-8\%$ reduction in the tension. This compares well 
with (\ref{alpha-rough}) 
--- recalling
that $\alpha \sim 6$ for $L_t=2$, although we should certainly 
not expect (\ref{alpha-rough}) to be accurate for such small $L_y$. 

However,
while we see that there are no significant $L_y$ corrections 
to the surface tension, this is certainly not the case for
the Polyakov loop mass, $am_{P}$. We see that not only does this
mass gap show large finite size corrections for the smallest
values of $L_y$, but that these corrections become noticeable
for values of $L_y$ that are not so small. In fact the 
pattern we see is consistent with a relative correction of the
form $\sim \exp(-am_{P} L_y)$.

The extensive finite-size studies we have carried out in this
section show that the potentially dangerous infra-red
effects are in fact under control and that the values of
the surface tension and mass gap that we shall be using
in the next Section, may be regarded as having been
obtained on an infinite system.

\subsection{Surface Tension and Debye Screening Mass}

Even without looking at the detailed numbers in our Tables, there are
two properties of the domain walls that are immediately apparent. One is 
that the probability of such a wall being produced at high $T$ is
very small. The other is that the walls have a finite width. Do
these qualitative features already teach us something?

Consider the finite width. This is significant because at tree level
a wall would have infinite width. That is to say, the width of the wall
would be $\propto L_z$ however large we made $L_z$. One can easily see
this by considering the minimum action configuration that interpolates
between the vacuum with all Polyakov loops +1 and the vacuum with all
loops $-1$. The fact that the walls we generate are of a finite width,
i.e. independent of $L_z$ once $L_z$ is sufficiently large, is
implicit in the finite volume studies of the previous section. 
However it is worth showing this explicitly. We define
the width of the wall, $w(90\%)$, as the distance, in lattice
units, from the value of $z$ where the Polyakov loop is
$90\%$ of its asymptotic value to the point where it is of
the same magnitude but of opposite sign. As a check that
there is nothing special about the choice of $90\%$,
we shall also define a width, $w(2/3)$, on the basis
of 2/3 of the asymptotic value. In Fig. 5 we
show how $w$ varies with $L_z$ in two of the cases where
we have made measurements for a wide range of lattice lengths.
What we see is that the width of the wall does not change with
increasing $L_z$ once $L_z > 2 w(90\%)$: the wall does indeed
have a finite fixed width. 

How does this width depend on $T$?
We extract $w$ from all our large volume calculations
(essentially those listed in Table 2) and plot the results
against $\gamma=\pi(\beta L_t)^{1/2}$ in Fig. 6. This variable, 
as we have seen, determines
the width of the wall in leading-order perturbation theory.  
As a matter of fact, $\gamma\approx w(0.97)$ to that order in 
the continuum limit,i.e. for $L_t=\infty$.
We see from Fig. 6 that for each value of $L_t$ the width varies linearly
with $\gamma$, with significant deviations
only at the very lowest values of $T$. The lines for different $L_t$
are close to each other, but do not coincide. This is 
what one expects in perturbation theory; the width has to
be proportional to the one scale, $\gamma=2\pi/g\sqrt T$, but the
constant of proportionality will suffer lattice spacing
corrections, i.e. will depend on $aT \equiv 1/L_t$. We show
in Fig. 6 the lines one gets in perturbation theory.
Clearly these are consistent.

The corrections to perturbation theory are governed by
$g^2/T = 4L_t/\beta$. We note that for $\beta=7$ at 
$L_t=2$, and for the corresponding values of $\beta$ at
other values of $L_t$, this is $\geq 1$. It is therefore 
remarkable that the calculated
widths deviate by no more than $\sim 10\%$ from the leading-order 
perturbative high $T$ expectations. It would seem that
not only does perturbation theory
work well where we might expect it to, but it even works well
where we have little reason to hope it might. We shall see other
instances of this later on.

The fact that wall-like quantum fluctuations
are very rare tells us that the action a wall
costs is positive: $S_w > 0$. Using (\ref{Fw-Sw})
and noting that, for fixed $L_t$ and $L_y$,
$\partial/\partial\beta \sim \partial/\partial T$
because $\beta \equiv 4/ag^2 = 4L_tT/g^2$, this tells us
that   
\begin{equation}
\label{eqn(13)}
{\partial \over {\partial T}}\left( {F_w \over T}\right) =
{\partial \over {\partial T}}\left({L_y \over L_t}{{\sigma(T)}
\over{T^2}}\right) \geq 0.
\end{equation}
Here we have used the definition of the domain wall surface tension:
\begin{equation}
\label{eqn(14)}
F_w = a L_y \sigma.
\end{equation}
 From (\ref{eqn(13)}) we immediately deduce that 
$\sigma(T) \propto T^\delta$, where $\delta\ge2$,
ignoring possible logs. At the same time we expect that it cannot
increase faster than $T^3$. The perturbative value for
the exponent is, as we have seen, $\delta = 2.5$. So we see
that our qualitative observations already constrain the 
temperature variation of the surface
tension to lie in the interval $\delta=2.5\pm0.5$. In
the quantitative comparisons below we shall attempt
to make the comparison with perturbation theory much more
precise. 

Before doing so it is interesting to ask what the above $T^2$
bound means physically. The following is a simple heuristic
interpretation. Let $m_D$ be the screening mass; then the
wall will have a thickness $O(1/m_D)$. A very crude 
expectation is that $\sigma \propto T^3/m_D$. Now if 
$m_D$ grows faster than $T$ then it cannot be thermally 
excited and the whole high-$T$ picture of a screened
plasma breaks down. The statement that $m_D$ grow no faster
than $T$ is equivalent to our bound that $\sigma$ grows at
least as fast as $T^2$.

We now turn to a quantitative analysis of the results displayed
in Table 4. We recall that the leading order perturbative result
is
\begin{equation}
\label{eqn(16)}
\sigma = {\alpha \over g} T^{2.5}
\end{equation}
The important energy scale is $T$ and the dimensionless
expansion parameter on this scale is $g^2/T$. Thus
the above leading order perturbative result should become
exact in the $T \to \infty$ limit. Naively one might expect 
finite temperature corrections to (\ref{eqn(16)}) to be $O(g^2/T)$
- as in 4 dimensions - but here in d=2+1
there might well be logarithms and the power
itself might be different. We shall see below that this uncertainty 
about the functional form of the leading corrections will limit 
the precision with which we can test perturbation theory. As we have
seen, (\ref{eqn(16)}) is valid both in the continuum and on the lattice,
except that in the latter case the value of the constant $\alpha$ 
will receive calculable lattice spacing corrections. Since T is the 
(largest) important physical energy scale, we expect that the
lattice spacing corrections should depend only on
$aT \equiv 1/L_t$. Moreover, since $\alpha$ is a dimensionless 
physical quantity we expect, on quite general grounds
for a pure gauge theory, that the corrections should
be $O(a^2 T^2) \sim O(1/{L_t}^2)$ for small enough $a$.
The detailed perturbative calculations do in fact bear out this
expectation.

 From (\ref{Fw-Sw},\ref{eqn(14)},\ref{eqn(16)}) we obtain
\begin{equation}
\label{eqn(17)}
S_w = {\partial \over {\partial \beta}}
\left( {F_w \over T} \right) =
{{\alpha L_y} \over {8{L_t}^2}}\left(g^2 \over T\right)^{1/2}
\end{equation}
to leading order in perturbation theory. We have calculated
the value of $\alpha$ as a function of $aT\equiv 1/L_t$, using lattice 
perturbation theory (\ref{alpha-lat}) and show a selection of these
values in Table 8. Since the lattice spacing corrections depend on $L_t$,
we shall mostly examine the $T$ dependence
at fixed $L_t$. Indeed, we have chosen our parameters 
with this in mind. However before doing so we briefly take the 
alternative approach of varying $L_t$ at fixed $\beta$.
For this purpose it is useful to rewrite (\ref{eqn(17)}) in the form
\begin{equation}
\label{eqn(18)}
{S_w \over L_y} = 
2{{\alpha (L_t)} \over \beta^2} {\left({T \over g^2}\right)}^{3/2}
\end{equation}
Now,  as we see in Table 4, it is only for 
$\beta=75.0$ that we have a usefully large range of $L_t$ values. 
So we plot these values of $S_w/L_y$ against
$T/g^2 \equiv \beta/4L_t$, in Fig. 7. We use logarithmic scales so that
a power dependence in $T$ will appear as a straight line. And, indeed,
the calculated values do fall on a straight line to a good approximation.
The slope suggests a variation $\propto T^{1.6}$ which is close
to the perturbative variation of $\propto T^{1.5}$. To carry the
comparison further we need to take into account the fact that
in addition to the predicted $T^{1.5}$ behaviour, there are
different $aT$ lattice corrections at the different values of $L_t$.
That is to say, the behaviour predicted by perturbation theory is 
$\propto \alpha(L_t) T^{1.5}$ and not just the power of T. In Fig. 7 we 
show the complete leading order perturbative prediction and we see
that its variation fits that of our data very well. The normalisation
is not exactly right --- but the difference is small and is
decreasing with increasing $T$ just as one would expect
from a higher order correction in $g^2/T$. If we do indeed try to fit
the data with a higher order correction we find that an
$O(g^2/T)$ correction will not work; but 
\begin{equation}
\label{eqn(19)}
\sigma = {\alpha \over g} T^{2.5} \left(1+0.13 
\left({g^2 \over T}\right)^{0.5}\right)
\end{equation}
fits perfectly well. The precise power of the correction is not to
be taken too seriously of course; it may be an effective power
that partially simulates the effects of logarithms in our limited
range of $T/g^2$ (this range being roughly 3 to 10). 

We conclude from the above comparison that at temperatures  
$T/g^2 \geq 3$ the $T$ dependence of $\sigma$ is 
very close to the perturbative expectation of $T^{2.5}$; indeed 
what we find is $\sim T^{2.6}$. Moreover this
slight difference almost entirely disappears when we include
the perturbatively calculated $aT$ corrections to $\alpha$.
The remnant discrepancy, a few percent, decreases as $T$
increases and so is consistent with being a higher order
perturbative correction, as, for example, in (\ref{eqn(19)}). 
Unfortunately the fact that we
do not know the precise functional form of this
correction, prevents us from carrying out the quantitative comparison
any further than this.

We turn now to consider the bulk of our calculations.
We shall consider the $T$ dependence at fixed values of
$L_t \equiv 1/aT$, so that the lattice spacing corrections do
not vary with $T$. At the same time, by performing 
calculations for several values of $aT$ we can
see whether the perturbative predictions show any
sign of failing as one approaches the continuum limit. 
To compare our results to the perturbative prediction
we define a quantity $\alpha_{\rm eff}$ by
\begin{equation}
\label{eqn(20)}
\alpha_{\rm eff} = {{\beta^2}\over 2}
{\left({g^2 \over T}\right)}^{3/2}
{{S_w}\over L_y}
\end{equation}
As we see from (\ref{eqn(18)}), to leading order in perturbation
theory $\alpha_{\rm eff}(L_t) = \alpha(L_t)$. In Fig. 8
we display our Monte Carlo results for $\alpha_{\rm eff}$
as a function of $g^2/T$. The perturbative value
of $\alpha$ is also shown, as a horizontal broken line,
in each case. We observe that the calculated surface tension 
indeed approaches the perturbative value as $T$ increases.
At the highest values of $T$ the discrepancy is no more
than a few percent. Our data is clearly compatible
with the leading perturbative result being exact in
the $T \to \infty$ limit. 

In Fig. 9 we plot the ratio $\alpha_{\rm eff}/\alpha$
for all our data. We see that to a good approximation
it is a function only of $g^2/T$. This
supports the idea that the small differences we
see between the full and perturbative surface
tensions are in fact due to higher order corrections 
in $g^2/T$. We note that these higher order corrections
are small over our whole range of $T$. Indeed, even
for $g^2/T \sim 1.1$ the correction is only about
$25\%$ of the leading term. Recall that this temperature
corresponds to only about twice the deconfining temperature.
It is quite extraordinary that lowest order perturbation
theory should still be so accurate at such low temperatures.

If we knew the functional form of the leading correction,
we would attempt to extrapolate our `measured' values
to $T = \infty$. Unfortunately we do not. In $d=3+1$ we
would expect the correction to be simply $\propto g^2/T$.
However in $d=2+1$ there are infrared logarithms which may also
resum into a power of $g^2/T$. So it seems 
reasonable that the correction should be some 
effective power of $g^2/T$ that lies between 0.5 and 1 in
our range of $T$. If we fit our data with a form
\begin{equation}
\label{eqn(21)}
\sigma = {\alpha \over g} T^{2.5} \left(1+ c 
{\left({{g^2} \over T}\right)}^{\varepsilon}\right)
\end{equation}
then we find that while $\varepsilon =1$ is excluded, powers 
near $\varepsilon =0.65$ work perfectly well, as we see
in Fig. 10. The intercepts of these fits are compatible
with the leading order perturbative predictions. 

The final question in this context is whether there
is any sign that this agreement with perturbation
theory breaks down as we approach the continuum limit.
The above detailed comparisons have involved 
reducing the lattice spacing by a factor of 2,
i.e. from $aT=1/2$ to $aT=1/4$. As we see in Figs 8,9
there is no sign of any lattice spacing dependence
other than that calculable in perturbation theory.
To go further we also  calculated the surface
tension with $aT=1/L_t=1/6$ at $\beta=75$. In Fig. 11
we show the value of $\alpha_{\rm eff}/\alpha(L_t)$
for this $L_t=6$ point as well as for other $L_t$ values 
at approximately the same temperature (using 
$T/g^2 = \beta/4L_t$). Note that since the $T$ dependence of
this quantity is very weak, as we have seen in Fig. 8, 
it is not important to get $T$ exactly the same at
different values of $L_t$. Note also that this is
a reasonably high value of $T$ at which to perform such
a test: the deviations from leading order perturbation
theory are only at the $7\%$ level or so.
We see from Fig. 11 that there is no significant
deviation from perturbation theory with decreasing
$a$ even down to $aT = 1/6$. It therefore seems extremely 
unlikely that the {\it continuum} surface tension
will not be equally well described by perturbation
theory.

So far we have focused on the surface tension of the domain wall.
However perturbation theory also makes predictions for 
more detailed aspects of the domain wall, such as its
profile. In Fig. 12 we show how these predictions
compare with our Monte Carlo calculated Polyakov loop profiles 
for several parameter values. We see very good agreement
with the main discrepancy arising from the fact that
the vacuum values of Polyakov loops are $\pm 1$ in leading
order perturbation theory. This is primarily an artifact 
arising from the fact that $\langle \theta\rangle =0$  does
not mean that $\langle \cos \theta \rangle = 1$. A less trivial,
but almost invisible difference is that the approach to
the vacuum at large distances is Gaussian for perturbation
theory, but exponential in the full theory.

In Fig. 13 we show some typical examples of profiles of
different components of the action density. Again
there is very good agreement with perturbation theory.

We turn now to the Debye screening mass, $m_D$.
When we expand the trace of the Polyakov loop the first
non-trivial term is $\sim {A_0}^2$ so that we expect
correlations of Polyakov loops to receive contributions
from  the exchange of pairs of screened electric gluons.  
However, as emphasised by \cite{arnold},
at higher order in the coupling, $g^2/T$,
they also receive contributions from the magnetic gluon
and larger numbers of gluons of both kinds. The question
naturally arises: which kind of contribution are we seeing
when we extract the masses displayed in Table 3? If
we expand the normalised correlation function, $C(z)$,
of Polyakov loop operators, $P$, in energy eigenstates
\begin{equation}
\label{eqn(22)}
{{C(z)} \over {C(0)}} = \sum_{n} c_n \exp 
{\left( - E_n z \right)}
\end{equation}
then the coefficients, $c_n$, are just the 
amplitudes squared
\begin{equation}
\label{eqn(23)}
c_n = {\vert  \langle {\rm vac} \vert P \vert n \rangle \vert}^2
\end{equation}
with normalisation $\sum c_n = 1$. From our above discussion
we expect $c_n$ to be largest for the state with
two screened electric gluons. Of course, for large enough
$z$, the correlation function will be dominated by
the lightest energy, $E_{min}$, irrespective of the value
of $c_n$ (as long as it is non-zero), and so we need
to check whether the masses we have listed in
Table 3 do indeed correspond to states for which
$c_n$ is large. In Table 9 we list the overlaps, $c_n$,
for the states whose masses are listed in Table 3.
We provide both the overlap onto the best smeared
Polyakov loop operator and the overlap onto
the simple, unsmeared Polyakov loop operator. We
do so for the extreme values of $T$ at each value of $L_t$.
We see that in all cases the normalised matrix
element squared is $\ge 79\%$. Given that 
we expect the magnetic gluon etc contributions to 
receive relative suppressions that are powers
of $g^2/T$, which is $ \sim 0.1$ at our highest $\beta$ values,
it is clear that the masses we have obtained
belong to states that have no such suppression. We
therefore claim that we can read off $m_D$
from Table 3 using $ am_D = 0.5 am_{P}$.

As we have already seen, the Debye screening mass is
infinite at 1-loop, due to an infrared divergence.
These divergences go away if we use non-zero
gluon masses in the diagrams and so one can try to
do a self-consistent calculation for the mass, $m_D$.
This has been done by D'Hoker \cite{DH82} who obtains
\begin{equation}
\label{eqn(24)}\label{D'Hoker}
{m_D}^2 = {{g^2 T} \over \pi} \left[
\ln \left({T \over {m_D}}\right) - 1
+ O\left(1 \over {(\ln(T/m_D))^\zeta}\right)\right]
\end{equation}
where $\zeta > 0$. We see from (\ref{eqn(24)}) that $m_D/g^2$ is
a function only of $T/g^2$. Naively we would, of course, have 
expected $m_D \sim g \sqrt T$ since, as we have seen, that is the
scale for the domain wall. And indeed this is the leading
$T$ dependence in (\ref{eqn(24)}), up to a weakly varying
logarithm. However the
correction term is down only by logarithms and so we
might not be surprised to find the comparison with
perturbation theory not as good as for the properties
of the domain wall, where the corrections are
powers of $g^2/T$. 

In Fig. 14 we plot our masses against $T/g^2$. What we 
actually choose to plot is ${m_D}^2/g^2 T$
since that way we factor out the supposedly dominant 
$g^2 T$ factor, and so expose the remaining variation more clearly.
We also show the leading perturbative prediction as obtained 
from the first term of (\ref{eqn(24)}). The first observation 
is that the dominant variation of $m_D$ is indeed $\sim g T^{1/2}$.
However there is a substantial additional variation which
is too strong to be due to corrections that are higher order
in $g^2/T$. Indeed if we try a fit of the form
\begin{equation}
\label{eqn(25)}
{m_D}^2 = g^2 T \left(c_0 + 
c_1 {\left({g^2 \over T}\right)}^{\varepsilon} \right)
\end{equation}
we find that it simply does not work,
even if we remove the lowest $T$ point from the fit.
Indeed, as we see, this additional $T$ variation is quite similar to 
that obtained from (\ref{eqn(24)}). However the normalisation is 
completely off. Even allowing for the $L_t$ dependence in our
values of $m_D$, there is a discrepancy of about
a factor of 3 with perturbation theory. Since the 
corrections in (\ref{eqn(24)}) are only logarithmic we cannot, of course, 
claim a contradiction with perturbation theory.
However it is worrying that there is no trend towards
a reduction of the discrepancy even at our highest values
of $T$ where $1/\ln(T/g^2) \sim 0.4$. This is in stark
contrast to other properties of the domain wall where we found
the corrections to leading order perturbation theory to be
small even for $g^2/T \sim 1$.

\section{Conclusions}

In this paper we have carried out extensive perturbative and
Monte Carlo calculations
of the high-$T$ domain walls  which are associated with
the spontaneous breaking of a $Z(N)$ symmetry in $SU(N)$ gauge
theories. As we argued, these walls can be viewed as
't Hooft disorder loops which become squeezed once the extent
in Euclidean time becomes small enough, as it does at high enough 
$T$. Our purpose has been to test high-$T$ perturbation
theory and to establish whether these unusual objects do
really exist in the Euclidean continuum theory.
In order to be able to obtain numerical results of sufficient
accuracy to be convincing, we have worked with the simplest
theory that one may consider as realistic in this context:
the $SU(2)$ gauge theory in 2+1 dimensions.

This kind of calculation is difficult for several reasons. Firstly
the potential problems are infrared and it is therefore
crucial to make sure that the volumes used are large enough.
This means not only doing detailed numerical finite-size
studies, but also calculating the appropriate finite-volume
corrections in perturbation theory. For example, in this paper
we have shown how one can calculate the effects of roughening 
on these domain walls. Secondly one-loop perturbation theory 
becomes exact, at best, only in the limit $T \to \infty$. 
Now at fixed $aT$ the size of the lattice will grow 
as $T^{1/2}$, in lattice units, 
simply because the width of the wall is $O(1/T^{1/2})$ in
$d=2+1$. This makes it difficult to simultaneously get
close to the continuum limit, where $aT$ is small, and to
reach very high values of $T$. For this reason we have
performed the perturbative calculations not only for the
continuum theory but also for the lattice theory. In this
way we can directly compare perturbation theory to the full
non-perturbative results one gets from Monte Carlo
simulations.

In practice we have carried out simulations for temperatures 
as high as $\sim 30T_c$, where $T_c$ is the deconfining 
temperature, and for lattice spacings as small as $1/6T$.
(This comparison with $1/T$ is appropriate, since $T$ is 
the largest important physical energy scale in the problem). 
At the highest values of $T$ our numerically obtained values of
the surface tension agree with the perturbative predictions
at the percent level and this is so at all our values of
$a$. Moreover there is agreement at the $25\%$ level or so,
even at temperatures as low as
$g^2/T \sim 1$. When we look at the variation
with $a$, over the range $a=1/2T$ to $1/6T$, we again
find excellent agreement with perturbation theory, with
not the slightest hint of any anomaly developing as 
$a \to 0$.

At the same time we have obtained perturbative
predictions for the more detailed properties of the
wall, such as the action density profile. These
calculations agree very well with our simulations. 
These profiles are interesting in themselves and
show, for example, that $\langle B^2 - E^2 \rangle$ ---
the thermal energy (in Euclidean space) --- becomes 
negative inside the wall (see Section \ref{sec:profs}).

The only quantity where we fail to find agreement
with perturbation theory is for the Debye screening
mass. However here the perturbative calculations are
not straightforward and the corrections
are expected to be a power of $1/\log(T/g^2)$ and not
of $g^2/T$. So there is no good reason to read too
much significance into this particular discrepancy.

Our conclusion is that these high-$T$ domain walls
are present in the Euclidean theory, exactly as predicted 
by perturbation theory, both on the lattice and in the 
continuum. Since these walls are quantum rather 
than semiclassical objects, they provide a severe testing 
ground for high-$T$ perturbation theory. Its success
here lends strong support to the usual pragmatic assumption 
that perturbation theory
reliably describes gauge theories at high temperatures.

\paragraph{Acknowledgement}

We are grateful for support under the Oxford Particle Theory
Grant GR/K/55752. In addition A.M. is grateful for support 
from Kobe Steel, the Queen's Trust, AFUW and IFUW; 
M.S. for a scholarship from Jesus College and support 
under the grant NSF-PHY92-00148; C.K.A. for the support 
of the Royal Society Exchange Program.
The numerical calculations were performed partly on the
RAL Cray YMP under PPARC grant GR/J21408, and partly
on alpha workstations in Oxford Theoretical Physics.

\pagestyle{empty}



\begin{table}[h]
\caption[]{Average plaquette on lattices without a twist.}
\vspace{10mm}
\small
\begin{center}
\begin{tabular}{|l|c|c|c|c|}
\hline
$L_t$ & $\beta$ & lattice & no. sweeps & $1 - s_{\rm nt}$ \\
\hline
\hline
2  &  100.0 &  30$\times$60   & 820,000  & 0.98998945(28)  \\
   &  75.0  &  26$\times$50   & 400,000  & 0.98664608(71)  \\
   &  50.0  &  50$\times$60   & 400,000  & 0.97995063(66)  \\
   &        &  40$\times$60   & 240,000  & 0.97994775(87)   \\
   &        &  30$\times$60   & 200,000  & 0.979994952(117)   \\
   &        &  20$\times$40   & 440,000  & 0.97995159(130)   \\
   &        &  16$\times$60   & 400,000  & 0.97994859(101)   \\
   &  25.0  &  30$\times$48   & 400,000  & 0.9597749(15)  \\
   &        &  20$\times$48   & 400,000  & 0.9597747(21)   \\
   &        &  12$\times$60   & 400,000  & 0.9597749(23)   \\
   &        &  12$\times$40   & 400,000  & 0.9597752(23)   \\
   &        &  12$\times$30   & 400,000  & 0.9597734(34)   \\
   &        &  12$\times$26   & 400,000  & 0.9597801(38)   \\
   &        &  12$\times$20   & 400,000  & 0.9597755(47)   \\
   &  15.0  &  12$\times$24   & 400,000  & 0.9326535(71)  \\
   &  7.0   &  12$\times$20   & 400,000  & 0.8533597(159)  \\
\hline
3  &  112.5 &  40$\times$80   & 800,000  & 0.99109673(14)   \\
   &  75.00 &  30$\times$80   & 360,000  & 0.98663259(35) \\
   &        &  30$\times$60   & 800,000  & 0.98663236(33) \\
   &        &  30$\times$50   & 196,000  & 0.98663197(77) \\
   &        &  24$\times$80   & 200,000  & 0.98663262(48) \\
   &        &  18$\times$80   & 200,000  & 0.98663250(37) \\
   &  37.47 &  18$\times$60   & 400,000  & 0.97317001(126) \\
   &        &  18$\times$46   & 400,000  & 0.97316810(130) \\
   &  22.45 &  18$\times$46   & 400,000  & 0.95504162(204) \\
   &  10.27 &  18$\times$32   & 400,000  & 0.9004471(62) \\
\hline
4  &  99.97 &  40$\times$80   & 800,000  & 0.98997622(15) \\
   &  49.95 &  24$\times$60   & 800,000  & 0.97989112(39) \\
   &  29.91 &  24$\times$48   & 800,000  & 0.96631709(87) \\
   &  13.81 &  24$\times$40   & 800,000  & 0.92634313(179) \\
\hline
5  &  62.40 &  30$\times$100  & 400,000  & 0.98391671(38) \\
   &  37.33 &  30$\times$80   & 400,000  & 0.97304743(54) \\
\hline
6  &  75.00 &  36$\times$90   & 840,000  & 0.98662597(12) \\
\hline
\end{tabular}
\end{center}
\end{table}



\begin{table}[h]
\caption[]{Average plaquette on lattices with a twist.}
\vspace{10mm}
\small
\begin{center}
\begin{tabular}{|l|c|c|c|c|}
\hline
$L_t$ & $\beta$ & lattice & no. sweeps & $1 - s_{\rm tw}$ \\
\hline
\hline
2  &  100.0 &  30$\times$80   & 800,000  & 0.98987388(26)  \\
   &  75.0  &  26$\times$70   & 400,000  & 0.98649347(50)  \\
   &  50.0  &  50$\times$60   & 400,000  & 0.97972630(65)  \\
   &        &  30$\times$60   & 200,000  & 0.97972542(91)   \\
   &        &  20$\times$60   & 400,000  & 0.97972801(91)   \\
   &        &  16$\times$60   & 400,000  & 0.97972622(124)   \\
   &  25.0  &  30$\times$48   & 400,000  & 0.9593664(15)  \\
   &        &  20$\times$48   & 400,000  & 0.9593643(21)   \\
   &        &  12$\times$60   & 400,000  & 0.9594461(23)   \\
   &        &  12$\times$48   & 400,000  & 0.9593680(29)   \\
   &        &  12$\times$40   & 400,000  & 0.9592867(25)   \\
   &  15.0  &  12$\times$36   & 400,000  & 0.9319060(58)  \\
   &  7.0   &  12$\times$30   & 400,000  & 0.8519115(137)  \\
   &        &  12$\times$20   & 400,000  & 0.8512002(174) \\
\hline
3  &  112.5 &  40$\times$100  & 650,000  & 0.99106642(14)   \\
   &  75.00 &  30$\times$100  & 350,000  & 0.98659536(35) \\
   &        &  30$\times$80   & 350,000  & 0.98658592(37) \\
   &  37.47 &  18$\times$60   & 400,000  & 0.97307998(106) \\
   &  22.45 &  18$\times$46   & 400,000  & 0.95488588(203) \\
   &  10.27 &  18$\times$32   & 400,000  & 0.9000855(59) \\
\hline
4  &  99.97 &  40$\times$120  & 800,000  & 0.98996380(18) \\
   &  49.95 &  24$\times$84   & 800,000  & 0.97986529(39) \\
   &  29.91 &  24$\times$64   & 800,000  & 0.96627028(56) \\
   &  13.81 &  24$\times$48   & 800,000  & 0.92624494(158) \\
   &        &  24$\times$40   & 400,000  & 0.92622498(271)\\
\hline
6  &  75.00 &  36$\times$130  & 600,000  & 0.98662125(13) \\
\hline
\end{tabular}
\end{center}
\end{table}



\begin{table}[h]
\caption[]{Masses obtained from correlations of Polyakov loops.} 
\vspace{10mm}
\small
\begin{center}
\begin{tabular}{|l|c|c|c|}
\hline
$L_t$ & $\beta$ & lattice & $a\:m_p$ \\
\hline
\hline
2  &  100.0 &  30$\times$60   & 0.315(7)  \\
   &  75.0  &  26$\times$50   & 0.352(4)  \\
   &  50.0  &  50$\times$60   & 0.414(5)  \\
   &        &  40$\times$60   & 0.418(6)   \\
   &        &  30$\times$60   & 0.413(6)   \\
   &        &  20$\times$40   & 0.409(6)   \\
   &        &  16$\times$60   & 0.410(5)   \\
   &  25.0  &  30$\times$60   & 0.526(6)  \\
   &        &  20$\times$48   & 0.529(6)   \\
   &        &  12$\times$60   & 0.529(6)   \\
   &        &  12$\times$40   & 0.525(7)   \\
   &        &  12$\times$30   & 0.520(9)   \\
   &        &  12$\times$26   & 0.533(9)   \\
   &        &  12$\times$20   & 0.522(10)   \\
   &  15.0  &  12$\times$24   & 0.632(8)  \\
   &  7.0   &  12$\times$20   & 0.771(7)  \\
\hline
3  &  112.5 &  40$\times$80   & 0.2188(44)   \\
   &  75.00 &  30$\times$80   & 0.2586(53) \\
   &        &  30$\times$60   & 0.2588(50) \\
   &        &  30$\times$50   & 0.2561(109) \\
   &        &  24$\times$80   & 0.2508(126) \\
   &        &  18$\times$80   & 0.2511(97) \\
   &  37.47 &  18$\times$60   & 0.3351(59) \\
   &        &  18$\times$46   & 0.3399(66) \\
   &  22.45 &  18$\times$46   & 0.395(6) \\
   &  10.27 &  18$\times$32   & 0.489(4) \\
\hline
4  &  99.97 &  40$\times$80   & 0.1897(18) \\
   &  49.95 &  24$\times$60   & 0.2420(19) \\
   &  29.91 &  24$\times$48   & 0.2802(23) \\
   &  13.81 &  24$\times$40   & 0.3411(40) \\
\hline
5  &  62.40 &  30$\times$100  & 0.177(7) \\
   &  37.33 &  30$\times$80   & 0.2226(42) \\
\hline
6  &  75.00 &  36$\times$90   & 0.1541(23) \\
\hline
\end{tabular}
\end{center}
\end{table}



\begin{table}[h]
\caption[]{The action density of the domain wall
per unit length and averaged Polyakov loop masses.}
\vspace{10mm}
\small
\begin{center}
\begin{tabular}{|l|c|c|c|}
\hline
$L_t$ & $\beta$ & $S_w/L_y$ & $a\:m_p$ \\
\hline
2  &  100.0 &  0.055474(183) & 0.315(7)  \\
   &  75.0  &  0.064096(365) & 0.352(4)  \\
   &  50.0  &  0.08048(22)   & 0.4123(25)  \\
   &  25.0  &  0.11773(38)   & 0.5269(27)  \\
   &  15.0  &  0.1615(20)    & 0.632(8)  \\
   &  7.0   &  0.2598(29)    & 0.771(7)  \\
\hline
3  &  112.5 &  0.02728(18)  & 0.2188(44)   \\
   &  75.00 &  0.03342(30)  & 0.2572(32) \\
   &  37.47 &  0.04811(76)  & 0.3372(44) \\
   &  22.45 &  0.06448(119) & 0.395(6) \\
   &  10.27 &  0.1041(25)   & 0.489(4) \\
\hline
4  &  99.97 &  0.01788(34)  & 0.1897(18) \\
   &  49.95 &  0.02604(56)  & 0.2420(19) \\
   &  29.91 &  0.03595(80)  & 0.2802(23) \\
   &  13.81 &  0.05661(119) & 0.3411(40) \\
\hline
5  &  62.40 &     -         & 0.177(7) \\
   &  37.33 &     -         & 0.223(4) \\
\hline
6  &  75.00 &  0.01104(41)  & 0.1541(23) \\
\hline
\end{tabular}
\end{center}
\end{table}


\begin{table}[h]
\caption[]{The action density of the domain wall
as a function of the lattice length, $L_z$, for selected
values of $\beta$ and $L_t$.} 
\vspace{10mm}
\small
\begin{center}
\begin{tabular}{|l|c|c|c|}
\hline
$\beta$ & $L_t$ & $L_z$ & $S_w/L_y$ \\
\hline
25 &  2  &  60 &  0.11840(101)  \\
   &     &  48 &  0.11722(96)  \\
   &     &  40 &  0.11719(71)  \\
   &     &  30 &  0.11792(84)  \\
   &     &  26 &  0.11497(77)  \\
   &     &  20 &  0.06947(84)  \\
\hline
50 &  2  &  60  & 0.07979(36) \\
   &     &  40  & 0.07877(36) \\
   &     &  34  & 0.07169(35) \\
   &     &  30  & 0.04981(52) \\
   &     &  26  & 0.01735(44) \\
\hline
75 &  3  &  100 & 0.03340(39) \\
   &     &  80  & 0.03352(32) \\
   &     &  60  & 0.03239(22) \\
   &     &  54  & 0.02940(35) \\
   &     &  50  & 0.02354(44) \\
   &     &  46  & 0.01577(44) \\
   &     &  40  & 0.00640(60) \\
\hline
\end{tabular}
\end{center}
\end{table}



\begin{table}[h]
\caption[]{Some additional values of the action density 
of the domain wall, for values of $L_z$ smaller than in Table 2.}
\vspace{10mm}
\small
\begin{center}
\begin{tabular}{|l|c|c|c|}
\hline
$L_t$ & $\beta$ & $L_z$ & $S_w/L_y$ \\
\hline
2  &  100.0  &  60 &  0.05498(18)  \\
   &  75.0   &  50 &  0.06391(28)  \\
   &  15.0   &  24 &  0.1603(14)  \\
\hline
3  &  112.5  &  60  & 0.02701(16) \\
   &  37.47  &  40  & 0.04863(66) \\
\hline
4  &  99.97   &  80  & 0.01688(26) \\
   &  49.95   &  60  & 0.02054(58) \\
   &  29.91   &  48  & 0.03557(82) \\
\hline
\end{tabular}
\end{center}
\end{table}



\begin{table}[h]
\caption[]{Variation of the action density of the domain wall
and Polyakov loop masses with the length, $L_y$, of the wall.}
\vspace{10mm}
\small
\begin{center}
\begin{tabular}{|l|c|c|c|c|}
\hline
$\beta$ & $L_t$ & $L_y$ & $S_w/L_y$ & $a\:m_p$ \\
\hline
25 &  2  &  30 &  0.11765(59)  & 0.529(3)  \\
   &     &  20 &  0.11820(86)  & 0.529(5)  \\
   &     &  12 &  0.1184(10)   & 0.529(6)  \\
   &     &  8  &  0.1162(15)   & 0.492(11) \\
   &     &  4  &  0.1091(21)   & $\leq$ 0.384(9) \\
\hline
50 &  2  &  50  & 0.08076(34)  & 0.396(12)  \\
   &     &  40  &     -        & 0.416(4)   \\
   &     &  30  & 0.08068(53)  & 0.410(10)  \\
   &     &  20  & 0.07991(61)  & 0.409(6)   \\
   &     &  16  & 0.08005(58)  & 0.386(15)  \\
   &     &  12  & 0.07937(64)  & 0.375(9)   \\
   &     &  8   & 0.07860(65)  & 0.335(8)   \\
   &     &  4   & 0.07870(179) & 0.272(12) \\
\hline
75 &  3  &  120 & 0.03271(54)  & -   \\
   &     &  60  & 0.03409(47)  & -   \\
   &     &  30  & 0.03347(28)  & -   \\
   &     &  24  &     -        & 0.251(13) \\
   &     &  18  &     -        & 0.251(7)  \\
   &     &  12  & 0.03303(65)  & 0.218(7)  \\
   &     &  6   &     -        & $\leq$ 0.183(6) \\
\hline
\end{tabular}
\end{center}
\end{table}



\begin{table}[hbt]
\caption[]{Perturbative values of the constant, $\alpha$,
in the interface tension.} 
\vspace{10mm}
\small
\begin{center}
\begin{tabular}{|l|l|}
\hline
$L_t$ & $\alpha$  \\
\hline \hline
2 & 6.024 \\
3 & 5.655 \\
4 & 5.409 \\
5 & 5.284 \\
6 & 5.221 \\
8 & 5.165 \\
10 & 5.142 \\
20 & 5.113 \\
$\infty$ & 5.104 \\
\hline 
\end{tabular}
\end{center}
\end{table}



\begin{table}[h]
\caption[]{Overlaps of Polyakov loop operators
onto the states coresponding to our values of $m_p$.}
\vspace{10mm}
\small
\begin{center}
\begin{tabular}{|l|c|c|c|}
\hline
$L_t$ & $\beta$ & $O_{\rm best}$ & $O_{\rm Bl=1}$ \\
\hline
\hline
2  &  100.0 &  0.88   & 0.81  \\
   &  25.0  &  0.93   & 0.82  \\
   &  7.0   &  0.98   & 0.90  \\
\hline
3  &  112.5 &  0.84   & 0.79   \\
   &  10.27 &  0.98   & 0.89 \\
\hline
4  &  99.97 &  0.92   & 0.81 \\
   &  13.81 &  0.97   & 0.86 \\
\hline
6  &  75.00 &  0.89   & 0.80 \\
\hline
\end{tabular}
\end{center}
\end{table}



\input epsf.sty

\pagestyle{empty}

\begin{figure}[h]
\caption[1]{Location of the twist. The plaquettes that are indicated 
will appear with a factor of $-1$ in the twisted action.}
\vskip 20pt
\epsfxsize 4in
\epsfbox{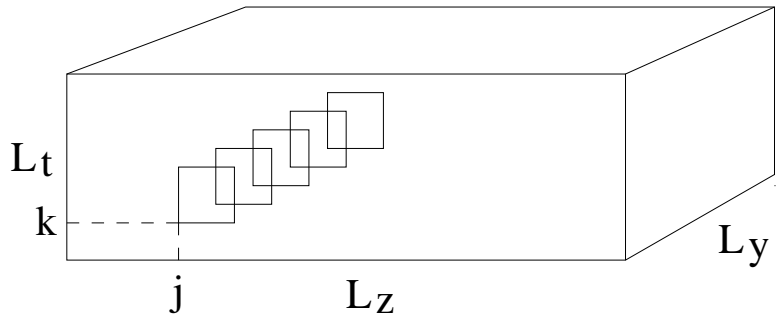}
\end{figure}

\begin{figure}[h]
\caption[2]{Values of Polyakov loops on typical field configurations 
with a domain wall: for $L_t=2$ at: (a) $\beta=100$, (b) $\beta=7$.}
\vskip 20pt
\epsfbox{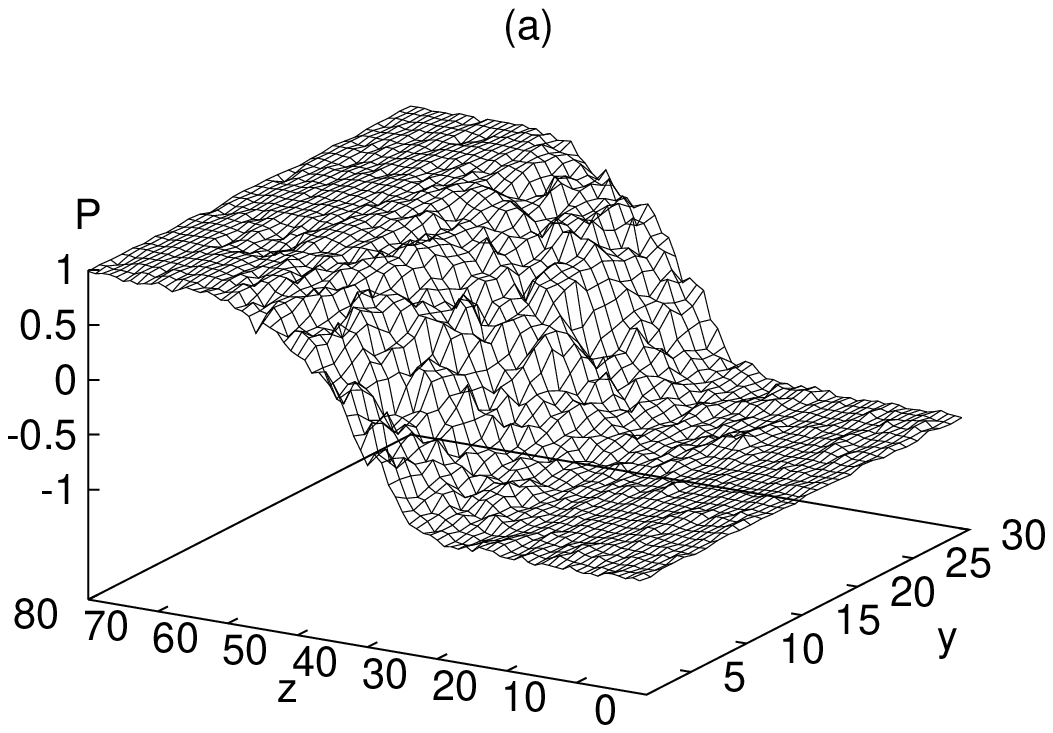}
\epsfbox{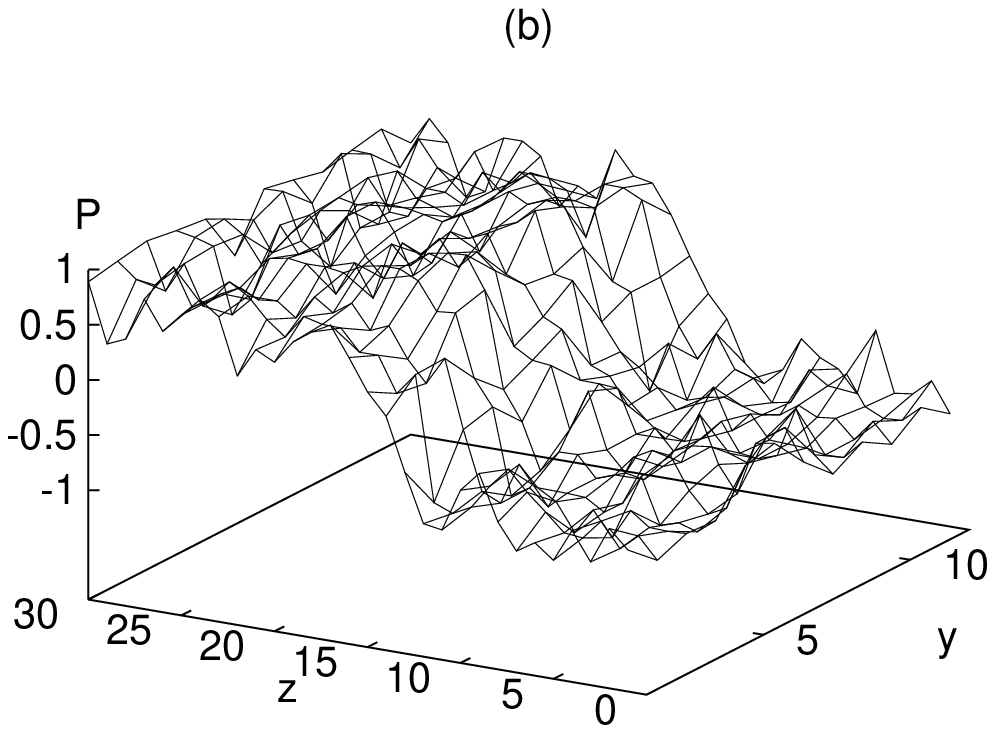}
\end{figure}

\begin{figure}[h]
\caption[3]{Effective masses from Polyakov loop correlations ($+$)
and from the tails of domain walls ($\diamond$) for
(a) $\beta=112.5$ with $L_t=3$, (b) $\beta=25$ with $L_t=2$.}
\vskip 20pt
\setlength{\unitlength}{0.1bp}
\special{!
/gnudict 40 dict def
gnudict begin
/Color false def
/Solid false def
/gnulinewidth 5.000 def
/vshift -33 def
/dl {10 mul} def
/hpt 31.5 def
/vpt 31.5 def
/M {moveto} bind def
/L {lineto} bind def
/R {rmoveto} bind def
/V {rlineto} bind def
/vpt2 vpt 2 mul def
/hpt2 hpt 2 mul def
/Lshow { currentpoint stroke M
  0 vshift R show } def
/Rshow { currentpoint stroke M
  dup stringwidth pop neg vshift R show } def
/Cshow { currentpoint stroke M
  dup stringwidth pop -2 div vshift R show } def
/DL { Color {setrgbcolor Solid {pop []} if 0 setdash }
 {pop pop pop Solid {pop []} if 0 setdash} ifelse } def
/BL { stroke gnulinewidth 2 mul setlinewidth } def
/AL { stroke gnulinewidth 2 div setlinewidth } def
/PL { stroke gnulinewidth setlinewidth } def
/LTb { BL [] 0 0 0 DL } def
/LTa { AL [1 dl 2 dl] 0 setdash 0 0 0 setrgbcolor } def
/LT0 { PL [] 0 1 0 DL } def
/LT1 { PL [4 dl 2 dl] 0 0 1 DL } def
/LT2 { PL [2 dl 3 dl] 1 0 0 DL } def
/LT3 { PL [1 dl 1.5 dl] 1 0 1 DL } def
/LT4 { PL [5 dl 2 dl 1 dl 2 dl] 0 1 1 DL } def
/LT5 { PL [4 dl 3 dl 1 dl 3 dl] 1 1 0 DL } def
/LT6 { PL [2 dl 2 dl 2 dl 4 dl] 0 0 0 DL } def
/LT7 { PL [2 dl 2 dl 2 dl 2 dl 2 dl 4 dl] 1 0.3 0 DL } def
/LT8 { PL [2 dl 2 dl 2 dl 2 dl 2 dl 2 dl 2 dl 4 dl] 0.5 0.5 0.5 DL } def
/P { stroke [] 0 setdash
  currentlinewidth 2 div sub M
  0 currentlinewidth V stroke } def
/D { stroke [] 0 setdash 2 copy vpt add M
  hpt neg vpt neg V hpt vpt neg V
  hpt vpt V hpt neg vpt V closepath stroke
  P } def
/A { stroke [] 0 setdash vpt sub M 0 vpt2 V
  currentpoint stroke M
  hpt neg vpt neg R hpt2 0 V stroke
  } def
/B { stroke [] 0 setdash 2 copy exch hpt sub exch vpt add M
  0 vpt2 neg V hpt2 0 V 0 vpt2 V
  hpt2 neg 0 V closepath stroke
  P } def
/C { stroke [] 0 setdash exch hpt sub exch vpt add M
  hpt2 vpt2 neg V currentpoint stroke M
  hpt2 neg 0 R hpt2 vpt2 V stroke } def
/T { stroke [] 0 setdash 2 copy vpt 1.12 mul add M
  hpt neg vpt -1.62 mul V
  hpt 2 mul 0 V
  hpt neg vpt 1.62 mul V closepath stroke
  P  } def
/S { 2 copy A C} def
end
}
\begin{picture}(2880,1728)(0,0)
\special{"
gnudict begin
gsave
50 50 translate
0.100 0.100 scale
0 setgray
/Helvetica findfont 100 scalefont setfont
newpath
-500.000000 -500.000000 translate
LTa
600 251 M
2097 0 V
600 251 M
0 1326 V
LTb
600 251 M
63 0 V
2034 0 R
-63 0 V
600 472 M
63 0 V
2034 0 R
-63 0 V
600 693 M
63 0 V
2034 0 R
-63 0 V
600 914 M
63 0 V
2034 0 R
-63 0 V
600 1135 M
63 0 V
2034 0 R
-63 0 V
600 1356 M
63 0 V
2034 0 R
-63 0 V
600 1577 M
63 0 V
2034 0 R
-63 0 V
600 251 M
0 63 V
0 1263 R
0 -63 V
810 251 M
0 63 V
0 1263 R
0 -63 V
1019 251 M
0 63 V
0 1263 R
0 -63 V
1229 251 M
0 63 V
0 1263 R
0 -63 V
1439 251 M
0 63 V
0 1263 R
0 -63 V
1649 251 M
0 63 V
0 1263 R
0 -63 V
1858 251 M
0 63 V
0 1263 R
0 -63 V
2068 251 M
0 63 V
0 1263 R
0 -63 V
2278 251 M
0 63 V
0 1263 R
0 -63 V
2487 251 M
0 63 V
0 1263 R
0 -63 V
2697 251 M
0 63 V
0 1263 R
0 -63 V
600 251 M
2097 0 V
0 1326 V
-2097 0 V
600 251 L
LT0
663 584 D
705 606 D
747 628 D
789 650 D
831 673 D
873 695 D
915 717 D
956 740 D
998 762 D
1040 784 D
1082 806 D
1124 827 D
1166 848 D
1208 869 D
1250 889 D
1292 909 D
1334 929 D
1376 949 D
1418 986 D
1460 987 D
1502 1004 D
1544 1019 D
1586 1037 D
1628 1051 D
1669 1067 D
1711 1080 D
1753 1092 D
1795 1108 D
1837 1120 D
1879 1127 D
1921 1133 D
1963 1145 D
2005 1161 D
2047 1160 D
2089 1181 D
2131 1182 D
2173 1196 D
2215 1205 D
2257 1214 D
2299 1224 D
2341 1238 D
2382 1282 D
2424 1252 D
2466 1223 D
2508 1258 D
2550 1296 D
2592 1273 D
2634 1283 D
663 583 M
0 1 V
-31 -1 R
62 0 V
-62 1 R
62 0 V
11 21 R
0 1 V
-31 -1 R
62 0 V
-62 1 R
62 0 V
11 21 R
0 1 V
-31 -1 R
62 0 V
-62 1 R
62 0 V
11 22 R
0 1 V
-31 -1 R
62 0 V
-62 1 R
62 0 V
11 21 R
0 1 V
-31 -1 R
62 0 V
-62 1 R
62 0 V
11 21 R
0 2 V
-31 -2 R
62 0 V
-62 2 R
62 0 V
11 20 R
0 2 V
-31 -2 R
62 0 V
-62 2 R
62 0 V
10 21 R
0 2 V
-31 -2 R
62 0 V
-62 2 R
62 0 V
11 20 R
0 2 V
-31 -2 R
62 0 V
-62 2 R
62 0 V
11 20 R
0 2 V
-31 -2 R
62 0 V
-62 2 R
62 0 V
11 19 R
0 3 V
-31 -3 R
62 0 V
-62 3 R
62 0 V
11 18 R
0 3 V
-31 -3 R
62 0 V
-62 3 R
62 0 V
11 19 R
0 3 V
-31 -3 R
62 0 V
-62 3 R
62 0 V
11 17 R
0 3 V
-31 -3 R
62 0 V
-62 3 R
62 0 V
11 18 R
0 3 V
-31 -3 R
62 0 V
-62 3 R
62 0 V
11 16 R
0 4 V
-31 -4 R
62 0 V
-62 4 R
62 0 V
11 16 R
0 4 V
-31 -4 R
62 0 V
-62 4 R
62 0 V
11 16 R
0 4 V
-31 -4 R
62 0 V
-62 4 R
62 0 V
11 33 R
0 4 V
-31 -4 R
62 0 V
-62 4 R
62 0 V
11 -4 R
0 5 V
-31 -5 R
62 0 V
-62 5 R
62 0 V
11 12 R
0 5 V
-31 -5 R
62 0 V
-62 5 R
62 0 V
11 11 R
0 5 V
-31 -5 R
62 0 V
-62 5 R
62 0 V
11 12 R
0 6 V
-31 -6 R
62 0 V
-62 6 R
62 0 V
11 8 R
0 6 V
-31 -6 R
62 0 V
-62 6 R
62 0 V
10 10 R
0 7 V
-31 -7 R
62 0 V
-62 7 R
62 0 V
11 6 R
0 7 V
-31 -7 R
62 0 V
-62 7 R
62 0 V
11 4 R
0 7 V
-31 -7 R
62 0 V
-62 7 R
62 0 V
11 9 R
0 8 V
-31 -8 R
62 0 V
-62 8 R
62 0 V
11 3 R
0 11 V
-31 -11 R
62 0 V
-62 11 R
62 0 V
11 -5 R
0 11 V
-31 -11 R
62 0 V
-62 11 R
62 0 V
11 -4 R
0 11 V
-31 -11 R
62 0 V
-62 11 R
62 0 V
11 -1 R
0 15 V
-31 -15 R
62 0 V
-62 15 R
62 0 V
11 0 R
0 16 V
-31 -16 R
62 0 V
-62 16 R
62 0 V
11 -19 R
0 20 V
-31 -20 R
62 0 V
-62 20 R
62 0 V
11 0 R
0 21 V
-31 -21 R
62 0 V
-62 21 R
62 0 V
11 -22 R
0 26 V
-31 -26 R
62 0 V
-62 26 R
62 0 V
11 -17 R
0 37 V
-31 -37 R
62 0 V
-62 37 R
62 0 V
11 -32 R
0 45 V
-31 -45 R
62 0 V
-62 45 R
62 0 V
11 -43 R
0 58 V
-31 -58 R
62 0 V
-62 58 R
62 0 V
11 -52 R
0 67 V
-31 -67 R
62 0 V
-62 67 R
62 0 V
11 -61 R
0 82 V
-31 -82 R
62 0 V
-62 82 R
62 0 V
10 -44 R
0 95 V
-31 -95 R
62 0 V
-62 95 R
62 0 V
11 -141 R
0 127 V
-31 -127 R
62 0 V
-62 127 R
62 0 V
11 -167 R
0 149 V
-31 -149 R
62 0 V
-62 149 R
62 0 V
11 -124 R
0 169 V
-31 -169 R
62 0 V
-62 169 R
62 0 V
11 -152 R
0 209 V
-31 -209 R
62 0 V
-62 209 R
62 0 V
11 -234 R
0 214 V
-31 -214 R
62 0 V
-62 214 R
62 0 V
11 -251 R
0 308 V
-31 -308 R
62 0 V
-62 308 R
62 0 V
LT1
642 1512 A
684 1394 A
726 1330 A
768 1292 A
810 1268 A
852 1246 A
894 1233 A
936 1229 A
977 1218 A
1019 1216 A
1061 1224 A
1103 1227 A
1145 1237 A
1187 1236 A
1229 1250 A
642 1505 M
0 14 V
-31 -14 R
62 0 V
-62 14 R
62 0 V
11 -132 R
0 14 V
-31 -14 R
62 0 V
-62 14 R
62 0 V
11 -79 R
0 16 V
-31 -16 R
62 0 V
-62 16 R
62 0 V
11 -55 R
0 18 V
-31 -18 R
62 0 V
-62 18 R
62 0 V
11 -44 R
0 22 V
-31 -22 R
62 0 V
-62 22 R
62 0 V
11 -45 R
0 24 V
-31 -24 R
62 0 V
-62 24 R
62 0 V
11 -40 R
0 30 V
-31 -30 R
62 0 V
-62 30 R
62 0 V
11 -36 R
0 34 V
-31 -34 R
62 0 V
-62 34 R
62 0 V
10 -47 R
0 39 V
-31 -39 R
62 0 V
-62 39 R
62 0 V
11 -45 R
0 46 V
-31 -46 R
62 0 V
-62 46 R
62 0 V
11 -39 R
0 48 V
-31 -48 R
62 0 V
-62 48 R
62 0 V
11 -52 R
0 61 V
-31 -61 R
62 0 V
-62 61 R
62 0 V
11 -56 R
0 72 V
-31 -72 R
62 0 V
-62 72 R
62 0 V
11 -85 R
0 96 V
-31 -96 R
62 0 V
-62 96 R
62 0 V
11 -91 R
0 114 V
-31 -114 R
62 0 V
-62 114 R
62 0 V
stroke
grestore
end
showpage
}
\put(1648,1677){\makebox(0,0){(a)}}
\put(1648,51){\makebox(0,0){\raisebox{-1em}{$z$}}}
\put(100,914){%
\special{ps: gsave currentpoint currentpoint translate
270 rotate neg exch neg exch translate}%
\makebox(0,0)[b]{\shortstack{\raisebox{-1em}{$am_{\rm eff}(z)$}}}%
\special{ps: currentpoint grestore moveto}%
}
\put(2697,151){\makebox(0,0){50}}
\put(2487,151){\makebox(0,0){45}}
\put(2278,151){\makebox(0,0){40}}
\put(2068,151){\makebox(0,0){35}}
\put(1858,151){\makebox(0,0){30}}
\put(1649,151){\makebox(0,0){25}}
\put(1439,151){\makebox(0,0){20}}
\put(1229,151){\makebox(0,0){15}}
\put(1019,151){\makebox(0,0){10}}
\put(810,151){\makebox(0,0){5}}
\put(600,151){\makebox(0,0){0}}
\put(540,1577){\makebox(0,0)[r]{0.3}}
\put(540,1356){\makebox(0,0)[r]{0.25}}
\put(540,1135){\makebox(0,0)[r]{0.2}}
\put(540,914){\makebox(0,0)[r]{0.15}}
\put(540,693){\makebox(0,0)[r]{0.1}}
\put(540,472){\makebox(0,0)[r]{0.05}}
\put(540,251){\makebox(0,0)[r]{0}}
\end{picture}
\vskip 20pt
\setlength{\unitlength}{0.1bp}
\special{!
/gnudict 40 dict def
gnudict begin
/Color false def
/Solid false def
/gnulinewidth 5.000 def
/vshift -33 def
/dl {10 mul} def
/hpt 31.5 def
/vpt 31.5 def
/M {moveto} bind def
/L {lineto} bind def
/R {rmoveto} bind def
/V {rlineto} bind def
/vpt2 vpt 2 mul def
/hpt2 hpt 2 mul def
/Lshow { currentpoint stroke M
  0 vshift R show } def
/Rshow { currentpoint stroke M
  dup stringwidth pop neg vshift R show } def
/Cshow { currentpoint stroke M
  dup stringwidth pop -2 div vshift R show } def
/DL { Color {setrgbcolor Solid {pop []} if 0 setdash }
 {pop pop pop Solid {pop []} if 0 setdash} ifelse } def
/BL { stroke gnulinewidth 2 mul setlinewidth } def
/AL { stroke gnulinewidth 2 div setlinewidth } def
/PL { stroke gnulinewidth setlinewidth } def
/LTb { BL [] 0 0 0 DL } def
/LTa { AL [1 dl 2 dl] 0 setdash 0 0 0 setrgbcolor } def
/LT0 { PL [] 0 1 0 DL } def
/LT1 { PL [4 dl 2 dl] 0 0 1 DL } def
/LT2 { PL [2 dl 3 dl] 1 0 0 DL } def
/LT3 { PL [1 dl 1.5 dl] 1 0 1 DL } def
/LT4 { PL [5 dl 2 dl 1 dl 2 dl] 0 1 1 DL } def
/LT5 { PL [4 dl 3 dl 1 dl 3 dl] 1 1 0 DL } def
/LT6 { PL [2 dl 2 dl 2 dl 4 dl] 0 0 0 DL } def
/LT7 { PL [2 dl 2 dl 2 dl 2 dl 2 dl 4 dl] 1 0.3 0 DL } def
/LT8 { PL [2 dl 2 dl 2 dl 2 dl 2 dl 2 dl 2 dl 4 dl] 0.5 0.5 0.5 DL } def
/P { stroke [] 0 setdash
  currentlinewidth 2 div sub M
  0 currentlinewidth V stroke } def
/D { stroke [] 0 setdash 2 copy vpt add M
  hpt neg vpt neg V hpt vpt neg V
  hpt vpt V hpt neg vpt V closepath stroke
  P } def
/A { stroke [] 0 setdash vpt sub M 0 vpt2 V
  currentpoint stroke M
  hpt neg vpt neg R hpt2 0 V stroke
  } def
/B { stroke [] 0 setdash 2 copy exch hpt sub exch vpt add M
  0 vpt2 neg V hpt2 0 V 0 vpt2 V
  hpt2 neg 0 V closepath stroke
  P } def
/C { stroke [] 0 setdash exch hpt sub exch vpt add M
  hpt2 vpt2 neg V currentpoint stroke M
  hpt2 neg 0 R hpt2 vpt2 V stroke } def
/T { stroke [] 0 setdash 2 copy vpt 1.12 mul add M
  hpt neg vpt -1.62 mul V
  hpt 2 mul 0 V
  hpt neg vpt 1.62 mul V closepath stroke
  P  } def
/S { 2 copy A C} def
end
}
\begin{picture}(2880,1728)(0,0)
\special{"
gnudict begin
gsave
50 50 translate
0.100 0.100 scale
0 setgray
/Helvetica findfont 100 scalefont setfont
newpath
-500.000000 -500.000000 translate
LTa
600 251 M
2097 0 V
600 251 M
0 1326 V
LTb
600 251 M
63 0 V
2034 0 R
-63 0 V
600 440 M
63 0 V
2034 0 R
-63 0 V
600 630 M
63 0 V
2034 0 R
-63 0 V
600 819 M
63 0 V
2034 0 R
-63 0 V
600 1009 M
63 0 V
2034 0 R
-63 0 V
600 1198 M
63 0 V
2034 0 R
-63 0 V
600 1388 M
63 0 V
2034 0 R
-63 0 V
600 1577 M
63 0 V
2034 0 R
-63 0 V
600 251 M
0 63 V
0 1263 R
0 -63 V
821 251 M
0 63 V
0 1263 R
0 -63 V
1041 251 M
0 63 V
0 1263 R
0 -63 V
1262 251 M
0 63 V
0 1263 R
0 -63 V
1483 251 M
0 63 V
0 1263 R
0 -63 V
1704 251 M
0 63 V
0 1263 R
0 -63 V
1924 251 M
0 63 V
0 1263 R
0 -63 V
2145 251 M
0 63 V
0 1263 R
0 -63 V
2366 251 M
0 63 V
0 1263 R
0 -63 V
2587 251 M
0 63 V
0 1263 R
0 -63 V
600 251 M
2097 0 V
0 1326 V
-2097 0 V
600 251 L
LT0
766 698 D
876 767 D
986 835 D
1097 900 D
1207 959 D
1317 1013 D
1428 1059 D
1538 1100 D
1649 1135 D
1759 1161 D
1869 1180 D
1980 1197 D
2090 1206 D
2200 1212 D
2311 1304 D
2421 1373 D
2531 1284 D
766 697 M
0 1 V
-31 -1 R
62 0 V
-62 1 R
62 0 V
79 68 R
0 1 V
-31 -1 R
62 0 V
-62 1 R
62 0 V
79 67 R
0 2 V
-31 -2 R
62 0 V
-62 2 R
62 0 V
80 63 R
0 2 V
-31 -2 R
62 0 V
-62 2 R
62 0 V
79 57 R
0 3 V
-31 -3 R
62 0 V
-62 3 R
62 0 V
79 50 R
0 3 V
-31 -3 R
62 0 V
-62 3 R
62 0 V
80 43 R
0 4 V
-31 -4 R
62 0 V
-62 4 R
62 0 V
79 36 R
0 5 V
-31 -5 R
62 0 V
-62 5 R
62 0 V
80 30 R
0 7 V
-31 -7 R
62 0 V
-62 7 R
62 0 V
79 17 R
0 10 V
-31 -10 R
62 0 V
-62 10 R
62 0 V
79 8 R
0 11 V
-31 -11 R
62 0 V
-62 11 R
62 0 V
80 4 R
0 17 V
-31 -17 R
62 0 V
-62 17 R
62 0 V
79 -11 R
0 23 V
-31 -23 R
62 0 V
-62 23 R
62 0 V
79 -25 R
0 38 V
-31 -38 R
62 0 V
-62 38 R
62 0 V
80 39 R
0 68 V
-31 -68 R
62 0 V
-62 68 R
62 0 V
79 -30 R
0 129 V
-31 -129 R
62 0 V
-62 129 R
62 0 V
79 -279 R
0 251 V
-31 -251 R
62 0 V
-62 251 R
62 0 V
LT1
710 1333 A
821 1283 A
931 1263 A
1041 1250 A
1152 1253 A
1262 1257 A
1373 1249 A
1483 1299 A
710 1331 M
0 4 V
-31 -4 R
62 0 V
-62 4 R
62 0 V
80 -55 R
0 6 V
-31 -6 R
62 0 V
-62 6 R
62 0 V
79 -27 R
0 8 V
-31 -8 R
62 0 V
-62 8 R
62 0 V
79 -24 R
0 14 V
-31 -14 R
62 0 V
-62 14 R
62 0 V
80 -15 R
0 22 V
-31 -22 R
62 0 V
-62 22 R
62 0 V
79 -26 R
0 37 V
-31 -37 R
62 0 V
-62 37 R
62 0 V
80 -53 R
0 53 V
-31 -53 R
62 0 V
-62 53 R
62 0 V
79 -22 R
0 93 V
-31 -93 R
62 0 V
-62 93 R
62 0 V
stroke
grestore
end
showpage
}
\put(1648,1677){\makebox(0,0){(b)}}
\put(1648,51){\makebox(0,0){\raisebox{-1em}{$z$}}}
\put(100,914){%
\special{ps: gsave currentpoint currentpoint translate
270 rotate neg exch neg exch translate}%
\makebox(0,0)[b]{\shortstack{\raisebox{-1em}{$am_{\rm eff}(z)$}}}%
\special{ps: currentpoint grestore moveto}%
}
\put(2587,151){\makebox(0,0){18}}
\put(2366,151){\makebox(0,0){16}}
\put(2145,151){\makebox(0,0){14}}
\put(1924,151){\makebox(0,0){12}}
\put(1704,151){\makebox(0,0){10}}
\put(1483,151){\makebox(0,0){8}}
\put(1262,151){\makebox(0,0){6}}
\put(1041,151){\makebox(0,0){4}}
\put(821,151){\makebox(0,0){2}}
\put(600,151){\makebox(0,0){0}}
\put(540,1577){\makebox(0,0)[r]{0.7}}
\put(540,1388){\makebox(0,0)[r]{0.6}}
\put(540,1198){\makebox(0,0)[r]{0.5}}
\put(540,1009){\makebox(0,0)[r]{0.4}}
\put(540,819){\makebox(0,0)[r]{0.3}}
\put(540,630){\makebox(0,0)[r]{0.2}}
\put(540,440){\makebox(0,0)[r]{0.1}}
\put(540,251){\makebox(0,0)[r]{0}}
\end{picture}
\end{figure}
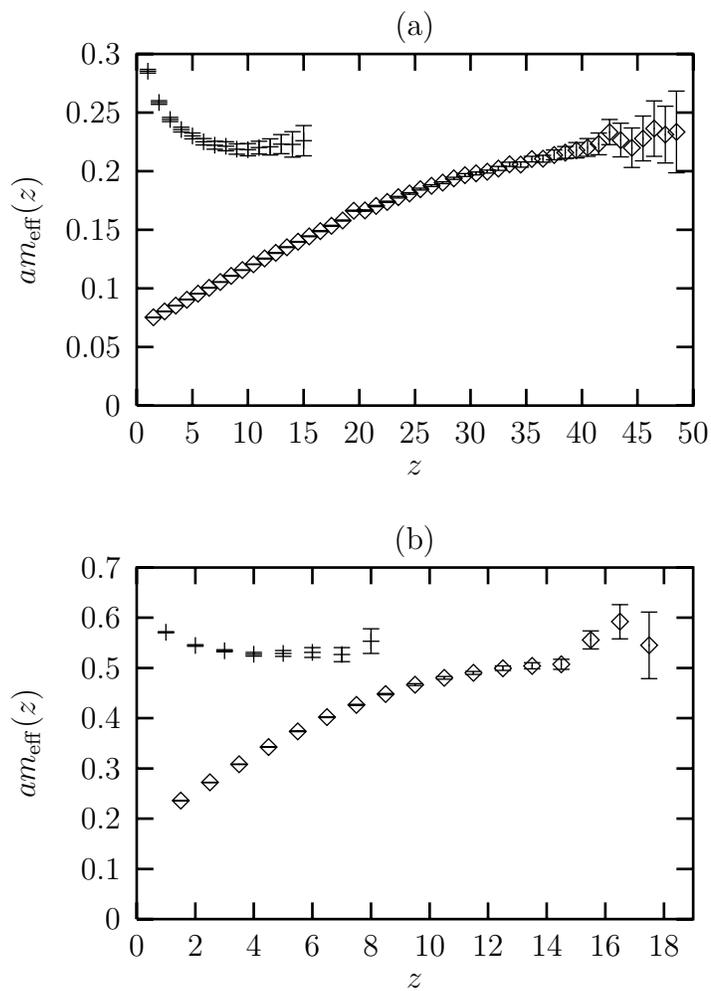

\begin{figure}[h]
\caption[4]{Dependence of the domain wall action density on the
(scaled) length of the lattice. Plotted are the values in
Table 5 and the lines are the leading order perturbation theory
expectation (see text).}
\vskip 20pt
\setlength{\unitlength}{0.1bp}
\special{!
/gnudict 40 dict def
gnudict begin
/Color false def
/Solid false def
/gnulinewidth 5.000 def
/vshift -33 def
/dl {10 mul} def
/hpt 31.5 def
/vpt 31.5 def
/M {moveto} bind def
/L {lineto} bind def
/R {rmoveto} bind def
/V {rlineto} bind def
/vpt2 vpt 2 mul def
/hpt2 hpt 2 mul def
/Lshow { currentpoint stroke M
  0 vshift R show } def
/Rshow { currentpoint stroke M
  dup stringwidth pop neg vshift R show } def
/Cshow { currentpoint stroke M
  dup stringwidth pop -2 div vshift R show } def
/DL { Color {setrgbcolor Solid {pop []} if 0 setdash }
 {pop pop pop Solid {pop []} if 0 setdash} ifelse } def
/BL { stroke gnulinewidth 2 mul setlinewidth } def
/AL { stroke gnulinewidth 2 div setlinewidth } def
/PL { stroke gnulinewidth setlinewidth } def
/LTb { BL [] 0 0 0 DL } def
/LTa { AL [1 dl 2 dl] 0 setdash 0 0 0 setrgbcolor } def
/LT0 { PL [] 0 1 0 DL } def
/LT1 { PL [4 dl 2 dl] 0 0 1 DL } def
/LT2 { PL [2 dl 3 dl] 1 0 0 DL } def
/LT3 { PL [1 dl 1.5 dl] 1 0 1 DL } def
/LT4 { PL [5 dl 2 dl 1 dl 2 dl] 0 1 1 DL } def
/LT5 { PL [4 dl 3 dl 1 dl 3 dl] 1 1 0 DL } def
/LT6 { PL [2 dl 2 dl 2 dl 4 dl] 0 0 0 DL } def
/LT7 { PL [2 dl 2 dl 2 dl 2 dl 2 dl 4 dl] 1 0.3 0 DL } def
/LT8 { PL [2 dl 2 dl 2 dl 2 dl 2 dl 2 dl 2 dl 4 dl] 0.5 0.5 0.5 DL } def
/P { stroke [] 0 setdash
  currentlinewidth 2 div sub M
  0 currentlinewidth V stroke } def
/D { stroke [] 0 setdash 2 copy vpt add M
  hpt neg vpt neg V hpt vpt neg V
  hpt vpt V hpt neg vpt V closepath stroke
  P } def
/A { stroke [] 0 setdash vpt sub M 0 vpt2 V
  currentpoint stroke M
  hpt neg vpt neg R hpt2 0 V stroke
  } def
/B { stroke [] 0 setdash 2 copy exch hpt sub exch vpt add M
  0 vpt2 neg V hpt2 0 V 0 vpt2 V
  hpt2 neg 0 V closepath stroke
  P } def
/C { stroke [] 0 setdash exch hpt sub exch vpt add M
  hpt2 vpt2 neg V currentpoint stroke M
  hpt2 neg 0 R hpt2 vpt2 V stroke } def
/T { stroke [] 0 setdash 2 copy vpt 1.12 mul add M
  hpt neg vpt -1.62 mul V
  hpt 2 mul 0 V
  hpt neg vpt 1.62 mul V closepath stroke
  P  } def
/S { 2 copy A C} def
end
}
\begin{picture}(2880,1728)(0,0)
\special{"
gnudict begin
gsave
50 50 translate
0.100 0.100 scale
0 setgray
/Helvetica findfont 100 scalefont setfont
newpath
-500.000000 -500.000000 translate
LTa
600 251 M
2097 0 V
600 251 M
0 1426 V
LTb
600 251 M
63 0 V
2034 0 R
-63 0 V
600 441 M
63 0 V
2034 0 R
-63 0 V
600 631 M
63 0 V
2034 0 R
-63 0 V
600 821 M
63 0 V
2034 0 R
-63 0 V
600 1012 M
63 0 V
2034 0 R
-63 0 V
600 1202 M
63 0 V
2034 0 R
-63 0 V
600 1392 M
63 0 V
2034 0 R
-63 0 V
600 1582 M
63 0 V
2034 0 R
-63 0 V
600 251 M
0 63 V
0 1363 R
0 -63 V
950 251 M
0 63 V
0 1363 R
0 -63 V
1299 251 M
0 63 V
0 1363 R
0 -63 V
1649 251 M
0 63 V
0 1363 R
0 -63 V
1998 251 M
0 63 V
0 1363 R
0 -63 V
2348 251 M
0 63 V
0 1363 R
0 -63 V
2697 251 M
0 63 V
0 1363 R
0 -63 V
600 251 M
2097 0 V
0 1426 V
-2097 0 V
600 251 L
LT0
2488 1377 D
2111 1365 D
1858 1365 D
1545 1372 D
1418 1344 D
1230 911 D
2488 1367 M
0 19 V
-31 -19 R
62 0 V
-62 19 R
62 0 V
-408 -30 R
0 18 V
-31 -18 R
62 0 V
-62 18 R
62 0 V
-284 -16 R
0 14 V
-31 -14 R
62 0 V
-62 14 R
62 0 V
-344 -8 R
0 16 V
-31 -16 R
62 0 V
-62 16 R
62 0 V
-158 -43 R
0 14 V
-31 -14 R
62 0 V
-62 14 R
62 0 V
1230 903 M
0 16 V
-31 -16 R
62 0 V
-62 16 R
62 0 V
LT1
1935 1010 A
1490 1000 A
1356 933 A
1267 725 A
1178 416 A
1935 1006 M
0 7 V
-31 -7 R
62 0 V
-62 7 R
62 0 V
1490 996 M
0 7 V
-31 -7 R
62 0 V
-62 7 R
62 0 V
1356 929 M
0 7 V
-31 -7 R
62 0 V
-62 7 R
62 0 V
1267 720 M
0 9 V
-31 -9 R
62 0 V
-62 9 R
62 0 V
1178 412 M
0 8 V
-31 -8 R
62 0 V
-62 8 R
62 0 V
LT2
2083 569 B
1787 570 B
1490 559 B
1401 530 B
1342 475 B
1282 401 B
1193 312 B
2083 565 M
0 7 V
-31 -7 R
62 0 V
-62 7 R
62 0 V
-327 -5 R
0 6 V
-31 -6 R
62 0 V
-62 6 R
62 0 V
1490 557 M
0 4 V
-31 -4 R
62 0 V
-62 4 R
62 0 V
1401 527 M
0 7 V
-31 -7 R
62 0 V
-62 7 R
62 0 V
-90 -63 R
0 8 V
-31 -8 R
62 0 V
-62 8 R
62 0 V
-91 -82 R
0 8 V
-31 -8 R
62 0 V
-62 8 R
62 0 V
1193 306 M
0 12 V
-31 -12 R
62 0 V
-62 12 R
62 0 V
LT2
2453 1263 M
-109 0 V
-107 0 V
-106 0 V
-103 0 V
-100 0 V
-105 0 V
-69 0 V
-179 -3 V
-86 -4 V
-56 -6 V
-42 -7 V
-32 -7 V
-26 -9 V
-22 -8 V
-19 -9 V
-16 -9 V
-13 -9 V
-12 -9 V
-10 -9 V
-9 -8 V
-8 -8 V
-7 -7 V
-6 -7 V
-5 -6 V
-4 -5 V
-3 -5 V
-3 -4 V
-3 -3 V
-1 -2 V
-1 -2 V
LT1
2453 967 M
-109 0 V
-107 0 V
-106 0 V
-103 0 V
-100 0 V
-105 0 V
-69 0 V
-179 -2 V
-86 -3 V
-56 -4 V
-42 -5 V
-32 -6 V
-26 -6 V
-22 -6 V
-19 -6 V
-16 -6 V
-13 -7 V
-12 -6 V
-10 -6 V
-9 -6 V
-8 -5 V
-7 -6 V
-6 -4 V
-5 -5 V
-4 -3 V
-3 -4 V
-3 -3 V
-3 -2 V
-1 -1 V
-1 -2 V
LT0
2548 550 M
-115 0 V
-114 0 V
-113 0 V
-111 0 V
-107 0 V
-104 0 V
-72 -1 V
-181 0 V
-87 -2 V
-57 -1 V
-42 -2 V
-33 -2 V
-27 -3 V
-22 -2 V
-19 -3 V
-16 -3 V
-15 -2 V
-12 -3 V
-11 -2 V
-9 -3 V
-8 -2 V
-7 -2 V
-6 -2 V
-6 -2 V
-4 -1 V
-4 -2 V
-3 -1 V
-2 -1 V
-2 0 V
-1 -1 V
stroke
grestore
end
showpage
}
\put(1648,51){\makebox(0,0){\raisebox{-1.5em}{$L_z/\gamma$}}}
\put(100,964){%
\special{ps: gsave currentpoint currentpoint translate
270 rotate neg exch neg exch translate}%
\makebox(0,0)[b]{\shortstack{\raisebox{-1em}{$S_w/L_y$}}}%
\special{ps: currentpoint grestore moveto}%
}
\put(2697,151){\makebox(0,0){3}}
\put(2348,151){\makebox(0,0){2.5}}
\put(1998,151){\makebox(0,0){2}}
\put(1649,151){\makebox(0,0){1.5}}
\put(1299,151){\makebox(0,0){1}}
\put(950,151){\makebox(0,0){0.5}}
\put(600,151){\makebox(0,0){0}}
\put(540,1582){\makebox(0,0)[r]{0.14}}
\put(540,1392){\makebox(0,0)[r]{0.12}}
\put(540,1202){\makebox(0,0)[r]{0.1}}
\put(540,1012){\makebox(0,0)[r]{0.08}}
\put(540,821){\makebox(0,0)[r]{0.06}}
\put(540,631){\makebox(0,0)[r]{0.04}}
\put(540,441){\makebox(0,0)[r]{0.02}}
\put(540,251){\makebox(0,0)[r]{0}}
\end{picture}
\end{figure}
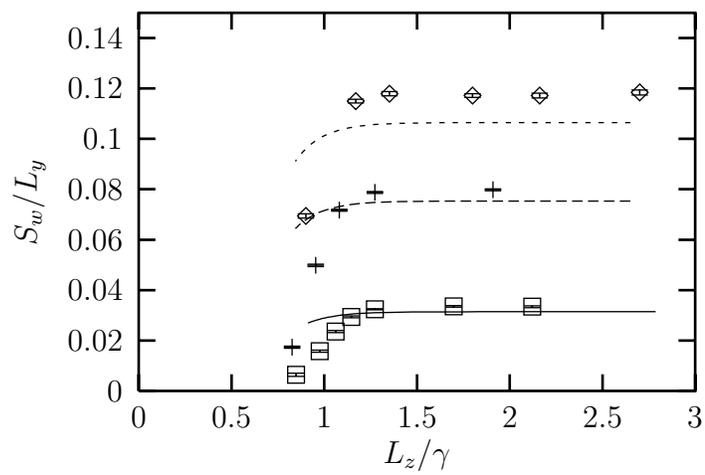

\begin{figure}[h]
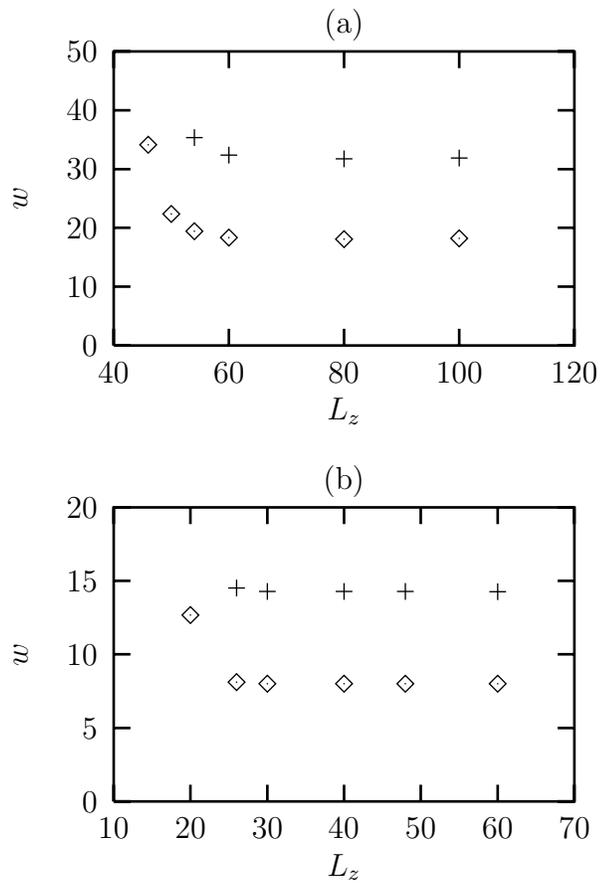

\caption[5]{The width of the domain wall as a function of the
length of the lattice for (a) $\beta=75$ and $L_t=3$,
(b) $\beta=25$ and $L_t=2$. Widths are calculated when the
Polyakov loop attains 2/3 ($\diamond$) and 9/10 ($+$) of
its vacuum value.}
\vskip 20pt
\input{fig5a.tex}
\vskip 20pt
\input{fig5b.tex}
\end{figure}

\begin{figure}[h]
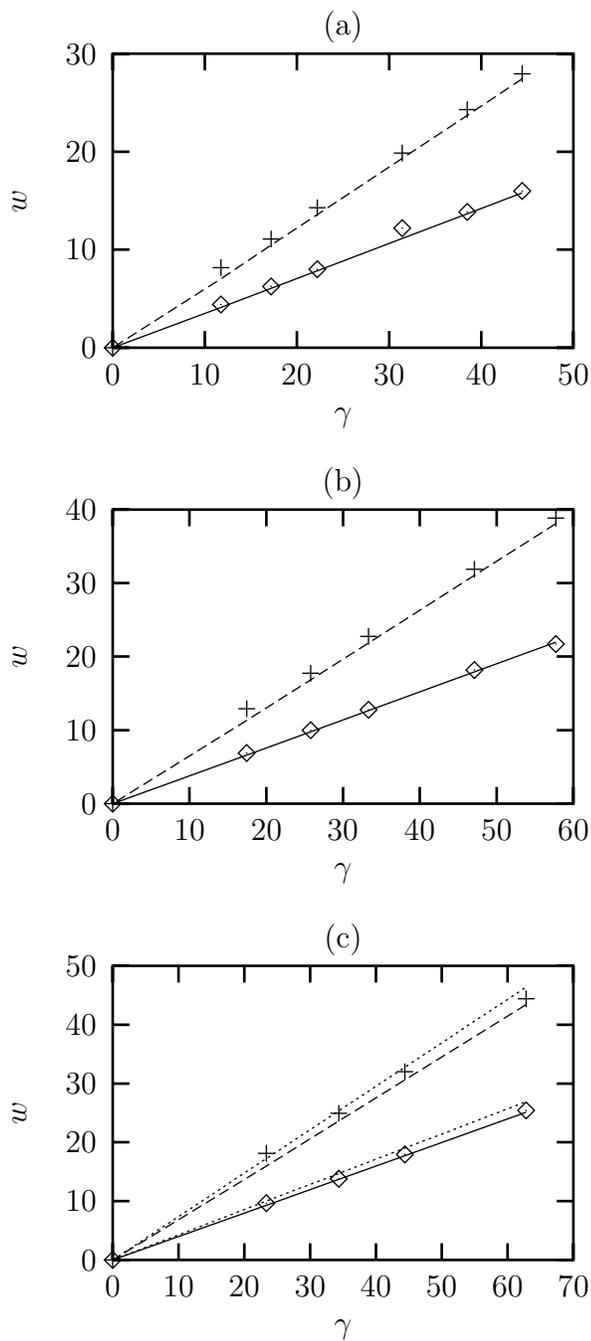

\caption[6]{Widths of the domain wall (defined as in Fig. 5) 
plotted against $\gamma=\pi(\beta L_t)^{1/2}$ for
(a) $L_t=2$, (b) $L_t=3$, (c) $L_t=4$. Lines are corresponding
perturbative predictions; the dotted lines in (c) are for the
continuum, $L_t = \infty$.}
\vskip 20pt
\input{fig6a.tex}
\vskip 20pt
\input{fig6b.tex}
\vskip 20pt
\setlength{\unitlength}{0.1bp}
\special{!
/gnudict 40 dict def
gnudict begin
/Color false def
/Solid false def
/gnulinewidth 5.000 def
/vshift -33 def
/dl {10 mul} def
/hpt 31.5 def
/vpt 31.5 def
/M {moveto} bind def
/L {lineto} bind def
/R {rmoveto} bind def
/V {rlineto} bind def
/vpt2 vpt 2 mul def
/hpt2 hpt 2 mul def
/Lshow { currentpoint stroke M
  0 vshift R show } def
/Rshow { currentpoint stroke M
  dup stringwidth pop neg vshift R show } def
/Cshow { currentpoint stroke M
  dup stringwidth pop -2 div vshift R show } def
/DL { Color {setrgbcolor Solid {pop []} if 0 setdash }
 {pop pop pop Solid {pop []} if 0 setdash} ifelse } def
/BL { stroke gnulinewidth 2 mul setlinewidth } def
/AL { stroke gnulinewidth 2 div setlinewidth } def
/PL { stroke gnulinewidth setlinewidth } def
/LTb { BL [] 0 0 0 DL } def
/LTa { AL [1 dl 2 dl] 0 setdash 0 0 0 setrgbcolor } def
/LT0 { PL [] 0 1 0 DL } def
/LT1 { PL [4 dl 2 dl] 0 0 1 DL } def
/LT2 { PL [2 dl 3 dl] 1 0 0 DL } def
/LT3 { PL [1 dl 1.5 dl] 1 0 1 DL } def
/LT4 { PL [5 dl 2 dl 1 dl 2 dl] 0 1 1 DL } def
/LT5 { PL [4 dl 3 dl 1 dl 3 dl] 1 1 0 DL } def
/LT6 { PL [2 dl 2 dl 2 dl 4 dl] 0 0 0 DL } def
/LT7 { PL [2 dl 2 dl 2 dl 2 dl 2 dl 4 dl] 1 0.3 0 DL } def
/LT8 { PL [2 dl 2 dl 2 dl 2 dl 2 dl 2 dl 2 dl 4 dl] 0.5 0.5 0.5 DL } def
/P { stroke [] 0 setdash
  currentlinewidth 2 div sub M
  0 currentlinewidth V stroke } def
/D { stroke [] 0 setdash 2 copy vpt add M
  hpt neg vpt neg V hpt vpt neg V
  hpt vpt V hpt neg vpt V closepath stroke
  P } def
/A { stroke [] 0 setdash vpt sub M 0 vpt2 V
  currentpoint stroke M
  hpt neg vpt neg R hpt2 0 V stroke
  } def
/B { stroke [] 0 setdash 2 copy exch hpt sub exch vpt add M
  0 vpt2 neg V hpt2 0 V 0 vpt2 V
  hpt2 neg 0 V closepath stroke
  P } def
/C { stroke [] 0 setdash exch hpt sub exch vpt add M
  hpt2 vpt2 neg V currentpoint stroke M
  hpt2 neg 0 R hpt2 vpt2 V stroke } def
/T { stroke [] 0 setdash 2 copy vpt 1.12 mul add M
  hpt neg vpt -1.62 mul V
  hpt 2 mul 0 V
  hpt neg vpt 1.62 mul V closepath stroke
  P  } def
/S { 2 copy A C} def
end
}
\begin{picture}(2519,1511)(0,0)
\special{"
gnudict begin
gsave
50 50 translate
0.100 0.100 scale
0 setgray
/Helvetica findfont 100 scalefont setfont
newpath
-500.000000 -500.000000 translate
LTa
600 251 M
1736 0 V
600 251 M
0 1109 V
LTb
600 251 M
63 0 V
1673 0 R
-63 0 V
600 473 M
63 0 V
1673 0 R
-63 0 V
600 695 M
63 0 V
1673 0 R
-63 0 V
600 916 M
63 0 V
1673 0 R
-63 0 V
600 1138 M
63 0 V
1673 0 R
-63 0 V
600 1360 M
63 0 V
1673 0 R
-63 0 V
600 251 M
0 63 V
0 1046 R
0 -63 V
848 251 M
0 63 V
0 1046 R
0 -63 V
1096 251 M
0 63 V
0 1046 R
0 -63 V
1344 251 M
0 63 V
0 1046 R
0 -63 V
1592 251 M
0 63 V
0 1046 R
0 -63 V
1840 251 M
0 63 V
0 1046 R
0 -63 V
2088 251 M
0 63 V
0 1046 R
0 -63 V
2336 251 M
0 63 V
0 1046 R
0 -63 V
600 251 M
1736 0 V
0 1109 V
-1736 0 V
600 251 L
LT0
2158 808 M
1701 644 L
1452 555 L
1179 457 L
600 251 L
LT1
2158 1214 M
1701 930 L
1452 775 L
1179 605 L
600 251 L
LT0
2158 815 D
1701 649 D
1452 557 D
1179 465 D
600 251 D
LT2
2158 1236 A
1701 961 A
1452 804 A
1179 653 A
600 251 A
LT3
600 251 M
78 30 V
78 30 V
78 29 V
78 30 V
78 30 V
77 30 V
78 30 V
78 30 V
78 30 V
78 29 V
78 30 V
78 30 V
78 30 V
78 30 V
78 30 V
78 29 V
77 30 V
78 30 V
78 30 V
78 30 V
600 251 M
78 51 V
78 52 V
78 51 V
78 52 V
78 51 V
77 52 V
78 51 V
78 51 V
78 52 V
78 51 V
78 52 V
78 51 V
78 52 V
78 51 V
78 52 V
78 51 V
77 52 V
78 51 V
78 51 V
78 52 V
stroke
grestore
end
showpage
}
\put(1468,1460){\makebox(0,0){(c)}}
\put(1468,51){\makebox(0,0){\raisebox{-1.5em}{$\gamma$}}}
\put(100,805){%
\special{ps: gsave currentpoint currentpoint translate
270 rotate neg exch neg exch translate}%
\makebox(0,0)[b]{\shortstack{\raisebox{-1.5em}{$w$}}}%
\special{ps: currentpoint grestore moveto}%
}
\put(2336,151){\makebox(0,0){70}}
\put(2088,151){\makebox(0,0){60}}
\put(1840,151){\makebox(0,0){50}}
\put(1592,151){\makebox(0,0){40}}
\put(1344,151){\makebox(0,0){30}}
\put(1096,151){\makebox(0,0){20}}
\put(848,151){\makebox(0,0){10}}
\put(600,151){\makebox(0,0){0}}
\put(540,1360){\makebox(0,0)[r]{50}}
\put(540,1138){\makebox(0,0)[r]{40}}
\put(540,916){\makebox(0,0)[r]{30}}
\put(540,695){\makebox(0,0)[r]{20}}
\put(540,473){\makebox(0,0)[r]{10}}
\put(540,251){\makebox(0,0)[r]{0}}
\end{picture}
\end{figure}

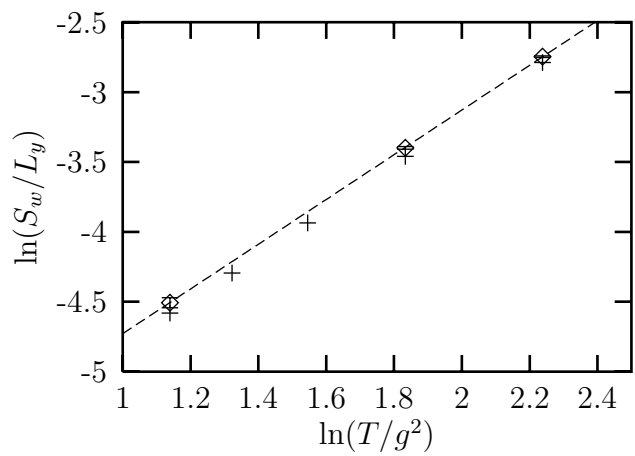
\begin{figure}[h]
\caption[7]{Action density of the domain wall ($\diamond$) for
$L_t=2,3,6$ at $\beta=75$. The line is a fit (see text).
Also shown are the perturbative values ($+$) for
$L_t=2,3,4,5,6$.}
\vskip 20pt
\setlength{\unitlength}{0.1bp}
\special{!
/gnudict 40 dict def
gnudict begin
/Color false def
/Solid false def
/gnulinewidth 5.000 def
/vshift -33 def
/dl {10 mul} def
/hpt 31.5 def
/vpt 31.5 def
/M {moveto} bind def
/L {lineto} bind def
/R {rmoveto} bind def
/V {rlineto} bind def
/vpt2 vpt 2 mul def
/hpt2 hpt 2 mul def
/Lshow { currentpoint stroke M
  0 vshift R show } def
/Rshow { currentpoint stroke M
  dup stringwidth pop neg vshift R show } def
/Cshow { currentpoint stroke M
  dup stringwidth pop -2 div vshift R show } def
/DL { Color {setrgbcolor Solid {pop []} if 0 setdash }
 {pop pop pop Solid {pop []} if 0 setdash} ifelse } def
/BL { stroke gnulinewidth 2 mul setlinewidth } def
/AL { stroke gnulinewidth 2 div setlinewidth } def
/PL { stroke gnulinewidth setlinewidth } def
/LTb { BL [] 0 0 0 DL } def
/LTa { AL [1 dl 2 dl] 0 setdash 0 0 0 setrgbcolor } def
/LT0 { PL [] 0 1 0 DL } def
/LT1 { PL [4 dl 2 dl] 0 0 1 DL } def
/LT2 { PL [2 dl 3 dl] 1 0 0 DL } def
/LT3 { PL [1 dl 1.5 dl] 1 0 1 DL } def
/LT4 { PL [5 dl 2 dl 1 dl 2 dl] 0 1 1 DL } def
/LT5 { PL [4 dl 3 dl 1 dl 3 dl] 1 1 0 DL } def
/LT6 { PL [2 dl 2 dl 2 dl 4 dl] 0 0 0 DL } def
/LT7 { PL [2 dl 2 dl 2 dl 2 dl 2 dl 4 dl] 1 0.3 0 DL } def
/LT8 { PL [2 dl 2 dl 2 dl 2 dl 2 dl 2 dl 2 dl 4 dl] 0.5 0.5 0.5 DL } def
/P { stroke [] 0 setdash
  currentlinewidth 2 div sub M
  0 currentlinewidth V stroke } def
/D { stroke [] 0 setdash 2 copy vpt add M
  hpt neg vpt neg V hpt vpt neg V
  hpt vpt V hpt neg vpt V closepath stroke
  P } def
/A { stroke [] 0 setdash vpt sub M 0 vpt2 V
  currentpoint stroke M
  hpt neg vpt neg R hpt2 0 V stroke
  } def
/B { stroke [] 0 setdash 2 copy exch hpt sub exch vpt add M
  0 vpt2 neg V hpt2 0 V 0 vpt2 V
  hpt2 neg 0 V closepath stroke
  P } def
/C { stroke [] 0 setdash exch hpt sub exch vpt add M
  hpt2 vpt2 neg V currentpoint stroke M
  hpt2 neg 0 R hpt2 vpt2 V stroke } def
/T { stroke [] 0 setdash 2 copy vpt 1.12 mul add M
  hpt neg vpt -1.62 mul V
  hpt 2 mul 0 V
  hpt neg vpt 1.62 mul V closepath stroke
  P  } def
/S { 2 copy A C} def
end
}
\begin{picture}(2700,1620)(0,0)
\special{"
gnudict begin
gsave
50 50 translate
0.100 0.100 scale
0 setgray
/Helvetica findfont 100 scalefont setfont
newpath
-500.000000 -500.000000 translate
LTa
LTb
600 251 M
63 0 V
1854 0 R
-63 0 V
600 515 M
63 0 V
1854 0 R
-63 0 V
600 778 M
63 0 V
1854 0 R
-63 0 V
600 1042 M
63 0 V
1854 0 R
-63 0 V
600 1305 M
63 0 V
1854 0 R
-63 0 V
600 1569 M
63 0 V
1854 0 R
-63 0 V
600 251 M
0 63 V
0 1255 R
0 -63 V
856 251 M
0 63 V
0 1255 R
0 -63 V
1111 251 M
0 63 V
0 1255 R
0 -63 V
1367 251 M
0 63 V
0 1255 R
0 -63 V
1622 251 M
0 63 V
0 1255 R
0 -63 V
1878 251 M
0 63 V
0 1255 R
0 -63 V
2134 251 M
0 63 V
0 1255 R
0 -63 V
2389 251 M
0 63 V
0 1255 R
0 -63 V
600 251 M
1917 0 V
0 1318 V
-1917 0 V
600 251 L
LT0
2182 1439 D
1665 1095 D
778 511 D
2182 1436 M
0 6 V
-31 -6 R
62 0 V
-62 6 R
62 0 V
1665 1090 M
0 10 V
-31 -10 R
62 0 V
-62 10 R
62 0 V
778 492 M
0 38 V
747 492 M
62 0 V
-62 38 R
62 0 V
LT1
600 394 M
19 13 V
20 13 V
19 13 V
19 12 V
20 13 V
19 13 V
20 13 V
19 12 V
19 13 V
20 13 V
19 13 V
19 13 V
20 12 V
19 13 V
19 13 V
20 13 V
19 12 V
20 13 V
19 13 V
19 13 V
20 13 V
19 12 V
19 13 V
20 13 V
19 13 V
19 12 V
20 13 V
19 13 V
20 13 V
19 13 V
19 12 V
20 13 V
19 13 V
19 13 V
20 13 V
19 12 V
19 13 V
20 13 V
19 13 V
20 12 V
19 13 V
19 13 V
20 13 V
19 13 V
19 12 V
20 13 V
19 13 V
19 13 V
20 12 V
19 13 V
20 13 V
19 13 V
19 13 V
20 12 V
19 13 V
19 13 V
20 13 V
19 12 V
19 13 V
20 13 V
19 13 V
20 13 V
19 12 V
19 13 V
20 13 V
19 13 V
19 12 V
20 13 V
19 13 V
19 13 V
20 13 V
19 12 V
20 13 V
19 13 V
19 13 V
20 13 V
19 12 V
19 13 V
20 13 V
19 13 V
19 12 V
20 13 V
19 13 V
20 13 V
19 13 V
19 12 V
20 13 V
19 13 V
19 13 V
20 12 V
19 13 V
18 12 V
LT2
2182 1417 A
1665 1063 A
1297 812 A
1012 623 A
778 472 A
stroke
grestore
end
showpage
}
\put(1558,51){\makebox(0,0){\raisebox{-1.5em}{$\ln(T/g^2)$}}}
\put(100,910){%
\special{ps: gsave currentpoint currentpoint translate
270 rotate neg exch neg exch translate}%
\makebox(0,0)[b]{\shortstack{\raisebox{-1.5em}{$\ln(S_w /L_y)$}}}%
\special{ps: currentpoint grestore moveto}%
}
\put(2389,151){\makebox(0,0){2.4}}
\put(2134,151){\makebox(0,0){2.2}}
\put(1878,151){\makebox(0,0){2}}
\put(1622,151){\makebox(0,0){1.8}}
\put(1367,151){\makebox(0,0){1.6}}
\put(1111,151){\makebox(0,0){1.4}}
\put(856,151){\makebox(0,0){1.2}}
\put(600,151){\makebox(0,0){1}}
\put(540,1569){\makebox(0,0)[r]{-2.5}}
\put(540,1305){\makebox(0,0)[r]{-3}}
\put(540,1042){\makebox(0,0)[r]{-3.5}}
\put(540,778){\makebox(0,0)[r]{-4}}
\put(540,515){\makebox(0,0)[r]{-4.5}}
\put(540,251){\makebox(0,0)[r]{-5}}
\end{picture}
\end{figure}

\begin{figure}[h]
\caption[8]{The interface tension (with the factor $T^{2.5}/g$
removed): numerical values ($\diamond$) compared to leading
order perturbation theory (dashed lines).}
\vskip 20pt
\setlength{\unitlength}{0.1bp}
\special{!
/gnudict 40 dict def
gnudict begin
/Color false def
/Solid false def
/gnulinewidth 5.000 def
/vshift -33 def
/dl {10 mul} def
/hpt 31.5 def
/vpt 31.5 def
/M {moveto} bind def
/L {lineto} bind def
/R {rmoveto} bind def
/V {rlineto} bind def
/vpt2 vpt 2 mul def
/hpt2 hpt 2 mul def
/Lshow { currentpoint stroke M
  0 vshift R show } def
/Rshow { currentpoint stroke M
  dup stringwidth pop neg vshift R show } def
/Cshow { currentpoint stroke M
  dup stringwidth pop -2 div vshift R show } def
/DL { Color {setrgbcolor Solid {pop []} if 0 setdash }
 {pop pop pop Solid {pop []} if 0 setdash} ifelse } def
/BL { stroke gnulinewidth 2 mul setlinewidth } def
/AL { stroke gnulinewidth 2 div setlinewidth } def
/PL { stroke gnulinewidth setlinewidth } def
/LTb { BL [] 0 0 0 DL } def
/LTa { AL [1 dl 2 dl] 0 setdash 0 0 0 setrgbcolor } def
/LT0 { PL [] 0 1 0 DL } def
/LT1 { PL [4 dl 2 dl] 0 0 1 DL } def
/LT2 { PL [2 dl 3 dl] 1 0 0 DL } def
/LT3 { PL [1 dl 1.5 dl] 1 0 1 DL } def
/LT4 { PL [5 dl 2 dl 1 dl 2 dl] 0 1 1 DL } def
/LT5 { PL [4 dl 3 dl 1 dl 3 dl] 1 1 0 DL } def
/LT6 { PL [2 dl 2 dl 2 dl 4 dl] 0 0 0 DL } def
/LT7 { PL [2 dl 2 dl 2 dl 2 dl 2 dl 4 dl] 1 0.3 0 DL } def
/LT8 { PL [2 dl 2 dl 2 dl 2 dl 2 dl 2 dl 2 dl 4 dl] 0.5 0.5 0.5 DL } def
/P { stroke [] 0 setdash
  currentlinewidth 2 div sub M
  0 currentlinewidth V stroke } def
/D { stroke [] 0 setdash 2 copy vpt add M
  hpt neg vpt neg V hpt vpt neg V
  hpt vpt V hpt neg vpt V closepath stroke
  P } def
/A { stroke [] 0 setdash vpt sub M 0 vpt2 V
  currentpoint stroke M
  hpt neg vpt neg R hpt2 0 V stroke
  } def
/B { stroke [] 0 setdash 2 copy exch hpt sub exch vpt add M
  0 vpt2 neg V hpt2 0 V 0 vpt2 V
  hpt2 neg 0 V closepath stroke
  P } def
/C { stroke [] 0 setdash exch hpt sub exch vpt add M
  hpt2 vpt2 neg V currentpoint stroke M
  hpt2 neg 0 R hpt2 vpt2 V stroke } def
/T { stroke [] 0 setdash 2 copy vpt 1.12 mul add M
  hpt neg vpt -1.62 mul V
  hpt 2 mul 0 V
  hpt neg vpt 1.62 mul V closepath stroke
  P  } def
/S { 2 copy A C} def
end
}
\begin{picture}(2700,1620)(0,0)
\special{"
gnudict begin
gsave
50 50 translate
0.100 0.100 scale
0 setgray
/Helvetica findfont 100 scalefont setfont
newpath
-500.000000 -500.000000 translate
LTa
600 251 M
1917 0 V
600 251 M
0 1218 V
LTb
600 251 M
63 0 V
1854 0 R
-63 0 V
600 495 M
63 0 V
1854 0 R
-63 0 V
600 738 M
63 0 V
1854 0 R
-63 0 V
600 982 M
63 0 V
1854 0 R
-63 0 V
600 1225 M
63 0 V
1854 0 R
-63 0 V
600 1469 M
63 0 V
1854 0 R
-63 0 V
600 251 M
0 63 V
0 1155 R
0 -63 V
856 251 M
0 63 V
0 1155 R
0 -63 V
1111 251 M
0 63 V
0 1155 R
0 -63 V
1367 251 M
0 63 V
0 1155 R
0 -63 V
1622 251 M
0 63 V
0 1155 R
0 -63 V
1878 251 M
0 63 V
0 1155 R
0 -63 V
2134 251 M
0 63 V
0 1155 R
0 -63 V
2389 251 M
0 63 V
0 1155 R
0 -63 V
600 251 M
1917 0 V
0 1218 V
-1917 0 V
600 251 L
LT0
702 1015 D
736 1016 D
804 1034 D
1009 1062 D
1282 1113 D
2061 1198 D
702 1013 M
0 5 V
-31 -5 R
62 0 V
-62 5 R
62 0 V
3 -6 R
0 8 V
-31 -8 R
62 0 V
-62 8 R
62 0 V
37 12 R
0 4 V
-31 -4 R
62 0 V
-62 4 R
62 0 V
174 24 R
0 5 V
-31 -5 R
62 0 V
-62 5 R
62 0 V
242 37 R
0 22 V
-31 -22 R
62 0 V
-62 22 R
62 0 V
748 65 R
0 18 V
-31 -18 R
62 0 V
-62 18 R
62 0 V
LT1
600 985 M
19 0 V
20 0 V
19 0 V
19 0 V
20 0 V
19 0 V
20 0 V
19 0 V
19 0 V
20 0 V
19 0 V
19 0 V
20 0 V
19 0 V
19 0 V
20 0 V
19 0 V
20 0 V
19 0 V
19 0 V
20 0 V
19 0 V
19 0 V
20 0 V
19 0 V
19 0 V
20 0 V
19 0 V
20 0 V
19 0 V
19 0 V
20 0 V
19 0 V
19 0 V
20 0 V
19 0 V
19 0 V
20 0 V
19 0 V
20 0 V
19 0 V
19 0 V
20 0 V
19 0 V
19 0 V
20 0 V
19 0 V
19 0 V
20 0 V
19 0 V
20 0 V
19 0 V
19 0 V
20 0 V
19 0 V
19 0 V
20 0 V
19 0 V
19 0 V
20 0 V
19 0 V
20 0 V
19 0 V
19 0 V
20 0 V
19 0 V
19 0 V
20 0 V
19 0 V
19 0 V
20 0 V
19 0 V
20 0 V
19 0 V
19 0 V
20 0 V
19 0 V
19 0 V
20 0 V
19 0 V
19 0 V
20 0 V
19 0 V
20 0 V
19 0 V
19 0 V
20 0 V
19 0 V
19 0 V
20 0 V
19 0 V
19 0 V
20 0 V
19 0 V
20 0 V
19 0 V
19 0 V
20 0 V
19 0 V
stroke
grestore
end
showpage
}
\put(1558,1569){\makebox(0,0){ (a) \  $L_t=2$}}
\put(1558,51){\makebox(0,0){\raisebox{-1.5em}{$g^2/T$}}}
\put(100,860){%
\special{ps: gsave currentpoint currentpoint translate
270 rotate neg exch neg exch translate}%
\makebox(0,0)[b]{\shortstack{\raisebox{-1.5em}{$\alpha_{\rm eff}$}}}%
\special{ps: currentpoint grestore moveto}%
}
\put(2389,151){\makebox(0,0){1.4}}
\put(2134,151){\makebox(0,0){1.2}}
\put(1878,151){\makebox(0,0){1}}
\put(1622,151){\makebox(0,0){0.8}}
\put(1367,151){\makebox(0,0){0.6}}
\put(1111,151){\makebox(0,0){0.4}}
\put(856,151){\makebox(0,0){0.2}}
\put(600,151){\makebox(0,0){0}}
\put(540,1469){\makebox(0,0)[r]{10}}
\put(540,1225){\makebox(0,0)[r]{8}}
\put(540,982){\makebox(0,0)[r]{6}}
\put(540,738){\makebox(0,0)[r]{4}}
\put(540,495){\makebox(0,0)[r]{2}}
\put(540,251){\makebox(0,0)[r]{0}}
\end{picture}
\vskip 20pt
\setlength{\unitlength}{0.1bp}
\special{!
/gnudict 40 dict def
gnudict begin
/Color false def
/Solid false def
/gnulinewidth 5.000 def
/vshift -33 def
/dl {10 mul} def
/hpt 31.5 def
/vpt 31.5 def
/M {moveto} bind def
/L {lineto} bind def
/R {rmoveto} bind def
/V {rlineto} bind def
/vpt2 vpt 2 mul def
/hpt2 hpt 2 mul def
/Lshow { currentpoint stroke M
  0 vshift R show } def
/Rshow { currentpoint stroke M
  dup stringwidth pop neg vshift R show } def
/Cshow { currentpoint stroke M
  dup stringwidth pop -2 div vshift R show } def
/DL { Color {setrgbcolor Solid {pop []} if 0 setdash }
 {pop pop pop Solid {pop []} if 0 setdash} ifelse } def
/BL { stroke gnulinewidth 2 mul setlinewidth } def
/AL { stroke gnulinewidth 2 div setlinewidth } def
/PL { stroke gnulinewidth setlinewidth } def
/LTb { BL [] 0 0 0 DL } def
/LTa { AL [1 dl 2 dl] 0 setdash 0 0 0 setrgbcolor } def
/LT0 { PL [] 0 1 0 DL } def
/LT1 { PL [4 dl 2 dl] 0 0 1 DL } def
/LT2 { PL [2 dl 3 dl] 1 0 0 DL } def
/LT3 { PL [1 dl 1.5 dl] 1 0 1 DL } def
/LT4 { PL [5 dl 2 dl 1 dl 2 dl] 0 1 1 DL } def
/LT5 { PL [4 dl 3 dl 1 dl 3 dl] 1 1 0 DL } def
/LT6 { PL [2 dl 2 dl 2 dl 4 dl] 0 0 0 DL } def
/LT7 { PL [2 dl 2 dl 2 dl 2 dl 2 dl 4 dl] 1 0.3 0 DL } def
/LT8 { PL [2 dl 2 dl 2 dl 2 dl 2 dl 2 dl 2 dl 4 dl] 0.5 0.5 0.5 DL } def
/P { stroke [] 0 setdash
  currentlinewidth 2 div sub M
  0 currentlinewidth V stroke } def
/D { stroke [] 0 setdash 2 copy vpt add M
  hpt neg vpt neg V hpt vpt neg V
  hpt vpt V hpt neg vpt V closepath stroke
  P } def
/A { stroke [] 0 setdash vpt sub M 0 vpt2 V
  currentpoint stroke M
  hpt neg vpt neg R hpt2 0 V stroke
  } def
/B { stroke [] 0 setdash 2 copy exch hpt sub exch vpt add M
  0 vpt2 neg V hpt2 0 V 0 vpt2 V
  hpt2 neg 0 V closepath stroke
  P } def
/C { stroke [] 0 setdash exch hpt sub exch vpt add M
  hpt2 vpt2 neg V currentpoint stroke M
  hpt2 neg 0 R hpt2 vpt2 V stroke } def
/T { stroke [] 0 setdash 2 copy vpt 1.12 mul add M
  hpt neg vpt -1.62 mul V
  hpt 2 mul 0 V
  hpt neg vpt 1.62 mul V closepath stroke
  P  } def
/S { 2 copy A C} def
end
}
\begin{picture}(2700,1620)(0,0)
\special{"
gnudict begin
gsave
50 50 translate
0.100 0.100 scale
0 setgray
/Helvetica findfont 100 scalefont setfont
newpath
-500.000000 -500.000000 translate
LTa
600 251 M
1917 0 V
600 251 M
0 1218 V
LTb
600 251 M
63 0 V
1854 0 R
-63 0 V
600 495 M
63 0 V
1854 0 R
-63 0 V
600 738 M
63 0 V
1854 0 R
-63 0 V
600 982 M
63 0 V
1854 0 R
-63 0 V
600 1225 M
63 0 V
1854 0 R
-63 0 V
600 1469 M
63 0 V
1854 0 R
-63 0 V
600 251 M
0 63 V
0 1155 R
0 -63 V
856 251 M
0 63 V
0 1155 R
0 -63 V
1111 251 M
0 63 V
0 1155 R
0 -63 V
1367 251 M
0 63 V
0 1155 R
0 -63 V
1622 251 M
0 63 V
0 1155 R
0 -63 V
1878 251 M
0 63 V
0 1155 R
0 -63 V
2134 251 M
0 63 V
0 1155 R
0 -63 V
2389 251 M
0 63 V
0 1155 R
0 -63 V
600 251 M
1917 0 V
0 1218 V
-1917 0 V
600 251 L
LT0
736 983 D
804 985 D
1009 996 D
1283 1024 D
2093 1096 D
736 979 M
0 9 V
-31 -9 R
62 0 V
-62 9 R
62 0 V
37 -9 R
0 12 V
773 979 M
62 0 V
-62 12 R
62 0 V
174 -6 R
0 23 V
978 985 M
62 0 V
-62 23 R
62 0 V
243 2 R
0 29 V
-31 -29 R
62 0 V
-62 29 R
62 0 V
779 36 R
0 41 V
-31 -41 R
62 0 V
-62 41 R
62 0 V
LT1
600 940 M
19 0 V
20 0 V
19 0 V
19 0 V
20 0 V
19 0 V
20 0 V
19 0 V
19 0 V
20 0 V
19 0 V
19 0 V
20 0 V
19 0 V
19 0 V
20 0 V
19 0 V
20 0 V
19 0 V
19 0 V
20 0 V
19 0 V
19 0 V
20 0 V
19 0 V
19 0 V
20 0 V
19 0 V
20 0 V
19 0 V
19 0 V
20 0 V
19 0 V
19 0 V
20 0 V
19 0 V
19 0 V
20 0 V
19 0 V
20 0 V
19 0 V
19 0 V
20 0 V
19 0 V
19 0 V
20 0 V
19 0 V
19 0 V
20 0 V
19 0 V
20 0 V
19 0 V
19 0 V
20 0 V
19 0 V
19 0 V
20 0 V
19 0 V
19 0 V
20 0 V
19 0 V
20 0 V
19 0 V
19 0 V
20 0 V
19 0 V
19 0 V
20 0 V
19 0 V
19 0 V
20 0 V
19 0 V
20 0 V
19 0 V
19 0 V
20 0 V
19 0 V
19 0 V
20 0 V
19 0 V
19 0 V
20 0 V
19 0 V
20 0 V
19 0 V
19 0 V
20 0 V
19 0 V
19 0 V
20 0 V
19 0 V
19 0 V
20 0 V
19 0 V
20 0 V
19 0 V
19 0 V
20 0 V
19 0 V
stroke
grestore
end
showpage
}
\put(1558,1569){\makebox(0,0){(b) \ $L_t=3$}}
\put(1558,51){\makebox(0,0){\raisebox{-1.5em}{$g^2/T$}}}
\put(100,860){%
\special{ps: gsave currentpoint currentpoint translate
270 rotate neg exch neg exch translate}%
\makebox(0,0)[b]{\shortstack{\raisebox{-1.5em}{$\alpha_{\rm eff}$}}}%
\special{ps: currentpoint grestore moveto}%
}
\put(2389,151){\makebox(0,0){1.4}}
\put(2134,151){\makebox(0,0){1.2}}
\put(1878,151){\makebox(0,0){1}}
\put(1622,151){\makebox(0,0){0.8}}
\put(1367,151){\makebox(0,0){0.6}}
\put(1111,151){\makebox(0,0){0.4}}
\put(856,151){\makebox(0,0){0.2}}
\put(600,151){\makebox(0,0){0}}
\put(540,1469){\makebox(0,0)[r]{10}}
\put(540,1225){\makebox(0,0)[r]{8}}
\put(540,982){\makebox(0,0)[r]{6}}
\put(540,738){\makebox(0,0)[r]{4}}
\put(540,495){\makebox(0,0)[r]{2}}
\put(540,251){\makebox(0,0)[r]{0}}
\end{picture}
\vskip 20pt
\setlength{\unitlength}{0.1bp}
\special{!
/gnudict 40 dict def
gnudict begin
/Color false def
/Solid false def
/gnulinewidth 5.000 def
/vshift -33 def
/dl {10 mul} def
/hpt 31.5 def
/vpt 31.5 def
/M {moveto} bind def
/L {lineto} bind def
/R {rmoveto} bind def
/V {rlineto} bind def
/vpt2 vpt 2 mul def
/hpt2 hpt 2 mul def
/Lshow { currentpoint stroke M
  0 vshift R show } def
/Rshow { currentpoint stroke M
  dup stringwidth pop neg vshift R show } def
/Cshow { currentpoint stroke M
  dup stringwidth pop -2 div vshift R show } def
/DL { Color {setrgbcolor Solid {pop []} if 0 setdash }
 {pop pop pop Solid {pop []} if 0 setdash} ifelse } def
/BL { stroke gnulinewidth 2 mul setlinewidth } def
/AL { stroke gnulinewidth 2 div setlinewidth } def
/PL { stroke gnulinewidth setlinewidth } def
/LTb { BL [] 0 0 0 DL } def
/LTa { AL [1 dl 2 dl] 0 setdash 0 0 0 setrgbcolor } def
/LT0 { PL [] 0 1 0 DL } def
/LT1 { PL [4 dl 2 dl] 0 0 1 DL } def
/LT2 { PL [2 dl 3 dl] 1 0 0 DL } def
/LT3 { PL [1 dl 1.5 dl] 1 0 1 DL } def
/LT4 { PL [5 dl 2 dl 1 dl 2 dl] 0 1 1 DL } def
/LT5 { PL [4 dl 3 dl 1 dl 3 dl] 1 1 0 DL } def
/LT6 { PL [2 dl 2 dl 2 dl 4 dl] 0 0 0 DL } def
/LT7 { PL [2 dl 2 dl 2 dl 2 dl 2 dl 4 dl] 1 0.3 0 DL } def
/LT8 { PL [2 dl 2 dl 2 dl 2 dl 2 dl 2 dl 2 dl 4 dl] 0.5 0.5 0.5 DL } def
/P { stroke [] 0 setdash
  currentlinewidth 2 div sub M
  0 currentlinewidth V stroke } def
/D { stroke [] 0 setdash 2 copy vpt add M
  hpt neg vpt neg V hpt vpt neg V
  hpt vpt V hpt neg vpt V closepath stroke
  P } def
/A { stroke [] 0 setdash vpt sub M 0 vpt2 V
  currentpoint stroke M
  hpt neg vpt neg R hpt2 0 V stroke
  } def
/B { stroke [] 0 setdash 2 copy exch hpt sub exch vpt add M
  0 vpt2 neg V hpt2 0 V 0 vpt2 V
  hpt2 neg 0 V closepath stroke
  P } def
/C { stroke [] 0 setdash exch hpt sub exch vpt add M
  hpt2 vpt2 neg V currentpoint stroke M
  hpt2 neg 0 R hpt2 vpt2 V stroke } def
/T { stroke [] 0 setdash 2 copy vpt 1.12 mul add M
  hpt neg vpt -1.62 mul V
  hpt 2 mul 0 V
  hpt neg vpt 1.62 mul V closepath stroke
  P  } def
/S { 2 copy A C} def
end
}
\begin{picture}(2700,1620)(0,0)
\special{"
gnudict begin
gsave
50 50 translate
0.100 0.100 scale
0 setgray
/Helvetica findfont 100 scalefont setfont
newpath
-500.000000 -500.000000 translate
LTa
600 251 M
1917 0 V
600 251 M
0 1218 V
LTb
600 251 M
63 0 V
1854 0 R
-63 0 V
600 495 M
63 0 V
1854 0 R
-63 0 V
600 738 M
63 0 V
1854 0 R
-63 0 V
600 982 M
63 0 V
1854 0 R
-63 0 V
600 1225 M
63 0 V
1854 0 R
-63 0 V
600 1469 M
63 0 V
1854 0 R
-63 0 V
600 251 M
0 63 V
0 1155 R
0 -63 V
856 251 M
0 63 V
0 1155 R
0 -63 V
1111 251 M
0 63 V
0 1155 R
0 -63 V
1367 251 M
0 63 V
0 1155 R
0 -63 V
1622 251 M
0 63 V
0 1155 R
0 -63 V
1878 251 M
0 63 V
0 1155 R
0 -63 V
2134 251 M
0 63 V
0 1155 R
0 -63 V
2389 251 M
0 63 V
0 1155 R
0 -63 V
600 251 M
1917 0 V
0 1218 V
-1917 0 V
600 251 L
LT0
804 948 D
1009 968 D
1284 1017 D
2081 1070 D
804 935 M
0 26 V
773 935 M
62 0 V
-62 26 R
62 0 V
174 -8 R
0 31 V
978 953 M
62 0 V
-62 31 R
62 0 V
244 16 R
0 34 V
-31 -34 R
62 0 V
-62 34 R
62 0 V
766 18 R
0 36 V
-31 -36 R
62 0 V
-62 36 R
62 0 V
LT1
600 910 M
19 0 V
20 0 V
19 0 V
19 0 V
20 0 V
19 0 V
20 0 V
19 0 V
19 0 V
20 0 V
19 0 V
19 0 V
20 0 V
19 0 V
19 0 V
20 0 V
19 0 V
20 0 V
19 0 V
19 0 V
20 0 V
19 0 V
19 0 V
20 0 V
19 0 V
19 0 V
20 0 V
19 0 V
20 0 V
19 0 V
19 0 V
20 0 V
19 0 V
19 0 V
20 0 V
19 0 V
19 0 V
20 0 V
19 0 V
20 0 V
19 0 V
19 0 V
20 0 V
19 0 V
19 0 V
20 0 V
19 0 V
19 0 V
20 0 V
19 0 V
20 0 V
19 0 V
19 0 V
20 0 V
19 0 V
19 0 V
20 0 V
19 0 V
19 0 V
20 0 V
19 0 V
20 0 V
19 0 V
19 0 V
20 0 V
19 0 V
19 0 V
20 0 V
19 0 V
19 0 V
20 0 V
19 0 V
20 0 V
19 0 V
19 0 V
20 0 V
19 0 V
19 0 V
20 0 V
19 0 V
19 0 V
20 0 V
19 0 V
20 0 V
19 0 V
19 0 V
20 0 V
19 0 V
19 0 V
20 0 V
19 0 V
19 0 V
20 0 V
19 0 V
20 0 V
19 0 V
19 0 V
20 0 V
19 0 V
stroke
grestore
end
showpage
}
\put(1558,1569){\makebox(0,0){(c) \ $L_t=4$}}
\put(1558,51){\makebox(0,0){\raisebox{-1.5em}{$g^2/T$}}}
\put(100,860){%
\special{ps: gsave currentpoint currentpoint translate
270 rotate neg exch neg exch translate}%
\makebox(0,0)[b]{\shortstack{\raisebox{-1.5em}{$\alpha_{\rm eff}$}}}%
\special{ps: currentpoint grestore moveto}%
}
\put(2389,151){\makebox(0,0){1.4}}
\put(2134,151){\makebox(0,0){1.2}}
\put(1878,151){\makebox(0,0){1}}
\put(1622,151){\makebox(0,0){0.8}}
\put(1367,151){\makebox(0,0){0.6}}
\put(1111,151){\makebox(0,0){0.4}}
\put(856,151){\makebox(0,0){0.2}}
\put(600,151){\makebox(0,0){0}}
\put(540,1469){\makebox(0,0)[r]{10}}
\put(540,1225){\makebox(0,0)[r]{8}}
\put(540,982){\makebox(0,0)[r]{6}}
\put(540,738){\makebox(0,0)[r]{4}}
\put(540,495){\makebox(0,0)[r]{2}}
\put(540,251){\makebox(0,0)[r]{0}}
\end{picture}
\end{figure}
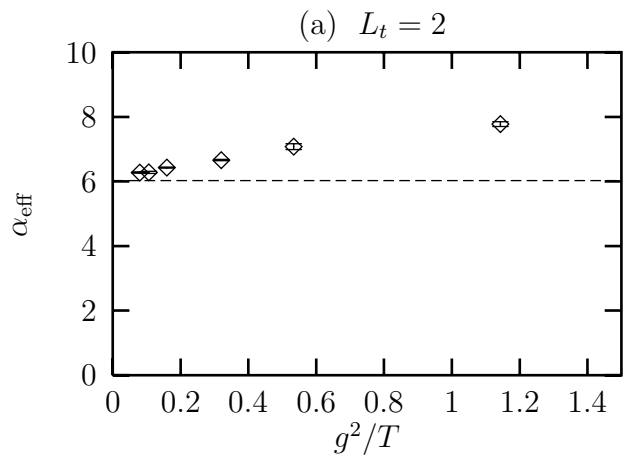

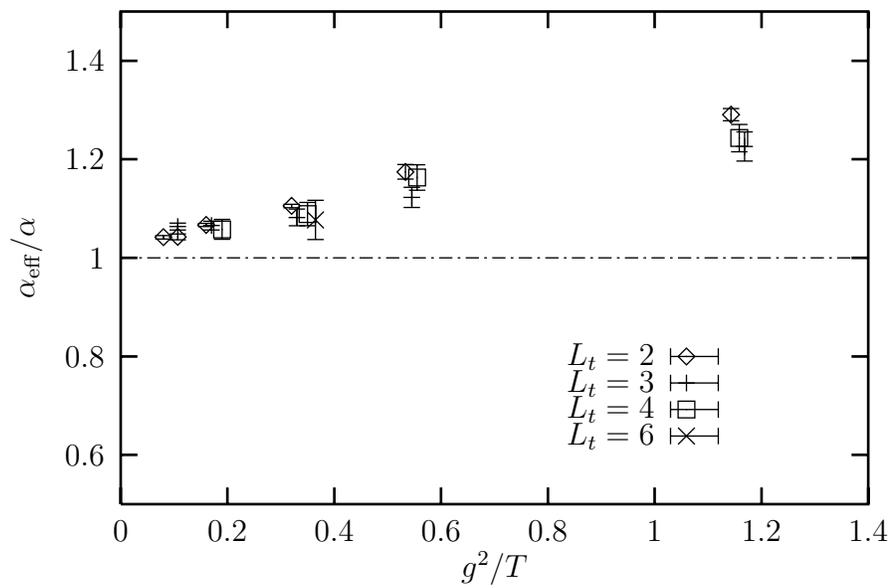
\begin{figure}[h]
\caption[9]{Ratio of numerical and perturbative interface tensions
versus the high-$T$ expansion parameter, $g^2/T$.}
\vskip 20pt
\setlength{\unitlength}{0.1bp}
\special{!
/gnudict 40 dict def
gnudict begin
/Color false def
/Solid false def
/gnulinewidth 5.000 def
/vshift -33 def
/dl {10 mul} def
/hpt 31.5 def
/vpt 31.5 def
/M {moveto} bind def
/L {lineto} bind def
/R {rmoveto} bind def
/V {rlineto} bind def
/vpt2 vpt 2 mul def
/hpt2 hpt 2 mul def
/Lshow { currentpoint stroke M
  0 vshift R show } def
/Rshow { currentpoint stroke M
  dup stringwidth pop neg vshift R show } def
/Cshow { currentpoint stroke M
  dup stringwidth pop -2 div vshift R show } def
/DL { Color {setrgbcolor Solid {pop []} if 0 setdash }
 {pop pop pop Solid {pop []} if 0 setdash} ifelse } def
/BL { stroke gnulinewidth 2 mul setlinewidth } def
/AL { stroke gnulinewidth 2 div setlinewidth } def
/PL { stroke gnulinewidth setlinewidth } def
/LTb { BL [] 0 0 0 DL } def
/LTa { AL [1 dl 2 dl] 0 setdash 0 0 0 setrgbcolor } def
/LT0 { PL [] 0 1 0 DL } def
/LT1 { PL [4 dl 2 dl] 0 0 1 DL } def
/LT2 { PL [2 dl 3 dl] 1 0 0 DL } def
/LT3 { PL [1 dl 1.5 dl] 1 0 1 DL } def
/LT4 { PL [5 dl 2 dl 1 dl 2 dl] 0 1 1 DL } def
/LT5 { PL [4 dl 3 dl 1 dl 3 dl] 1 1 0 DL } def
/LT6 { PL [2 dl 2 dl 2 dl 4 dl] 0 0 0 DL } def
/LT7 { PL [2 dl 2 dl 2 dl 2 dl 2 dl 4 dl] 1 0.3 0 DL } def
/LT8 { PL [2 dl 2 dl 2 dl 2 dl 2 dl 2 dl 2 dl 4 dl] 0.5 0.5 0.5 DL } def
/P { stroke [] 0 setdash
  currentlinewidth 2 div sub M
  0 currentlinewidth V stroke } def
/D { stroke [] 0 setdash 2 copy vpt add M
  hpt neg vpt neg V hpt vpt neg V
  hpt vpt V hpt neg vpt V closepath stroke
  P } def
/A { stroke [] 0 setdash vpt sub M 0 vpt2 V
  currentpoint stroke M
  hpt neg vpt neg R hpt2 0 V stroke
  } def
/B { stroke [] 0 setdash 2 copy exch hpt sub exch vpt add M
  0 vpt2 neg V hpt2 0 V 0 vpt2 V
  hpt2 neg 0 V closepath stroke
  P } def
/C { stroke [] 0 setdash exch hpt sub exch vpt add M
  hpt2 vpt2 neg V currentpoint stroke M
  hpt2 neg 0 R hpt2 vpt2 V stroke } def
/T { stroke [] 0 setdash 2 copy vpt 1.12 mul add M
  hpt neg vpt -1.62 mul V
  hpt 2 mul 0 V
  hpt neg vpt 1.62 mul V closepath stroke
  P  } def
/S { 2 copy A C} def
end
}
\begin{picture}(3600,2160)(0,0)
\special{"
gnudict begin
gsave
50 50 translate
0.100 0.100 scale
0 setgray
/Helvetica findfont 100 scalefont setfont
newpath
-500.000000 -500.000000 translate
LTa
600 251 M
0 1858 V
LTb
600 437 M
63 0 V
2754 0 R
-63 0 V
600 808 M
63 0 V
2754 0 R
-63 0 V
600 1180 M
63 0 V
2754 0 R
-63 0 V
600 1552 M
63 0 V
2754 0 R
-63 0 V
600 1923 M
63 0 V
2754 0 R
-63 0 V
600 251 M
0 63 V
0 1795 R
0 -63 V
1002 251 M
0 63 V
0 1795 R
0 -63 V
1405 251 M
0 63 V
0 1795 R
0 -63 V
1807 251 M
0 63 V
0 1795 R
0 -63 V
2210 251 M
0 63 V
0 1795 R
0 -63 V
2612 251 M
0 63 V
0 1795 R
0 -63 V
3015 251 M
0 63 V
0 1795 R
0 -63 V
3417 251 M
0 63 V
0 1795 R
0 -63 V
600 251 M
2817 0 V
0 1858 V
-2817 0 V
600 251 L
LT0
2732 808 D
761 1258 D
815 1259 D
922 1304 D
1244 1376 D
1673 1504 D
2900 1720 D
2672 808 M
180 0 V
-180 31 R
0 -62 V
180 62 R
0 -62 V
761 1251 M
0 13 V
-31 -13 R
62 0 V
-62 13 R
62 0 V
23 -16 R
0 22 V
-31 -22 R
62 0 V
-62 22 R
62 0 V
76 28 R
0 12 V
-31 -12 R
62 0 V
-62 12 R
62 0 V
291 59 R
0 13 V
-31 -13 R
62 0 V
-62 13 R
62 0 V
398 95 R
0 55 V
-31 -55 R
62 0 V
-62 55 R
62 0 V
1196 165 R
0 46 V
-31 -46 R
62 0 V
-62 46 R
62 0 V
LT1
2732 708 A
815 1298 A
942 1302 A
1264 1332 A
1697 1408 A
2951 1600 A
2672 708 M
180 0 V
-180 31 R
0 -62 V
180 62 R
0 -62 V
815 1285 M
0 26 V
-31 -26 R
62 0 V
-62 26 R
62 0 V
96 -26 R
0 33 V
-31 -33 R
62 0 V
-62 33 R
62 0 V
291 -17 R
0 63 V
-31 -63 R
62 0 V
-62 63 R
62 0 V
402 6 R
0 76 V
-31 -76 R
62 0 V
-62 76 R
62 0 V
1223 99 R
0 110 V
-31 -110 R
62 0 V
-62 110 R
62 0 V
LT2
2732 608 B
982 1287 B
1304 1345 B
1717 1483 B
2931 1632 B
2672 608 M
180 0 V
-180 31 R
0 -62 V
180 62 R
0 -62 V
982 1250 M
0 75 V
-31 -75 R
62 0 V
-62 75 R
62 0 V
291 -24 R
0 88 V
-31 -88 R
62 0 V
-62 88 R
62 0 V
382 46 R
0 96 V
-31 -96 R
62 0 V
-62 96 R
62 0 V
1183 49 R
0 103 V
-31 -103 R
62 0 V
-62 103 R
62 0 V
LT3
2732 508 C
1334 1323 C
2672 508 M
180 0 V
-180 31 R
0 -62 V
180 62 R
0 -62 V
1334 1249 M
0 148 V
-31 -148 R
62 0 V
-62 148 R
62 0 V
LT4
600 1180 M
28 0 V
29 0 V
28 0 V
29 0 V
28 0 V
29 0 V
28 0 V
29 0 V
28 0 V
29 0 V
28 0 V
28 0 V
29 0 V
28 0 V
29 0 V
28 0 V
29 0 V
28 0 V
29 0 V
28 0 V
29 0 V
28 0 V
28 0 V
29 0 V
28 0 V
29 0 V
28 0 V
29 0 V
28 0 V
29 0 V
28 0 V
29 0 V
28 0 V
28 0 V
29 0 V
28 0 V
29 0 V
28 0 V
29 0 V
28 0 V
29 0 V
28 0 V
29 0 V
28 0 V
28 0 V
29 0 V
28 0 V
29 0 V
28 0 V
29 0 V
28 0 V
29 0 V
28 0 V
29 0 V
28 0 V
28 0 V
29 0 V
28 0 V
29 0 V
28 0 V
29 0 V
28 0 V
29 0 V
28 0 V
29 0 V
28 0 V
28 0 V
29 0 V
28 0 V
29 0 V
28 0 V
29 0 V
28 0 V
29 0 V
28 0 V
29 0 V
28 0 V
28 0 V
29 0 V
28 0 V
29 0 V
28 0 V
29 0 V
28 0 V
29 0 V
28 0 V
29 0 V
28 0 V
28 0 V
29 0 V
28 0 V
29 0 V
28 0 V
29 0 V
28 0 V
29 0 V
28 0 V
29 0 V
28 0 V
stroke
grestore
end
showpage
}
\put(2612,508){\makebox(0,0)[r]{$L_t=6$}}
\put(2612,608){\makebox(0,0)[r]{$L_t=4$}}
\put(2612,708){\makebox(0,0)[r]{$L_t=3$}}
\put(2612,808){\makebox(0,0)[r]{$L_t=2$}}
\put(2008,51){\makebox(0,0){\raisebox{-1.5em}{$g^2/T$}}}
\put(100,1180){%
\special{ps: gsave currentpoint currentpoint translate
270 rotate neg exch neg exch translate}%
\makebox(0,0)[b]{\shortstack{\raisebox{-1.5em}{$\alpha_{\rm eff}/\alpha$}}}%
\special{ps: currentpoint grestore moveto}%
}
\put(3417,151){\makebox(0,0){1.4}}
\put(3015,151){\makebox(0,0){1.2}}
\put(2612,151){\makebox(0,0){1}}
\put(2210,151){\makebox(0,0){0.8}}
\put(1807,151){\makebox(0,0){0.6}}
\put(1405,151){\makebox(0,0){0.4}}
\put(1002,151){\makebox(0,0){0.2}}
\put(600,151){\makebox(0,0){0}}
\put(540,1923){\makebox(0,0)[r]{1.4}}
\put(540,1552){\makebox(0,0)[r]{1.2}}
\put(540,1180){\makebox(0,0)[r]{1}}
\put(540,808){\makebox(0,0)[r]{0.8}}
\put(540,437){\makebox(0,0)[r]{0.6}}
\end{picture}
\end{figure}

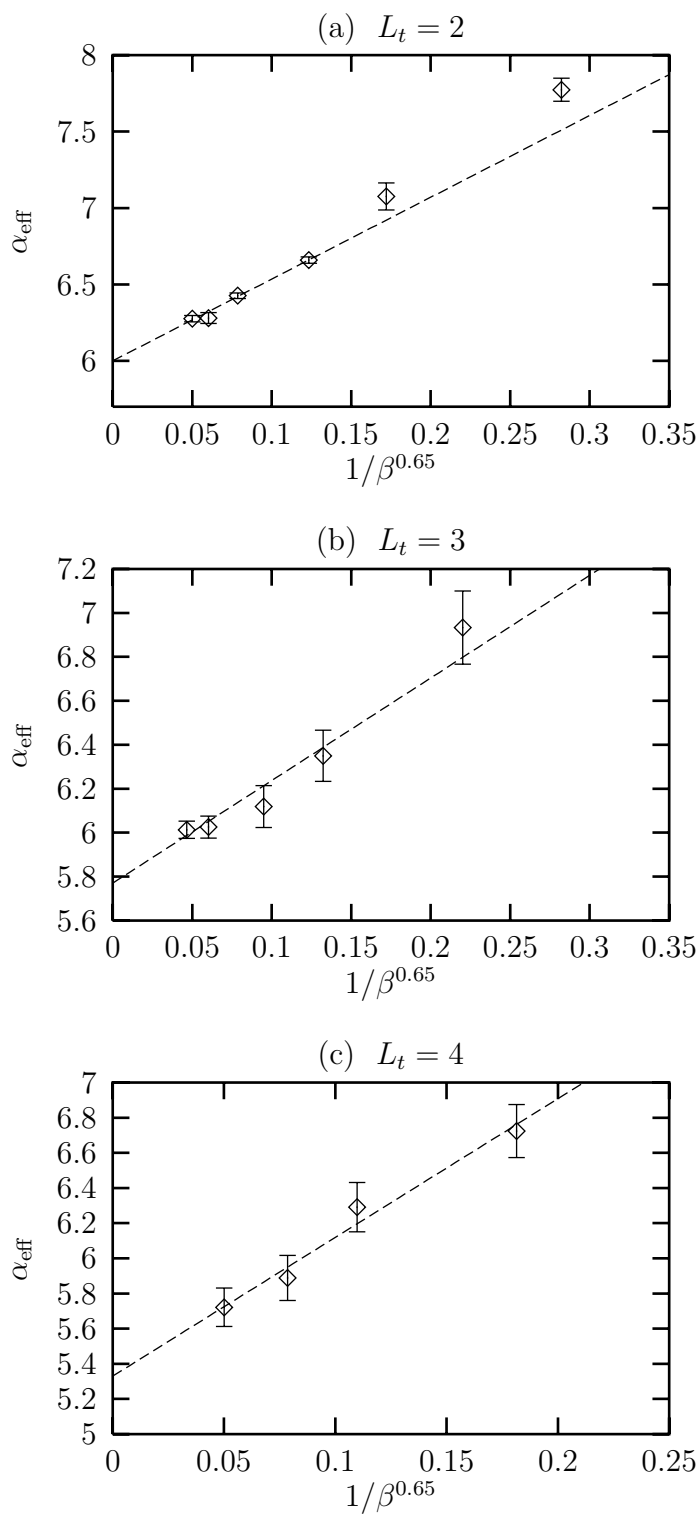
\begin{figure}[h]
\caption[10]{Trial extrapolations to the $T=\infty$ limit of
the calculated interface tensions.}
\vskip 20pt
\setlength{\unitlength}{0.1bp}
\special{!
/gnudict 40 dict def
gnudict begin
/Color false def
/Solid false def
/gnulinewidth 5.000 def
/vshift -33 def
/dl {10 mul} def
/hpt 31.5 def
/vpt 31.5 def
/M {moveto} bind def
/L {lineto} bind def
/R {rmoveto} bind def
/V {rlineto} bind def
/vpt2 vpt 2 mul def
/hpt2 hpt 2 mul def
/Lshow { currentpoint stroke M
  0 vshift R show } def
/Rshow { currentpoint stroke M
  dup stringwidth pop neg vshift R show } def
/Cshow { currentpoint stroke M
  dup stringwidth pop -2 div vshift R show } def
/DL { Color {setrgbcolor Solid {pop []} if 0 setdash }
 {pop pop pop Solid {pop []} if 0 setdash} ifelse } def
/BL { stroke gnulinewidth 2 mul setlinewidth } def
/AL { stroke gnulinewidth 2 div setlinewidth } def
/PL { stroke gnulinewidth setlinewidth } def
/LTb { BL [] 0 0 0 DL } def
/LTa { AL [1 dl 2 dl] 0 setdash 0 0 0 setrgbcolor } def
/LT0 { PL [] 0 1 0 DL } def
/LT1 { PL [4 dl 2 dl] 0 0 1 DL } def
/LT2 { PL [2 dl 3 dl] 1 0 0 DL } def
/LT3 { PL [1 dl 1.5 dl] 1 0 1 DL } def
/LT4 { PL [5 dl 2 dl 1 dl 2 dl] 0 1 1 DL } def
/LT5 { PL [4 dl 3 dl 1 dl 3 dl] 1 1 0 DL } def
/LT6 { PL [2 dl 2 dl 2 dl 4 dl] 0 0 0 DL } def
/LT7 { PL [2 dl 2 dl 2 dl 2 dl 2 dl 4 dl] 1 0.3 0 DL } def
/LT8 { PL [2 dl 2 dl 2 dl 2 dl 2 dl 2 dl 2 dl 4 dl] 0.5 0.5 0.5 DL } def
/P { stroke [] 0 setdash
  currentlinewidth 2 div sub M
  0 currentlinewidth V stroke } def
/D { stroke [] 0 setdash 2 copy vpt add M
  hpt neg vpt neg V hpt vpt neg V
  hpt vpt V hpt neg vpt V closepath stroke
  P } def
/A { stroke [] 0 setdash vpt sub M 0 vpt2 V
  currentpoint stroke M
  hpt neg vpt neg R hpt2 0 V stroke
  } def
/B { stroke [] 0 setdash 2 copy exch hpt sub exch vpt add M
  0 vpt2 neg V hpt2 0 V 0 vpt2 V
  hpt2 neg 0 V closepath stroke
  P } def
/C { stroke [] 0 setdash exch hpt sub exch vpt add M
  hpt2 vpt2 neg V currentpoint stroke M
  hpt2 neg 0 R hpt2 vpt2 V stroke } def
/T { stroke [] 0 setdash 2 copy vpt 1.12 mul add M
  hpt neg vpt -1.62 mul V
  hpt 2 mul 0 V
  hpt neg vpt 1.62 mul V closepath stroke
  P  } def
/S { 2 copy A C} def
end
}
\begin{picture}(2880,1728)(0,0)
\special{"
gnudict begin
gsave
50 50 translate
0.100 0.100 scale
0 setgray
/Helvetica findfont 100 scalefont setfont
newpath
-500.000000 -500.000000 translate
LTa
600 251 M
0 1326 V
LTb
600 424 M
63 0 V
2034 0 R
-63 0 V
600 712 M
63 0 V
2034 0 R
-63 0 V
600 1000 M
63 0 V
2034 0 R
-63 0 V
600 1289 M
63 0 V
2034 0 R
-63 0 V
600 1577 M
63 0 V
2034 0 R
-63 0 V
600 251 M
0 63 V
0 1263 R
0 -63 V
900 251 M
0 63 V
0 1263 R
0 -63 V
1199 251 M
0 63 V
0 1263 R
0 -63 V
1499 251 M
0 63 V
0 1263 R
0 -63 V
1798 251 M
0 63 V
0 1263 R
0 -63 V
2098 251 M
0 63 V
0 1263 R
0 -63 V
2397 251 M
0 63 V
0 1263 R
0 -63 V
2697 251 M
0 63 V
0 1263 R
0 -63 V
600 251 M
2097 0 V
0 1326 V
-2097 0 V
600 251 L
LT0
900 583 D
961 585 D
1071 670 D
1339 804 D
1631 1044 D
2291 1446 D
900 571 M
0 24 V
869 571 M
62 0 V
-62 24 R
62 0 V
30 -30 R
0 41 V
930 565 M
62 0 V
-62 41 R
62 0 V
79 53 R
0 21 V
-31 -21 R
62 0 V
-62 21 R
62 0 V
237 112 R
0 24 V
-31 -24 R
62 0 V
-62 24 R
62 0 V
261 177 R
0 102 V
1600 993 M
62 0 V
-62 102 R
62 0 V
629 308 R
0 87 V
-31 -87 R
62 0 V
-62 87 R
62 0 V
LT1
600 424 M
21 11 V
21 11 V
22 11 V
21 11 V
21 10 V
21 11 V
21 11 V
21 11 V
22 11 V
21 11 V
21 11 V
21 11 V
21 11 V
22 11 V
21 11 V
21 10 V
21 11 V
21 11 V
21 11 V
22 11 V
21 11 V
21 11 V
21 11 V
21 11 V
22 11 V
21 10 V
21 11 V
21 11 V
21 11 V
21 11 V
22 11 V
21 11 V
21 11 V
21 11 V
21 11 V
22 11 V
21 10 V
21 11 V
21 11 V
21 11 V
21 11 V
22 11 V
21 11 V
21 11 V
21 11 V
21 11 V
22 10 V
21 11 V
21 11 V
21 11 V
21 11 V
21 11 V
22 11 V
21 11 V
21 11 V
21 11 V
21 11 V
22 10 V
21 11 V
21 11 V
21 11 V
21 11 V
21 11 V
22 11 V
21 11 V
21 11 V
21 11 V
21 10 V
22 11 V
21 11 V
21 11 V
21 11 V
21 11 V
21 11 V
22 11 V
21 11 V
21 11 V
21 11 V
21 10 V
22 11 V
21 11 V
21 11 V
21 11 V
21 11 V
21 11 V
22 11 V
21 11 V
21 11 V
21 10 V
21 11 V
22 11 V
21 11 V
21 11 V
21 11 V
21 11 V
21 11 V
22 11 V
21 11 V
21 10 V
stroke
grestore
end
showpage
}
\put(1648,1677){\makebox(0,0){(a) \ $L_t = 2$}}
\put(1648,51){\makebox(0,0){\raisebox{-1.5em}{$1/\beta^{0.65}$}}}
\put(100,914){%
\special{ps: gsave currentpoint currentpoint translate
270 rotate neg exch neg exch translate}%
\makebox(0,0)[b]{\shortstack{\raisebox{-1.5em}{$\alpha_{\rm eff}$}}}%
\special{ps: currentpoint grestore moveto}%
}
\put(2697,151){\makebox(0,0){0.35}}
\put(2397,151){\makebox(0,0){0.3}}
\put(2098,151){\makebox(0,0){0.25}}
\put(1798,151){\makebox(0,0){0.2}}
\put(1499,151){\makebox(0,0){0.15}}
\put(1199,151){\makebox(0,0){0.1}}
\put(900,151){\makebox(0,0){0.05}}
\put(600,151){\makebox(0,0){0}}
\put(540,1577){\makebox(0,0)[r]{8}}
\put(540,1289){\makebox(0,0)[r]{7.5}}
\put(540,1000){\makebox(0,0)[r]{7}}
\put(540,712){\makebox(0,0)[r]{6.5}}
\put(540,424){\makebox(0,0)[r]{6}}
\end{picture}
\vskip 20pt
\setlength{\unitlength}{0.1bp}
\special{!
/gnudict 40 dict def
gnudict begin
/Color false def
/Solid false def
/gnulinewidth 5.000 def
/vshift -33 def
/dl {10 mul} def
/hpt 31.5 def
/vpt 31.5 def
/M {moveto} bind def
/L {lineto} bind def
/R {rmoveto} bind def
/V {rlineto} bind def
/vpt2 vpt 2 mul def
/hpt2 hpt 2 mul def
/Lshow { currentpoint stroke M
  0 vshift R show } def
/Rshow { currentpoint stroke M
  dup stringwidth pop neg vshift R show } def
/Cshow { currentpoint stroke M
  dup stringwidth pop -2 div vshift R show } def
/DL { Color {setrgbcolor Solid {pop []} if 0 setdash }
 {pop pop pop Solid {pop []} if 0 setdash} ifelse } def
/BL { stroke gnulinewidth 2 mul setlinewidth } def
/AL { stroke gnulinewidth 2 div setlinewidth } def
/PL { stroke gnulinewidth setlinewidth } def
/LTb { BL [] 0 0 0 DL } def
/LTa { AL [1 dl 2 dl] 0 setdash 0 0 0 setrgbcolor } def
/LT0 { PL [] 0 1 0 DL } def
/LT1 { PL [4 dl 2 dl] 0 0 1 DL } def
/LT2 { PL [2 dl 3 dl] 1 0 0 DL } def
/LT3 { PL [1 dl 1.5 dl] 1 0 1 DL } def
/LT4 { PL [5 dl 2 dl 1 dl 2 dl] 0 1 1 DL } def
/LT5 { PL [4 dl 3 dl 1 dl 3 dl] 1 1 0 DL } def
/LT6 { PL [2 dl 2 dl 2 dl 4 dl] 0 0 0 DL } def
/LT7 { PL [2 dl 2 dl 2 dl 2 dl 2 dl 4 dl] 1 0.3 0 DL } def
/LT8 { PL [2 dl 2 dl 2 dl 2 dl 2 dl 2 dl 2 dl 4 dl] 0.5 0.5 0.5 DL } def
/P { stroke [] 0 setdash
  currentlinewidth 2 div sub M
  0 currentlinewidth V stroke } def
/D { stroke [] 0 setdash 2 copy vpt add M
  hpt neg vpt neg V hpt vpt neg V
  hpt vpt V hpt neg vpt V closepath stroke
  P } def
/A { stroke [] 0 setdash vpt sub M 0 vpt2 V
  currentpoint stroke M
  hpt neg vpt neg R hpt2 0 V stroke
  } def
/B { stroke [] 0 setdash 2 copy exch hpt sub exch vpt add M
  0 vpt2 neg V hpt2 0 V 0 vpt2 V
  hpt2 neg 0 V closepath stroke
  P } def
/C { stroke [] 0 setdash exch hpt sub exch vpt add M
  hpt2 vpt2 neg V currentpoint stroke M
  hpt2 neg 0 R hpt2 vpt2 V stroke } def
/T { stroke [] 0 setdash 2 copy vpt 1.12 mul add M
  hpt neg vpt -1.62 mul V
  hpt 2 mul 0 V
  hpt neg vpt 1.62 mul V closepath stroke
  P  } def
/S { 2 copy A C} def
end
}
\begin{picture}(2880,1728)(0,0)
\special{"
gnudict begin
gsave
50 50 translate
0.100 0.100 scale
0 setgray
/Helvetica findfont 100 scalefont setfont
newpath
-500.000000 -500.000000 translate
LTa
600 251 M
0 1326 V
LTb
600 251 M
63 0 V
2034 0 R
-63 0 V
600 417 M
63 0 V
2034 0 R
-63 0 V
600 583 M
63 0 V
2034 0 R
-63 0 V
600 748 M
63 0 V
2034 0 R
-63 0 V
600 914 M
63 0 V
2034 0 R
-63 0 V
600 1080 M
63 0 V
2034 0 R
-63 0 V
600 1246 M
63 0 V
2034 0 R
-63 0 V
600 1411 M
63 0 V
2034 0 R
-63 0 V
600 1577 M
63 0 V
2034 0 R
-63 0 V
600 251 M
0 63 V
0 1263 R
0 -63 V
900 251 M
0 63 V
0 1263 R
0 -63 V
1199 251 M
0 63 V
0 1263 R
0 -63 V
1499 251 M
0 63 V
0 1263 R
0 -63 V
1798 251 M
0 63 V
0 1263 R
0 -63 V
2098 251 M
0 63 V
0 1263 R
0 -63 V
2397 251 M
0 63 V
0 1263 R
0 -63 V
2697 251 M
0 63 V
0 1263 R
0 -63 V
600 251 M
2097 0 V
0 1326 V
-2097 0 V
600 251 L
LT0
879 593 D
961 604 D
1169 681 D
1393 872 D
1919 1356 D
879 561 M
0 65 V
848 561 M
62 0 V
-62 65 R
62 0 V
51 -64 R
0 83 V
930 562 M
62 0 V
-62 83 R
62 0 V
177 -43 R
0 158 V
1138 602 M
62 0 V
-62 158 R
62 0 V
193 16 R
0 193 V
1362 776 M
62 0 V
-62 193 R
62 0 V
495 249 R
0 276 V
-31 -276 R
62 0 V
-62 276 R
62 0 V
LT1
600 392 M
21 14 V
21 13 V
22 14 V
21 14 V
21 13 V
21 14 V
21 14 V
21 13 V
22 14 V
21 14 V
21 13 V
21 14 V
21 14 V
22 13 V
21 14 V
21 14 V
21 13 V
21 14 V
21 14 V
22 14 V
21 13 V
21 14 V
21 14 V
21 13 V
22 14 V
21 14 V
21 13 V
21 14 V
21 14 V
21 13 V
22 14 V
21 14 V
21 13 V
21 14 V
21 14 V
22 13 V
21 14 V
21 14 V
21 14 V
21 13 V
21 14 V
22 14 V
21 13 V
21 14 V
21 14 V
21 13 V
22 14 V
21 14 V
21 13 V
21 14 V
21 14 V
21 13 V
22 14 V
21 14 V
21 13 V
21 14 V
21 14 V
22 13 V
21 14 V
21 14 V
21 14 V
21 13 V
21 14 V
22 14 V
21 13 V
21 14 V
21 14 V
21 13 V
22 14 V
21 14 V
21 13 V
21 14 V
21 14 V
21 13 V
22 14 V
21 14 V
21 13 V
21 14 V
21 14 V
22 14 V
21 13 V
21 14 V
21 14 V
21 13 V
21 14 V
22 14 V
13 8 V
stroke
grestore
end
showpage
}
\put(1648,1677){\makebox(0,0){(b) \ $L_t = 3$}}
\put(1648,51){\makebox(0,0){\raisebox{-1.5em}{$1/\beta^{0.65}$}}}
\put(100,914){%
\special{ps: gsave currentpoint currentpoint translate
270 rotate neg exch neg exch translate}%
\makebox(0,0)[b]{\shortstack{\raisebox{-1.5em}{$\alpha_{\rm eff}$}}}%
\special{ps: currentpoint grestore moveto}%
}
\put(2697,151){\makebox(0,0){0.35}}
\put(2397,151){\makebox(0,0){0.3}}
\put(2098,151){\makebox(0,0){0.25}}
\put(1798,151){\makebox(0,0){0.2}}
\put(1499,151){\makebox(0,0){0.15}}
\put(1199,151){\makebox(0,0){0.1}}
\put(900,151){\makebox(0,0){0.05}}
\put(600,151){\makebox(0,0){0}}
\put(540,1577){\makebox(0,0)[r]{7.2}}
\put(540,1411){\makebox(0,0)[r]{7}}
\put(540,1246){\makebox(0,0)[r]{6.8}}
\put(540,1080){\makebox(0,0)[r]{6.6}}
\put(540,914){\makebox(0,0)[r]{6.4}}
\put(540,748){\makebox(0,0)[r]{6.2}}
\put(540,583){\makebox(0,0)[r]{6}}
\put(540,417){\makebox(0,0)[r]{5.8}}
\put(540,251){\makebox(0,0)[r]{5.6}}
\end{picture}
\vskip 20pt
\setlength{\unitlength}{0.1bp}
\special{!
/gnudict 40 dict def
gnudict begin
/Color false def
/Solid false def
/gnulinewidth 5.000 def
/vshift -33 def
/dl {10 mul} def
/hpt 31.5 def
/vpt 31.5 def
/M {moveto} bind def
/L {lineto} bind def
/R {rmoveto} bind def
/V {rlineto} bind def
/vpt2 vpt 2 mul def
/hpt2 hpt 2 mul def
/Lshow { currentpoint stroke M
  0 vshift R show } def
/Rshow { currentpoint stroke M
  dup stringwidth pop neg vshift R show } def
/Cshow { currentpoint stroke M
  dup stringwidth pop -2 div vshift R show } def
/DL { Color {setrgbcolor Solid {pop []} if 0 setdash }
 {pop pop pop Solid {pop []} if 0 setdash} ifelse } def
/BL { stroke gnulinewidth 2 mul setlinewidth } def
/AL { stroke gnulinewidth 2 div setlinewidth } def
/PL { stroke gnulinewidth setlinewidth } def
/LTb { BL [] 0 0 0 DL } def
/LTa { AL [1 dl 2 dl] 0 setdash 0 0 0 setrgbcolor } def
/LT0 { PL [] 0 1 0 DL } def
/LT1 { PL [4 dl 2 dl] 0 0 1 DL } def
/LT2 { PL [2 dl 3 dl] 1 0 0 DL } def
/LT3 { PL [1 dl 1.5 dl] 1 0 1 DL } def
/LT4 { PL [5 dl 2 dl 1 dl 2 dl] 0 1 1 DL } def
/LT5 { PL [4 dl 3 dl 1 dl 3 dl] 1 1 0 DL } def
/LT6 { PL [2 dl 2 dl 2 dl 4 dl] 0 0 0 DL } def
/LT7 { PL [2 dl 2 dl 2 dl 2 dl 2 dl 4 dl] 1 0.3 0 DL } def
/LT8 { PL [2 dl 2 dl 2 dl 2 dl 2 dl 2 dl 2 dl 4 dl] 0.5 0.5 0.5 DL } def
/P { stroke [] 0 setdash
  currentlinewidth 2 div sub M
  0 currentlinewidth V stroke } def
/D { stroke [] 0 setdash 2 copy vpt add M
  hpt neg vpt neg V hpt vpt neg V
  hpt vpt V hpt neg vpt V closepath stroke
  P } def
/A { stroke [] 0 setdash vpt sub M 0 vpt2 V
  currentpoint stroke M
  hpt neg vpt neg R hpt2 0 V stroke
  } def
/B { stroke [] 0 setdash 2 copy exch hpt sub exch vpt add M
  0 vpt2 neg V hpt2 0 V 0 vpt2 V
  hpt2 neg 0 V closepath stroke
  P } def
/C { stroke [] 0 setdash exch hpt sub exch vpt add M
  hpt2 vpt2 neg V currentpoint stroke M
  hpt2 neg 0 R hpt2 vpt2 V stroke } def
/T { stroke [] 0 setdash 2 copy vpt 1.12 mul add M
  hpt neg vpt -1.62 mul V
  hpt 2 mul 0 V
  hpt neg vpt 1.62 mul V closepath stroke
  P  } def
/S { 2 copy A C} def
end
}
\begin{picture}(2880,1728)(0,0)
\special{"
gnudict begin
gsave
50 50 translate
0.100 0.100 scale
0 setgray
/Helvetica findfont 100 scalefont setfont
newpath
-500.000000 -500.000000 translate
LTa
600 251 M
0 1326 V
LTb
600 251 M
63 0 V
2034 0 R
-63 0 V
600 384 M
63 0 V
2034 0 R
-63 0 V
600 516 M
63 0 V
2034 0 R
-63 0 V
600 649 M
63 0 V
2034 0 R
-63 0 V
600 781 M
63 0 V
2034 0 R
-63 0 V
600 914 M
63 0 V
2034 0 R
-63 0 V
600 1047 M
63 0 V
2034 0 R
-63 0 V
600 1179 M
63 0 V
2034 0 R
-63 0 V
600 1312 M
63 0 V
2034 0 R
-63 0 V
600 1444 M
63 0 V
2034 0 R
-63 0 V
600 1577 M
63 0 V
2034 0 R
-63 0 V
600 251 M
0 63 V
0 1263 R
0 -63 V
1019 251 M
0 63 V
0 1263 R
0 -63 V
1439 251 M
0 63 V
0 1263 R
0 -63 V
1858 251 M
0 63 V
0 1263 R
0 -63 V
2278 251 M
0 63 V
0 1263 R
0 -63 V
2697 251 M
0 63 V
0 1263 R
0 -63 V
600 251 M
2097 0 V
0 1326 V
-2097 0 V
600 251 L
LT0
1020 729 D
1259 840 D
1521 1107 D
2122 1394 D
1020 657 M
0 145 V
989 657 M
62 0 V
989 802 M
62 0 V
208 -47 R
0 170 V
1228 755 M
62 0 V
-62 170 R
62 0 V
231 89 R
0 186 V
-31 -186 R
62 0 V
-62 186 R
62 0 V
570 94 R
0 200 V
-31 -200 R
62 0 V
-62 200 R
62 0 V
LT1
600 470 M
21 13 V
21 13 V
22 13 V
21 14 V
21 13 V
21 13 V
21 13 V
21 13 V
22 14 V
21 13 V
21 13 V
21 13 V
21 14 V
22 13 V
21 13 V
21 13 V
21 13 V
21 14 V
21 13 V
22 13 V
21 13 V
21 13 V
21 14 V
21 13 V
22 13 V
21 13 V
21 13 V
21 14 V
21 13 V
21 13 V
22 13 V
21 14 V
21 13 V
21 13 V
21 13 V
22 13 V
21 14 V
21 13 V
21 13 V
21 13 V
21 13 V
22 14 V
21 13 V
21 13 V
21 13 V
21 13 V
22 14 V
21 13 V
21 13 V
21 13 V
21 13 V
21 14 V
22 13 V
21 13 V
21 13 V
21 14 V
21 13 V
22 13 V
21 13 V
21 13 V
21 14 V
21 13 V
21 13 V
22 13 V
21 13 V
21 14 V
21 13 V
21 13 V
22 13 V
21 13 V
21 14 V
21 13 V
21 13 V
21 13 V
22 14 V
21 13 V
21 13 V
21 13 V
21 13 V
22 14 V
21 13 V
21 13 V
21 13 V
17 11 V
stroke
grestore
end
showpage
}
\put(1648,1677){\makebox(0,0){(c) \ $L_t = 4$}}
\put(1648,51){\makebox(0,0){\raisebox{-1.5em}{$1/\beta^{0.65}$}}}
\put(100,914){%
\special{ps: gsave currentpoint currentpoint translate
270 rotate neg exch neg exch translate}%
\makebox(0,0)[b]{\shortstack{\raisebox{-1.5em}{$\alpha_{\rm eff}$}}}%
\special{ps: currentpoint grestore moveto}%
}
\put(2697,151){\makebox(0,0){0.25}}
\put(2278,151){\makebox(0,0){0.2}}
\put(1858,151){\makebox(0,0){0.15}}
\put(1439,151){\makebox(0,0){0.1}}
\put(1019,151){\makebox(0,0){0.05}}
\put(600,151){\makebox(0,0){0}}
\put(540,1577){\makebox(0,0)[r]{7}}
\put(540,1444){\makebox(0,0)[r]{6.8}}
\put(540,1312){\makebox(0,0)[r]{6.6}}
\put(540,1179){\makebox(0,0)[r]{6.4}}
\put(540,1047){\makebox(0,0)[r]{6.2}}
\put(540,914){\makebox(0,0)[r]{6}}
\put(540,781){\makebox(0,0)[r]{5.8}}
\put(540,649){\makebox(0,0)[r]{5.6}}
\put(540,516){\makebox(0,0)[r]{5.4}}
\put(540,384){\makebox(0,0)[r]{5.2}}
\put(540,251){\makebox(0,0)[r]{5}}
\end{picture}
\end{figure}

\begin{figure}[h]
\caption[11]{Ratio, at fixed $T$, of numerical and (leading-order) 
perturbative interface tensions for decreasing lattice
spacing ($a \equiv 1/L_t T$).}
\vskip 20pt
\setlength{\unitlength}{0.1bp}
\special{!
/gnudict 40 dict def
gnudict begin
/Color false def
/Solid false def
/gnulinewidth 5.000 def
/vshift -33 def
/dl {10 mul} def
/hpt 31.5 def
/vpt 31.5 def
/M {moveto} bind def
/L {lineto} bind def
/R {rmoveto} bind def
/V {rlineto} bind def
/vpt2 vpt 2 mul def
/hpt2 hpt 2 mul def
/Lshow { currentpoint stroke M
  0 vshift R show } def
/Rshow { currentpoint stroke M
  dup stringwidth pop neg vshift R show } def
/Cshow { currentpoint stroke M
  dup stringwidth pop -2 div vshift R show } def
/DL { Color {setrgbcolor Solid {pop []} if 0 setdash }
 {pop pop pop Solid {pop []} if 0 setdash} ifelse } def
/BL { stroke gnulinewidth 2 mul setlinewidth } def
/AL { stroke gnulinewidth 2 div setlinewidth } def
/PL { stroke gnulinewidth setlinewidth } def
/LTb { BL [] 0 0 0 DL } def
/LTa { AL [1 dl 2 dl] 0 setdash 0 0 0 setrgbcolor } def
/LT0 { PL [] 0 1 0 DL } def
/LT1 { PL [4 dl 2 dl] 0 0 1 DL } def
/LT2 { PL [2 dl 3 dl] 1 0 0 DL } def
/LT3 { PL [1 dl 1.5 dl] 1 0 1 DL } def
/LT4 { PL [5 dl 2 dl 1 dl 2 dl] 0 1 1 DL } def
/LT5 { PL [4 dl 3 dl 1 dl 3 dl] 1 1 0 DL } def
/LT6 { PL [2 dl 2 dl 2 dl 4 dl] 0 0 0 DL } def
/LT7 { PL [2 dl 2 dl 2 dl 2 dl 2 dl 4 dl] 1 0.3 0 DL } def
/LT8 { PL [2 dl 2 dl 2 dl 2 dl 2 dl 2 dl 2 dl 4 dl] 0.5 0.5 0.5 DL } def
/P { stroke [] 0 setdash
  currentlinewidth 2 div sub M
  0 currentlinewidth V stroke } def
/D { stroke [] 0 setdash 2 copy vpt add M
  hpt neg vpt neg V hpt vpt neg V
  hpt vpt V hpt neg vpt V closepath stroke
  P } def
/A { stroke [] 0 setdash vpt sub M 0 vpt2 V
  currentpoint stroke M
  hpt neg vpt neg R hpt2 0 V stroke
  } def
/B { stroke [] 0 setdash 2 copy exch hpt sub exch vpt add M
  0 vpt2 neg V hpt2 0 V 0 vpt2 V
  hpt2 neg 0 V closepath stroke
  P } def
/C { stroke [] 0 setdash exch hpt sub exch vpt add M
  hpt2 vpt2 neg V currentpoint stroke M
  hpt2 neg 0 R hpt2 vpt2 V stroke } def
/T { stroke [] 0 setdash 2 copy vpt 1.12 mul add M
  hpt neg vpt -1.62 mul V
  hpt 2 mul 0 V
  hpt neg vpt 1.62 mul V closepath stroke
  P  } def
/S { 2 copy A C} def
end
}
\begin{picture}(2519,1511)(0,0)
\special{"
gnudict begin
gsave
50 50 translate
0.100 0.100 scale
0 setgray
/Helvetica findfont 100 scalefont setfont
newpath
-500.000000 -500.000000 translate
LTa
600 251 M
1736 0 V
600 251 M
0 1209 V
LTb
600 251 M
63 0 V
1673 0 R
-63 0 V
600 412 M
63 0 V
1673 0 R
-63 0 V
600 573 M
63 0 V
1673 0 R
-63 0 V
600 735 M
63 0 V
1673 0 R
-63 0 V
600 896 M
63 0 V
1673 0 R
-63 0 V
600 1057 M
63 0 V
1673 0 R
-63 0 V
600 1218 M
63 0 V
1673 0 R
-63 0 V
600 1379 M
63 0 V
1673 0 R
-63 0 V
600 251 M
0 63 V
0 1146 R
0 -63 V
817 251 M
0 63 V
0 1146 R
0 -63 V
1034 251 M
0 63 V
0 1146 R
0 -63 V
1251 251 M
0 63 V
0 1146 R
0 -63 V
1468 251 M
0 63 V
0 1146 R
0 -63 V
1685 251 M
0 63 V
0 1146 R
0 -63 V
1902 251 M
0 63 V
0 1146 R
0 -63 V
2119 251 M
0 63 V
0 1146 R
0 -63 V
2336 251 M
0 63 V
0 1146 R
0 -63 V
600 251 M
1736 0 V
0 1209 V
-1736 0 V
600 251 L
LT0
1034 1142 D
1251 1123 D
1468 1128 D
1902 1119 D
1034 1139 M
0 6 V
-31 -6 R
62 0 V
-62 6 R
62 0 V
186 -35 R
0 27 V
-31 -27 R
62 0 V
-62 27 R
62 0 V
186 -28 R
0 38 V
-31 -38 R
62 0 V
-62 38 R
62 0 V
403 -60 R
0 64 V
-31 -64 R
62 0 V
-62 64 R
62 0 V
LT1
600 1057 M
18 0 V
17 0 V
18 0 V
17 0 V
18 0 V
17 0 V
18 0 V
17 0 V
18 0 V
17 0 V
18 0 V
17 0 V
18 0 V
17 0 V
18 0 V
18 0 V
17 0 V
18 0 V
17 0 V
18 0 V
17 0 V
18 0 V
17 0 V
18 0 V
17 0 V
18 0 V
17 0 V
18 0 V
18 0 V
17 0 V
18 0 V
17 0 V
18 0 V
17 0 V
18 0 V
17 0 V
18 0 V
17 0 V
18 0 V
17 0 V
18 0 V
17 0 V
18 0 V
18 0 V
17 0 V
18 0 V
17 0 V
18 0 V
17 0 V
18 0 V
17 0 V
18 0 V
17 0 V
18 0 V
17 0 V
18 0 V
18 0 V
17 0 V
18 0 V
17 0 V
18 0 V
17 0 V
18 0 V
17 0 V
18 0 V
17 0 V
18 0 V
17 0 V
18 0 V
17 0 V
18 0 V
18 0 V
17 0 V
18 0 V
17 0 V
18 0 V
17 0 V
18 0 V
17 0 V
18 0 V
17 0 V
18 0 V
17 0 V
18 0 V
18 0 V
17 0 V
18 0 V
17 0 V
18 0 V
17 0 V
18 0 V
17 0 V
18 0 V
17 0 V
18 0 V
17 0 V
18 0 V
17 0 V
18 0 V
stroke
grestore
end
showpage
}
\put(1468,51){\makebox(0,0){\raisebox{-1.5em}{$L_t$}}}
\put(100,855){%
\special{ps: gsave currentpoint currentpoint translate
270 rotate neg exch neg exch translate}%
\makebox(0,0)[b]{\shortstack{\raisebox{-1.5em}{$\alpha_{\rm eff}/\alpha$}}}%
\special{ps: currentpoint grestore moveto}%
}
\put(2336,151){\makebox(0,0){8}}
\put(2119,151){\makebox(0,0){7}}
\put(1902,151){\makebox(0,0){6}}
\put(1685,151){\makebox(0,0){5}}
\put(1468,151){\makebox(0,0){4}}
\put(1251,151){\makebox(0,0){3}}
\put(1034,151){\makebox(0,0){2}}
\put(817,151){\makebox(0,0){1}}
\put(600,151){\makebox(0,0){0}}
\put(540,1379){\makebox(0,0)[r]{1.4}}
\put(540,1218){\makebox(0,0)[r]{1.2}}
\put(540,1057){\makebox(0,0)[r]{1}}
\put(540,896){\makebox(0,0)[r]{0.8}}
\put(540,735){\makebox(0,0)[r]{0.6}}
\put(540,573){\makebox(0,0)[r]{0.4}}
\put(540,412){\makebox(0,0)[r]{0.2}}
\put(540,251){\makebox(0,0)[r]{0}}
\end{picture}
\end{figure}
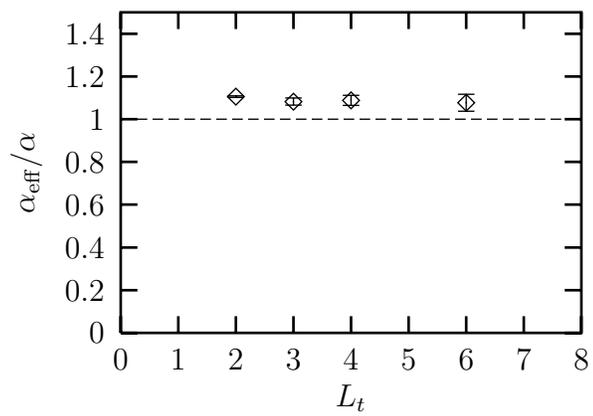

\begin{figure}[h]
\caption[12]{Profile of the Polyakov loop within the domain wall
(centred at $z=0$): numerical values (solid lines) and
perturbation theory (dashed lines).}
\vskip 20pt
\input{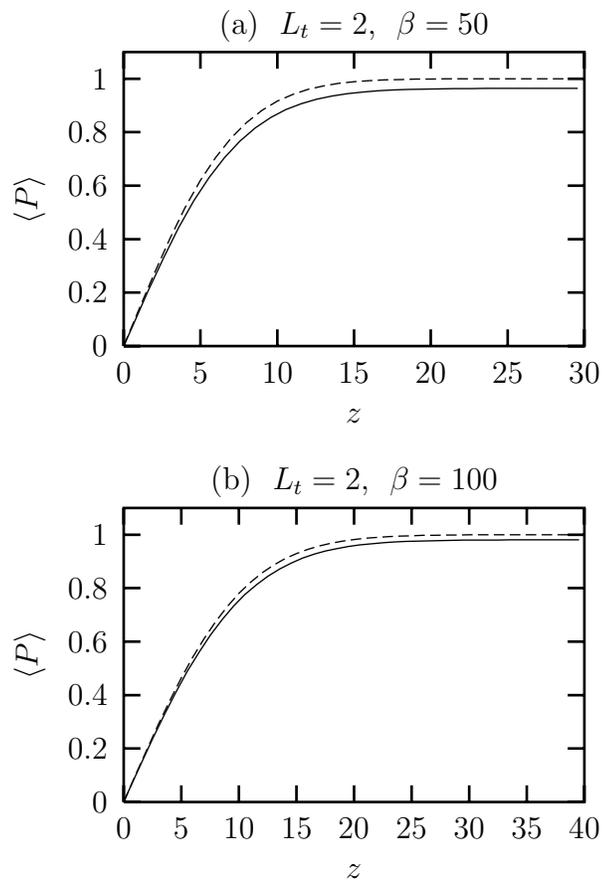}
\vskip 20pt
\setlength{\unitlength}{0.1bp}
\special{!
/gnudict 40 dict def
gnudict begin
/Color false def
/Solid false def
/gnulinewidth 5.000 def
/vshift -33 def
/dl {10 mul} def
/hpt 31.5 def
/vpt 31.5 def
/M {moveto} bind def
/L {lineto} bind def
/R {rmoveto} bind def
/V {rlineto} bind def
/vpt2 vpt 2 mul def
/hpt2 hpt 2 mul def
/Lshow { currentpoint stroke M
  0 vshift R show } def
/Rshow { currentpoint stroke M
  dup stringwidth pop neg vshift R show } def
/Cshow { currentpoint stroke M
  dup stringwidth pop -2 div vshift R show } def
/DL { Color {setrgbcolor Solid {pop []} if 0 setdash }
 {pop pop pop Solid {pop []} if 0 setdash} ifelse } def
/BL { stroke gnulinewidth 2 mul setlinewidth } def
/AL { stroke gnulinewidth 2 div setlinewidth } def
/PL { stroke gnulinewidth setlinewidth } def
/LTb { BL [] 0 0 0 DL } def
/LTa { AL [1 dl 2 dl] 0 setdash 0 0 0 setrgbcolor } def
/LT0 { PL [] 0 1 0 DL } def
/LT1 { PL [4 dl 2 dl] 0 0 1 DL } def
/LT2 { PL [2 dl 3 dl] 1 0 0 DL } def
/LT3 { PL [1 dl 1.5 dl] 1 0 1 DL } def
/LT4 { PL [5 dl 2 dl 1 dl 2 dl] 0 1 1 DL } def
/LT5 { PL [4 dl 3 dl 1 dl 3 dl] 1 1 0 DL } def
/LT6 { PL [2 dl 2 dl 2 dl 4 dl] 0 0 0 DL } def
/LT7 { PL [2 dl 2 dl 2 dl 2 dl 2 dl 4 dl] 1 0.3 0 DL } def
/LT8 { PL [2 dl 2 dl 2 dl 2 dl 2 dl 2 dl 2 dl 4 dl] 0.5 0.5 0.5 DL } def
/P { stroke [] 0 setdash
  currentlinewidth 2 div sub M
  0 currentlinewidth V stroke } def
/D { stroke [] 0 setdash 2 copy vpt add M
  hpt neg vpt neg V hpt vpt neg V
  hpt vpt V hpt neg vpt V closepath stroke
  P } def
/A { stroke [] 0 setdash vpt sub M 0 vpt2 V
  currentpoint stroke M
  hpt neg vpt neg R hpt2 0 V stroke
  } def
/B { stroke [] 0 setdash 2 copy exch hpt sub exch vpt add M
  0 vpt2 neg V hpt2 0 V 0 vpt2 V
  hpt2 neg 0 V closepath stroke
  P } def
/C { stroke [] 0 setdash exch hpt sub exch vpt add M
  hpt2 vpt2 neg V currentpoint stroke M
  hpt2 neg 0 R hpt2 vpt2 V stroke } def
/T { stroke [] 0 setdash 2 copy vpt 1.12 mul add M
  hpt neg vpt -1.62 mul V
  hpt 2 mul 0 V
  hpt neg vpt 1.62 mul V closepath stroke
  P  } def
/S { 2 copy A C} def
end
}
\begin{picture}(2519,1511)(0,0)
\special{"
gnudict begin
gsave
50 50 translate
0.100 0.100 scale
0 setgray
/Helvetica findfont 100 scalefont setfont
newpath
-500.000000 -500.000000 translate
LTa
600 251 M
1736 0 V
600 251 M
0 1109 V
LTb
600 251 M
63 0 V
1673 0 R
-63 0 V
600 453 M
63 0 V
1673 0 R
-63 0 V
600 654 M
63 0 V
1673 0 R
-63 0 V
600 856 M
63 0 V
1673 0 R
-63 0 V
600 1058 M
63 0 V
1673 0 R
-63 0 V
600 1259 M
63 0 V
1673 0 R
-63 0 V
600 251 M
0 63 V
0 1046 R
0 -63 V
817 251 M
0 63 V
0 1046 R
0 -63 V
1034 251 M
0 63 V
0 1046 R
0 -63 V
1251 251 M
0 63 V
0 1046 R
0 -63 V
1468 251 M
0 63 V
0 1046 R
0 -63 V
1685 251 M
0 63 V
0 1046 R
0 -63 V
1902 251 M
0 63 V
0 1046 R
0 -63 V
2119 251 M
0 63 V
0 1046 R
0 -63 V
2336 251 M
0 63 V
0 1046 R
0 -63 V
600 251 M
1736 0 V
0 1109 V
-1736 0 V
600 251 L
LT0
600 251 M
23 52 V
44 96 V
43 93 V
44 90 V
43 84 V
43 79 V
44 71 V
43 65 V
44 58 V
43 51 V
43 44 V
44 37 V
43 32 V
44 27 V
43 22 V
43 19 V
44 15 V
43 12 V
44 9 V
43 8 V
43 6 V
44 4 V
43 4 V
44 3 V
43 2 V
43 1 V
44 1 V
43 1 V
44 1 V
43 1 V
43 0 V
44 0 V
43 0 V
44 1 V
43 0 V
43 0 V
44 0 V
43 0 V
44 0 V
43 0 V
LT1
600 251 M
43 99 V
44 98 V
43 95 V
44 91 V
43 85 V
43 78 V
44 72 V
43 63 V
44 57 V
43 49 V
43 41 V
44 36 V
43 29 V
44 25 V
43 19 V
43 16 V
44 13 V
43 10 V
44 8 V
43 6 V
43 5 V
44 3 V
43 3 V
44 2 V
43 1 V
43 2 V
44 0 V
43 1 V
44 0 V
43 1 V
43 0 V
44 0 V
43 0 V
44 0 V
43 0 V
43 0 V
44 0 V
43 0 V
44 0 V
43 0 V
stroke
grestore
end
showpage
}
\put(1468,1460){\makebox(0,0){(b) \  $L_t=2$, \  $\beta=100$}}
\put(1468,51){\makebox(0,0){\raisebox{-1.5em}{$z$}}}
\put(100,805){%
\special{ps: gsave currentpoint currentpoint translate
270 rotate neg exch neg exch translate}%
\makebox(0,0)[b]{\shortstack{\raisebox{-1.5em}{$\langle P\rangle$}}}%
\special{ps: currentpoint grestore moveto}%
}
\put(2336,151){\makebox(0,0){40}}
\put(2119,151){\makebox(0,0){35}}
\put(1902,151){\makebox(0,0){30}}
\put(1685,151){\makebox(0,0){25}}
\put(1468,151){\makebox(0,0){20}}
\put(1251,151){\makebox(0,0){15}}
\put(1034,151){\makebox(0,0){10}}
\put(817,151){\makebox(0,0){5}}
\put(600,151){\makebox(0,0){0}}
\put(540,1259){\makebox(0,0)[r]{1}}
\put(540,1058){\makebox(0,0)[r]{0.8}}
\put(540,856){\makebox(0,0)[r]{0.6}}
\put(540,654){\makebox(0,0)[r]{0.4}}
\put(540,453){\makebox(0,0)[r]{0.2}}
\put(540,251){\makebox(0,0)[r]{0}}
\end{picture}
\end{figure}

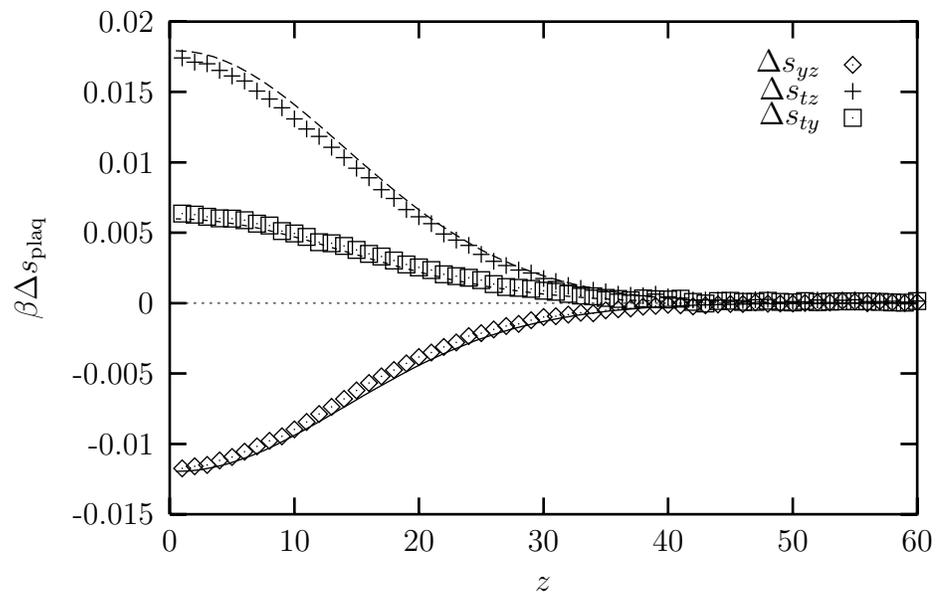
\begin{figure}[h]
\caption[13]{Vacuum-subtracted local action densities within the domain
wall: numerical calculations at $\beta=99.97$ and $L_t=4$,
compared with perturbation theory (lines).}
\vskip 20pt
\setlength{\unitlength}{0.1bp}
\special{!
/gnudict 40 dict def
gnudict begin
/Color false def
/Solid false def
/gnulinewidth 5.000 def
/vshift -33 def
/dl {10 mul} def
/hpt 31.5 def
/vpt 31.5 def
/M {moveto} bind def
/L {lineto} bind def
/R {rmoveto} bind def
/V {rlineto} bind def
/vpt2 vpt 2 mul def
/hpt2 hpt 2 mul def
/Lshow { currentpoint stroke M
  0 vshift R show } def
/Rshow { currentpoint stroke M
  dup stringwidth pop neg vshift R show } def
/Cshow { currentpoint stroke M
  dup stringwidth pop -2 div vshift R show } def
/DL { Color {setrgbcolor Solid {pop []} if 0 setdash }
 {pop pop pop Solid {pop []} if 0 setdash} ifelse } def
/BL { stroke gnulinewidth 2 mul setlinewidth } def
/AL { stroke gnulinewidth 2 div setlinewidth } def
/PL { stroke gnulinewidth setlinewidth } def
/LTb { BL [] 0 0 0 DL } def
/LTa { AL [1 dl 2 dl] 0 setdash 0 0 0 setrgbcolor } def
/LT0 { PL [] 0 1 0 DL } def
/LT1 { PL [4 dl 2 dl] 0 0 1 DL } def
/LT2 { PL [2 dl 3 dl] 1 0 0 DL } def
/LT3 { PL [1 dl 1.5 dl] 1 0 1 DL } def
/LT4 { PL [5 dl 2 dl 1 dl 2 dl] 0 1 1 DL } def
/LT5 { PL [4 dl 3 dl 1 dl 3 dl] 1 1 0 DL } def
/LT6 { PL [2 dl 2 dl 2 dl 4 dl] 0 0 0 DL } def
/LT7 { PL [2 dl 2 dl 2 dl 2 dl 2 dl 4 dl] 1 0.3 0 DL } def
/LT8 { PL [2 dl 2 dl 2 dl 2 dl 2 dl 2 dl 2 dl 4 dl] 0.5 0.5 0.5 DL } def
/P { stroke [] 0 setdash
  currentlinewidth 2 div sub M
  0 currentlinewidth V stroke } def
/D { stroke [] 0 setdash 2 copy vpt add M
  hpt neg vpt neg V hpt vpt neg V
  hpt vpt V hpt neg vpt V closepath stroke
  P } def
/A { stroke [] 0 setdash vpt sub M 0 vpt2 V
  currentpoint stroke M
  hpt neg vpt neg R hpt2 0 V stroke
  } def
/B { stroke [] 0 setdash 2 copy exch hpt sub exch vpt add M
  0 vpt2 neg V hpt2 0 V 0 vpt2 V
  hpt2 neg 0 V closepath stroke
  P } def
/C { stroke [] 0 setdash exch hpt sub exch vpt add M
  hpt2 vpt2 neg V currentpoint stroke M
  hpt2 neg 0 R hpt2 vpt2 V stroke } def
/T { stroke [] 0 setdash 2 copy vpt 1.12 mul add M
  hpt neg vpt -1.62 mul V
  hpt 2 mul 0 V
  hpt neg vpt 1.62 mul V closepath stroke
  P  } def
/S { 2 copy A C} def
end
}
\begin{picture}(3600,2160)(0,0)
\special{"
gnudict begin
gsave
50 50 translate
0.100 0.100 scale
0 setgray
/Helvetica findfont 100 scalefont setfont
newpath
-500.000000 -500.000000 translate
LTa
600 1047 M
2817 0 V
600 251 M
0 1858 V
LTb
600 251 M
63 0 V
2754 0 R
-63 0 V
600 516 M
63 0 V
2754 0 R
-63 0 V
600 782 M
63 0 V
2754 0 R
-63 0 V
600 1047 M
63 0 V
2754 0 R
-63 0 V
600 1313 M
63 0 V
2754 0 R
-63 0 V
600 1578 M
63 0 V
2754 0 R
-63 0 V
600 1844 M
63 0 V
2754 0 R
-63 0 V
600 2109 M
63 0 V
2754 0 R
-63 0 V
600 251 M
0 63 V
0 1795 R
0 -63 V
1070 251 M
0 63 V
0 1795 R
0 -63 V
1539 251 M
0 63 V
0 1795 R
0 -63 V
2009 251 M
0 63 V
0 1795 R
0 -63 V
2478 251 M
0 63 V
0 1795 R
0 -63 V
2948 251 M
0 63 V
0 1795 R
0 -63 V
3417 251 M
0 63 V
0 1795 R
0 -63 V
600 251 M
2817 0 V
0 1858 V
-2817 0 V
600 251 L
LT0
623 413 M
47 1 V
47 5 V
47 9 V
47 11 V
47 15 V
47 17 V
47 19 V
47 22 V
47 24 V
47 25 V
47 26 V
47 27 V
47 28 V
47 27 V
47 28 V
47 27 V
47 26 V
47 26 V
47 25 V
46 23 V
47 22 V
47 21 V
47 19 V
47 19 V
47 16 V
47 15 V
47 14 V
47 12 V
47 11 V
47 10 V
47 9 V
47 8 V
47 7 V
47 6 V
47 6 V
47 4 V
47 4 V
47 4 V
47 3 V
46 2 V
47 2 V
47 2 V
47 1 V
47 1 V
47 1 V
47 1 V
47 1 V
47 0 V
47 1 V
47 0 V
47 0 V
47 0 V
47 1 V
47 0 V
47 0 V
47 0 V
47 0 V
47 0 V
LT1
3174 1946 D
647 424 D
694 432 D
741 437 D
788 454 D
835 467 D
882 487 D
929 507 D
976 528 D
1023 545 D
1070 571 D
1116 599 D
1163 629 D
1210 656 D
1257 686 D
1304 718 D
1351 746 D
1398 771 D
1445 795 D
1492 820 D
1539 844 D
1586 861 D
1633 882 D
1680 899 D
1727 923 D
1774 934 D
1821 948 D
1868 961 D
1915 969 D
1962 983 D
2009 995 D
2055 998 D
2102 1004 D
2149 1011 D
2196 1009 D
2243 1020 D
2290 1024 D
2337 1027 D
2384 1034 D
2431 1037 D
2478 1041 D
2525 1039 D
2572 1035 D
2619 1039 D
2666 1042 D
2713 1044 D
2760 1044 D
2807 1050 D
2854 1045 D
2901 1046 D
2948 1046 D
2994 1048 D
3041 1050 D
3088 1050 D
3135 1056 D
3182 1053 D
3229 1047 D
3276 1054 D
3323 1050 D
3370 1048 D
3417 1051 D
LT1
623 1999 M
47 -2 V
47 -8 V
47 -12 V
47 -18 V
47 -21 V
47 -26 V
47 -29 V
47 -33 V
47 -35 V
47 -38 V
47 -39 V
47 -41 V
47 -41 V
47 -42 V
47 -41 V
47 -40 V
47 -40 V
47 -39 V
47 -37 V
46 -35 V
47 -33 V
47 -31 V
47 -29 V
47 -28 V
47 -25 V
47 -22 V
47 -21 V
47 -18 V
47 -17 V
47 -15 V
47 -13 V
47 -12 V
47 -10 V
47 -10 V
47 -8 V
47 -7 V
47 -6 V
47 -5 V
47 -4 V
46 -4 V
47 -3 V
47 -3 V
47 -2 V
47 -1 V
47 -2 V
47 -1 V
47 -1 V
47 -1 V
47 0 V
47 -1 V
47 0 V
47 0 V
47 -1 V
47 0 V
47 0 V
47 0 V
47 0 V
47 0 V
LT3
3174 1846 A
647 1972 A
694 1956 A
741 1950 A
788 1925 A
835 1904 A
882 1885 A
929 1847 A
976 1817 A
1023 1784 A
1070 1742 A
1116 1704 A
1163 1676 A
1210 1635 A
1257 1596 A
1304 1556 A
1351 1520 A
1398 1475 A
1445 1442 A
1492 1400 A
1539 1373 A
1586 1347 A
1633 1308 A
1680 1285 A
1727 1265 A
1774 1231 A
1821 1205 A
1868 1189 A
1915 1172 A
1962 1161 A
2009 1140 A
2055 1129 A
2102 1114 A
2149 1103 A
2196 1096 A
2243 1090 A
2290 1081 A
2337 1077 A
2384 1075 A
2431 1081 A
2478 1071 A
2525 1066 A
2572 1059 A
2619 1059 A
2666 1059 A
2713 1058 A
2760 1057 A
2807 1060 A
2854 1061 A
2901 1050 A
2948 1053 A
2994 1052 A
3041 1059 A
3088 1049 A
3135 1058 A
3182 1060 A
3229 1053 A
3276 1053 A
3323 1054 A
3370 1051 A
3417 1055 A
LT2
623 1365 M
47 -1 V
47 -3 V
47 -4 V
47 -6 V
47 -7 V
47 -8 V
47 -10 V
47 -11 V
47 -12 V
47 -12 V
47 -13 V
47 -14 V
47 -14 V
47 -14 V
47 -13 V
47 -14 V
47 -13 V
47 -13 V
47 -12 V
46 -12 V
47 -11 V
47 -11 V
47 -9 V
47 -9 V
47 -9 V
47 -7 V
47 -7 V
47 -6 V
47 -6 V
47 -5 V
47 -4 V
47 -4 V
47 -4 V
47 -3 V
47 -2 V
47 -3 V
47 -2 V
47 -1 V
47 -2 V
46 -1 V
47 -1 V
47 -1 V
47 -1 V
47 0 V
47 -1 V
47 0 V
47 0 V
47 -1 V
47 0 V
47 0 V
47 0 V
47 0 V
47 0 V
47 0 V
47 0 V
47 -1 V
47 0 V
47 0 V
LT5
3174 1746 B
647 1385 B
694 1381 B
741 1373 B
788 1367 B
835 1366 B
882 1360 B
929 1347 B
976 1340 B
1023 1319 B
1070 1310 B
1116 1297 B
1163 1277 B
1210 1273 B
1257 1263 B
1304 1248 B
1351 1233 B
1398 1223 B
1445 1208 B
1492 1193 B
1539 1182 B
1586 1172 B
1633 1156 B
1680 1149 B
1727 1143 B
1774 1134 B
1821 1119 B
1868 1108 B
1915 1105 B
1962 1101 B
2009 1094 B
2055 1090 B
2102 1082 B
2149 1076 B
2196 1073 B
2243 1067 B
2290 1063 B
2337 1058 B
2384 1065 B
2431 1062 B
2478 1061 B
2525 1064 B
2572 1052 B
2619 1046 B
2666 1055 B
2713 1056 B
2760 1052 B
2807 1052 B
2854 1056 B
2901 1052 B
2948 1050 B
2994 1055 B
3041 1051 B
3088 1053 B
3135 1054 B
3182 1054 B
3229 1054 B
3276 1052 B
3323 1050 B
3370 1049 B
3417 1055 B
stroke
grestore
end
showpage
}
\put(3054,1746){\makebox(0,0)[r]{$\Delta s_{ty}$}}
\put(3054,1846){\makebox(0,0)[r]{$\Delta s_{tz}$}}
\put(3054,1946){\makebox(0,0)[r]{$\Delta s_{yz}$}}
\put(2008,51){\makebox(0,0){\raisebox{-1.5em}{$z$}}}
\put(100,1180){%
\special{ps: gsave currentpoint currentpoint translate
270 rotate neg exch neg exch translate}%
\makebox(0,0)[b]{\shortstack{$\beta\Delta s_{\rm plaq}$}}%
\special{ps: currentpoint grestore moveto}%
}
\put(3417,151){\makebox(0,0){60}}
\put(2948,151){\makebox(0,0){50}}
\put(2478,151){\makebox(0,0){40}}
\put(2009,151){\makebox(0,0){30}}
\put(1539,151){\makebox(0,0){20}}
\put(1070,151){\makebox(0,0){10}}
\put(600,151){\makebox(0,0){0}}
\put(540,2109){\makebox(0,0)[r]{0.02}}
\put(540,1844){\makebox(0,0)[r]{0.015}}
\put(540,1578){\makebox(0,0)[r]{0.01}}
\put(540,1313){\makebox(0,0)[r]{0.005}}
\put(540,1047){\makebox(0,0)[r]{0}}
\put(540,782){\makebox(0,0)[r]{-0.005}}
\put(540,516){\makebox(0,0)[r]{-0.01}}
\put(540,251){\makebox(0,0)[r]{-0.015}}
\end{picture}
\end{figure}

\begin{figure}[h]
\caption[14]{Numerically calculated values of the Debye mass,
compared to D'Hoker's self-consistent perturbative prediction.}
\vskip 20pt
\setlength{\unitlength}{0.1bp}
\special{!
/gnudict 40 dict def
gnudict begin
/Color false def
/Solid false def
/gnulinewidth 5.000 def
/vshift -33 def
/dl {10 mul} def
/hpt 31.5 def
/vpt 31.5 def
/M {moveto} bind def
/L {lineto} bind def
/R {rmoveto} bind def
/V {rlineto} bind def
/vpt2 vpt 2 mul def
/hpt2 hpt 2 mul def
/Lshow { currentpoint stroke M
  0 vshift R show } def
/Rshow { currentpoint stroke M
  dup stringwidth pop neg vshift R show } def
/Cshow { currentpoint stroke M
  dup stringwidth pop -2 div vshift R show } def
/DL { Color {setrgbcolor Solid {pop []} if 0 setdash }
 {pop pop pop Solid {pop []} if 0 setdash} ifelse } def
/BL { stroke gnulinewidth 2 mul setlinewidth } def
/AL { stroke gnulinewidth 2 div setlinewidth } def
/PL { stroke gnulinewidth setlinewidth } def
/LTb { BL [] 0 0 0 DL } def
/LTa { AL [1 dl 2 dl] 0 setdash 0 0 0 setrgbcolor } def
/LT0 { PL [] 0 1 0 DL } def
/LT1 { PL [4 dl 2 dl] 0 0 1 DL } def
/LT2 { PL [2 dl 3 dl] 1 0 0 DL } def
/LT3 { PL [1 dl 1.5 dl] 1 0 1 DL } def
/LT4 { PL [5 dl 2 dl 1 dl 2 dl] 0 1 1 DL } def
/LT5 { PL [4 dl 3 dl 1 dl 3 dl] 1 1 0 DL } def
/LT6 { PL [2 dl 2 dl 2 dl 4 dl] 0 0 0 DL } def
/LT7 { PL [2 dl 2 dl 2 dl 2 dl 2 dl 4 dl] 1 0.3 0 DL } def
/LT8 { PL [2 dl 2 dl 2 dl 2 dl 2 dl 2 dl 2 dl 4 dl] 0.5 0.5 0.5 DL } def
/P { stroke [] 0 setdash
  currentlinewidth 2 div sub M
  0 currentlinewidth V stroke } def
/D { stroke [] 0 setdash 2 copy vpt add M
  hpt neg vpt neg V hpt vpt neg V
  hpt vpt V hpt neg vpt V closepath stroke
  P } def
/A { stroke [] 0 setdash vpt sub M 0 vpt2 V
  currentpoint stroke M
  hpt neg vpt neg R hpt2 0 V stroke
  } def
/B { stroke [] 0 setdash 2 copy exch hpt sub exch vpt add M
  0 vpt2 neg V hpt2 0 V 0 vpt2 V
  hpt2 neg 0 V closepath stroke
  P } def
/C { stroke [] 0 setdash exch hpt sub exch vpt add M
  hpt2 vpt2 neg V currentpoint stroke M
  hpt2 neg 0 R hpt2 vpt2 V stroke } def
/T { stroke [] 0 setdash 2 copy vpt 1.12 mul add M
  hpt neg vpt -1.62 mul V
  hpt 2 mul 0 V
  hpt neg vpt 1.62 mul V closepath stroke
  P  } def
/S { 2 copy A C} def
end
}
\begin{picture}(3239,1943)(0,0)
\special{"
gnudict begin
gsave
50 50 translate
0.100 0.100 scale
0 setgray
/Helvetica findfont 100 scalefont setfont
newpath
-500.000000 -500.000000 translate
LTa
600 251 M
2456 0 V
600 251 M
0 1641 V
LTb
600 251 M
63 0 V
2393 0 R
-63 0 V
600 485 M
63 0 V
2393 0 R
-63 0 V
600 720 M
63 0 V
2393 0 R
-63 0 V
600 954 M
63 0 V
2393 0 R
-63 0 V
600 1189 M
63 0 V
2393 0 R
-63 0 V
600 1423 M
63 0 V
2393 0 R
-63 0 V
600 1658 M
63 0 V
2393 0 R
-63 0 V
600 1892 M
63 0 V
2393 0 R
-63 0 V
600 251 M
0 63 V
0 1578 R
0 -63 V
907 251 M
0 63 V
0 1578 R
0 -63 V
1214 251 M
0 63 V
0 1578 R
0 -63 V
1521 251 M
0 63 V
0 1578 R
0 -63 V
1828 251 M
0 63 V
0 1578 R
0 -63 V
2135 251 M
0 63 V
0 1578 R
0 -63 V
2442 251 M
0 63 V
0 1578 R
0 -63 V
2749 251 M
0 63 V
0 1578 R
0 -63 V
3056 251 M
0 63 V
0 1578 R
0 -63 V
600 251 M
2456 0 V
0 1641 V
-2456 0 V
600 251 L
LT0
2562 1247 D
2519 1705 D
2039 1613 D
1559 1495 D
1080 1267 D
888 1134 D
734 861 D
2502 1247 M
180 0 V
-180 31 R
0 -62 V
180 62 R
0 -62 V
-163 443 R
0 92 V
-31 -92 R
62 0 V
-62 92 R
62 0 V
2039 1591 M
0 43 V
-31 -43 R
62 0 V
-62 43 R
62 0 V
1559 1477 M
0 35 V
-31 -35 R
62 0 V
-62 35 R
62 0 V
1080 1262 M
0 11 V
-31 -11 R
62 0 V
-62 11 R
62 0 V
888 1127 M
0 15 V
-31 -15 R
62 0 V
-62 15 R
62 0 V
734 853 M
0 15 V
703 853 M
62 0 V
-62 15 R
62 0 V
LT1
2562 1147 A
2039 1461 A
1559 1340 A
1079 1189 A
887 1009 A
731 791 A
2502 1147 M
180 0 V
-180 31 R
0 -62 V
180 62 R
0 -62 V
-643 315 R
0 59 V
-31 -59 R
62 0 V
-62 59 R
62 0 V
1559 1312 M
0 55 V
-31 -55 R
62 0 V
-62 55 R
62 0 V
1079 1173 M
0 32 V
-31 -32 R
62 0 V
-62 32 R
62 0 V
887 987 M
0 44 V
856 987 M
62 0 V
-62 44 R
62 0 V
731 785 M
0 12 V
700 785 M
62 0 V
-62 12 R
62 0 V
LT2
2562 1047 B
1559 1305 B
1079 1110 B
887 939 B
732 704 B
2502 1047 M
180 0 V
-180 31 R
0 -62 V
180 62 R
0 -62 V
1559 1291 M
0 28 V
-31 -28 R
62 0 V
-62 28 R
62 0 V
1079 1102 M
0 16 V
-31 -16 R
62 0 V
-62 16 R
62 0 V
887 931 M
0 16 V
856 931 M
62 0 V
-62 16 R
62 0 V
732 694 M
0 19 V
701 694 M
62 0 V
-62 19 R
62 0 V
LT3
2562 947 C
887 930 C
2502 947 M
180 0 V
-180 31 R
0 -62 V
180 62 R
0 -62 V
887 915 M
0 30 V
856 915 M
62 0 V
-62 30 R
62 0 V
LT4
2562 847 T
1080 1034 T
2502 847 M
180 0 V
-180 31 R
0 -62 V
180 62 R
0 -62 V
1080 1017 M
0 33 V
-31 -33 R
62 0 V
-62 33 R
62 0 V
LT1
2502 747 M
180 0 V
612 263 M
14 11 V
15 12 V
17 12 V
20 12 V
21 11 V
24 12 V
27 12 V
30 11 V
33 12 V
36 12 V
40 12 V
45 11 V
49 12 V
54 12 V
59 12 V
65 11 V
72 12 V
78 12 V
86 11 V
94 12 V
103 12 V
113 12 V
123 11 V
134 12 V
147 12 V
159 11 V
175 12 V
189 12 V
207 12 V
stroke
grestore
end
showpage
}
\put(2442,747){\makebox(0,0)[r]{D'Hoker}}
\put(2442,847){\makebox(0,0)[r]{$L_t=6$}}
\put(2442,947){\makebox(0,0)[r]{$L_t=5$}}
\put(2442,1047){\makebox(0,0)[r]{$L_t=4$}}
\put(2442,1147){\makebox(0,0)[r]{$L_t=3$}}
\put(2442,1247){\makebox(0,0)[r]{$L_t=2$}}
\put(1828,51){\makebox(0,0){\raisebox{-1.5em}{$T/g^2$}}}
\put(100,1071){%
\special{ps: gsave currentpoint currentpoint translate
270 rotate neg exch neg exch translate}%
\makebox(0,0)[b]{\shortstack{\raisebox{-1.5em}{$m_D^2/g^2 T$}}}%
\special{ps: currentpoint grestore moveto}%
}
\put(3056,151){\makebox(0,0){16}}
\put(2749,151){\makebox(0,0){14}}
\put(2442,151){\makebox(0,0){12}}
\put(2135,151){\makebox(0,0){10}}
\put(1828,151){\makebox(0,0){8}}
\put(1521,151){\makebox(0,0){6}}
\put(1214,151){\makebox(0,0){4}}
\put(907,151){\makebox(0,0){2}}
\put(600,151){\makebox(0,0){0}}
\put(540,1892){\makebox(0,0)[r]{1.4}}
\put(540,1658){\makebox(0,0)[r]{1.2}}
\put(540,1423){\makebox(0,0)[r]{1}}
\put(540,1189){\makebox(0,0)[r]{0.8}}
\put(540,954){\makebox(0,0)[r]{0.6}}
\put(540,720){\makebox(0,0)[r]{0.4}}
\put(540,485){\makebox(0,0)[r]{0.2}}
\put(540,251){\makebox(0,0)[r]{0}}
\end{picture}
\end{figure}
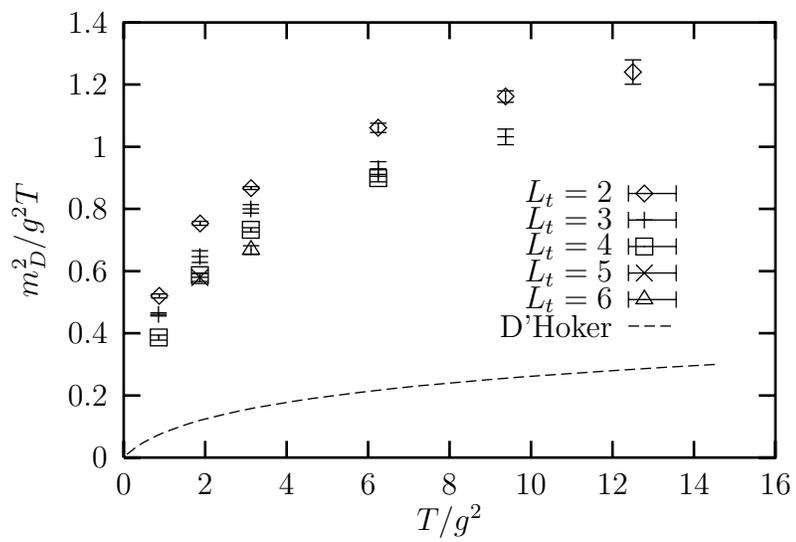



\newpage

\begin{thebibliography}{99}

\bibitem{Su79} L.Susskind, Phys. Rev. {\bf D20} (1979) 2610.

\bibitem{Po78} A.M.Polyakov, Phys. Lett. {\bf 72B} (1978) 477.

\bibitem{BhGo91B} T.Bhattacharya, A.Gocksch, C.P.Korthals Altes 
and R.D. Pisarski, Phys. Rev. Lett. {\bf 66} (1991) 998.

\bibitem{BeKo92} V.M.Belyaev, I.I.Kogan, G.W.Semenoff and N.Weiss, 
Phys. Lett. {\bf B277} (1992) 331.

\bibitem{LeKo94} 
		K.Lee, C.P.Korthals Altes and R.D.Pisarski, 
		Phys. Rev. Lett. {\bf 73} (1994) 1754;
		C.P.Korthals Altes and N.Watson, 
		Phys. Rev. Lett. {\bf 75} (1995) 2799.

\bibitem{KoMi95} C.P.Korthals Altes, A.Michels, M.Stephanov and M.Teper,
Nucl. Phys. (Proc. Suppl.) {\bf B42} (1995) 517, and A.Michels, D.Phil.
Thesis, Oxford, 1995.

\bibitem{WeWh96} S.T. West and J.F. Wheater, Oxford preprint OUTP-96-21P.

\bibitem{KaKa91} K.Kajantie, L.Karkainen and K.Rummukainen, 
Nucl. Phys. {\bf B357} (1991) 693.

\bibitem{Sm94} A. Smilga, Ann. of Phys. {\bf 234} (1994) 1.

\bibitem{tH79} G.'t Hooft, Nucl. Phys. {\bf B153} (1979) 141.

\bibitem{GoSh82}  Y. Y. Goldschmidt and J. Shigemitsu, 
			Nucl. Phys. {\bf B200}[FS4] (1982) 149.

\bibitem{BhGo92} T.Bhattacharya, A.Gocksch, C.P.Korthals Altes and R.D.
Pisarski, Nucl. Phys. {\bf B383} (1992) 497.

\bibitem{BhGo91} T. Bhattacharya, A.Gocksch, C.P.Korthals Altes 
and R.D. Pisarski, Nucl. Phys. (Proc. Suppl.) {B20} (1991) 305.

\bibitem{Ki95} 
		J.Kiskis, 
		Phys. Rev. {\bf D51} (1995) 3781, preprint UCD 95-14.

\bibitem{Ko94} C.P.Korthals Altes, Nucl. Phys. {\bf B420} (1994) 637.

\bibitem{We81} N. Weiss, Phys. Rev. {\bf D24} (1981) 475. 

\bibitem{GrPi81} D.J.Gross, R.D.Pisarski and L.G.Yaffe, Rev. Mod. Phys. 
{\bf 53} (1981) 43.

\bibitem{DH82} E. D'Hoker, Nucl. Phys. {\bf B201} (1982) 401.

\bibitem{Wi74} K.G.Wilson, Phys. Rev. {\bf D10} (1974) 2445.

\bibitem{KaRu91}
		L. K\"arkk\"ainen and K. Rummukainen,
		Nucl. Phys. B (Proc. Suppl.) {\bf 20} (1991) 309.

\bibitem{Lu81} 
		M. L\"uscher,
		Nucl. Phys. {\bf B180} (1981) 317.

\bibitem{StSt81} J. Stack and M. Stone, Phys. Lett. {\bf 100B} (1981)476.

\bibitem{GJA} J. Groeneveld, J. Jurkiewicz, and C. Korthals Altes,
		 Physica Scripta {\bf 23} (1981) 1022.

\bibitem{tH78} G.'t Hooft, Nucl. Phys. {\bf B138} (1978) 1.

\bibitem{Tep1} M. Teper, Phys. Lett. {\bf B289} (1992) 115, 
		{\bf B311} (1993) 223, {\bf B313} (1993) 417.

\bibitem{arnold} P. Arnold, L. Yaffe, Phys. Rev. {\bf D49} (1994) 3003. 


\end{thebibliography}
\end{document}